\documentclass[a4paper,twoside,11pt]{book}

\makeatletter
\newcommand{\iffinal}[2][{}]{\ifx\draftversion\@undefined #2\else #1\fi}
\makeatother

\usepackage{t1enc}
\usepackage[latin1]{inputenc}
\usepackage[round,longnamesfirst]{natbib}
\usepackage[italian,french,spanish,english]{babel}

\usepackage{amsmath,amssymb,latexsym}

\usepackage{pslatex}

\usepackage{bm,mathrsfs}

\usepackage{epsfig,enumerate}

\usepackage{makeidx,float,index}
\usepackage[sl,ruled]{caption}

\usepackage[a4paper,twoside,scale={0.7,0.8}]{geometry}
\usepackage{fancyhdr}

\iffinal[{\usepackage[defenv,algflt]{santini}}]{\usepackage[defenv,algflt,final]{santini}}
\usepackage{mychapter}

\RCSfooter$Id: m.tex,v 2.5 1999/11/01 11:30:53 santini Exp $

\newcommand{\tc}{\tilde{c}}

\newcommand{\tp}{\tilde{p}}

\newcommand{\sA}{\ensuremath{\textup{\textsf{A}}}\xspace}
\newcommand{\sB}{\ensuremath{\textup{\textsf{B}}}\xspace}

\newcommand{\bP}{\ensuremath{\bm{P}}\xspace}
\newcommand{\bS}{\ensuremath{\bm{S}}\xspace}
\newcommand{\bT}{\ensuremath{\bm{T}}\xspace}
\newcommand{\bU}{\ensuremath{\bm{U}}\xspace}
\newcommand{\bG}{\ensuremath{\bm{G}}\xspace}
\newcommand{\bX}{\ensuremath{\bm{X}}\xspace}

\newcommand{\bx}{\ensuremath{\bm{x}}\xspace}
\newcommand{\by}{\ensuremath{\bm{y}}\xspace}
\newcommand{\ba}{\ensuremath{\bm{a}}\xspace}
\newcommand{\be}{\ensuremath{\bm{e}}\xspace}

\newcommand{\bN}{\ensuremath{\mathbf{N}}\xspace}
\newcommand{\bQ}{\ensuremath{\mathbf{Q}}\xspace}
\newcommand{\bR}{\ensuremath{\mathbf{R}}\xspace}
\newcommand{\bZ}{\ensuremath{\mathbf{Z}}\xspace}

\newcommand{\cA}{\ensuremath{\mathscr{A}}\xspace}
\newcommand{\cB}{\ensuremath{\mathscr{B}}\xspace}
\newcommand{\cC}{\ensuremath{\mathscr{C}}\xspace}
\newcommand{\cE}{\ensuremath{\mathscr{E}}\xspace}
\newcommand{\cF}{\ensuremath{\mathscr{F}}\xspace}
\newcommand{\cR}{\ensuremath{\mathscr{R}}\xspace}
\newcommand{\cS}{\ensuremath{\mathscr{S}}\xspace}

\newcommand{\bbM}{\ensuremath{\mathbb{M}}\xspace}

\newcommand{\ras}{\mbox{r.a.s.}\xspace}

\newcommand{\rec}{\mbox{r.e.c.}\xspace}
\newcommand{\urg}{\mbox{u.r.g.}\xspace}
\newcommand{\cf}{\mbox{c.f.}\xspace}
\newcommand{\ie}{\mbox{i.e.,}\xspace}

\newcommand{\cP}{\textmd{\textrm{\textup{P}}}\xspace}
\newcommand{\cFP}{\textmd{\textrm{\textup{FP}}}\xspace}
\newcommand{\cRP}{\textmd{\textrm{\textup{RP}}}\xspace}
\newcommand{\cNC}{\textmd{\textrm{\textup{NC}}}\xspace}
\newcommand{\cnP}{\textmd{\textrm{\textup{\ensuremath{\sharp}P}}}\xspace}
\newcommand{\cNP}{\textmd{\textrm{\textup{NP}}}\xspace}
\newcommand{\cNPO}{\textmd{\textrm{\textup{NPO}}}\xspace}
\newcommand{\cGLO}{\textmd{\textrm{\textup{GLO}}}\xspace}
\newcommand{\cEG}{\textmd{\textrm{\textup{EG}}}\xspace}
\newcommand{\cAPX}{\textmd{\textrm{\textup{APX}}}\xspace}
\newcommand{\cPTAS}{\textmd{\textrm{\textup{PTAS}}}\xspace}
\newcommand{\cFPTAS}{\textmd{\textrm{\textup{FPTAS}}}\xspace}
\newcommand{\cMaxNP}{\textmd{\textrm{\textup{MaxNP}}}\xspace}
\newcommand{\cMaxSNP}{\textmd{\textrm{\textup{MaxSNP}}}\xspace}

\newcommand{\mPrRAM}{\textmd{\textrm{\textup{PrRAM}}}\xspace}
\newcommand{\mRAM}{\textmd{\textrm{\textup{RAM}}}\xspace}
\newcommand{\mTM}{\textmd{\textrm{\textup{TM}}}\xspace}
\newcommand{\mNTM}{\textmd{\textrm{\textup{NTM}}}\xspace}
\newcommand{\moAuxPDA}{\textmd{\textrm{\textup{\ensuremath{1}-AuxPDA}}}\xspace}
\newcommand{\moNAuxPDA}{\textmd{\textrm{\textup{\ensuremath{1}-NAuxPDA}}}\xspace}
\newcommand{\moUAuxPDA}{\textmd{\textrm{\textup{\ensuremath{1}-UAuxPDA}}}\xspace}

\newcommand{\combs}[1]{\ensuremath{\langle \bm #1,|\cdot|\rangle}}

\newcommand{\sex}[1]{%
  \smallskip
  \begin{center}
    \parbox{.9\textwidth}{\small #1}
  \end{center}
  \smallskip}

\newenvironment{mytabular}[2]{%
  \begin{tabular}{lp{9cm}}
    \hline\\[-8pt]\textsl{#1}&\textsl{#2}\\[6pt]\hline\\[-8pt]}{%
    \\[6pt]\hline
  \end{tabular}}

\newcommand{\mycite}[4]{%
  \begin{otherlanguage}{#1}
    \iflanguage{spanish}{\shorthandon{"}}{}
    \iflanguage{french}{\shorthandon{:!?;}}{}
    \begin{quotation} 
      \emph{#2}
      \begin{flushright} 
        #3, \emph{#4}
      \end{flushright}
    \end{quotation}
  \end{otherlanguage}
  \shorthandoff{":!?;}
  \vspace{1cm}}

\newindex{aut}{adx}{and}{Citations Index}

\newcommand{\idx}[1]{#1\index{#1}}                   
\newcommand{\midx}[1]{\index{#1}}                    
\newcommand{\eidx}[1]{\emph{#1}\index{#1}}           

\newcommand{\mridx}[2]{\index{#1|see{#2}}}      
 
\newcommand{\sidx}[2]{#1\index{#2!#1}}               
\newcommand{\esidx}[2]{\emph{#1}\index{#2!#1}}       
\newcommand{\msidx}[2]{\index{#2!#1}}   

\newcommand{\edidx}[2]{\emph{#1}\index{#1!#2}}       
\newcommand{\mdidx}[2]{\index{#1!#2}}   

\newcommand{\irpR}{p@$p$-relation}
\newcommand{\irP}{polynomial}
\newcommand{\irNS}{neighborhood structure}
\newcommand{\irAC}{arithmetic circuit}
\newcommand{\irLS}{local search}
\newcommand{\irNPO}{NP optimization problem}
\newcommand{\irTl}{trace language}
\newcommand{\irCS}{combinatorial structure}
\newcommand{\irCFg}{context-free grammar}
\newcommand{\irCFl}{context-free language}
\newcommand{\irRAM}{random access machine}
\newcommand{\irPrRAM}{probabilistic random access machine}
\newcommand{\irTM}{Turing machine}

\newcommand{\ireTM}{extended Turing machine}
\newcommand{\irAuxPDA}{one-way auxiliary pushdown automata}
\newcommand{\ireAuxPDA}{extended one-way auxiliary pushdown automata}

\DeclareMathSymbol{\ew}{\mathord}{letters}{"0F}
\renewcommand{\epsilon}{\varepsilon}
\renewcommand{\emptyset}{\varnothing}
\newcommand{\sep}{\lozenge}

\newcommand{\lexa}{\ensuremath{\preccurlyeq}\xspace}
\newcommand{\lex}{\ensuremath{\mathrel \lexa_{\textsc{lex}}}\xspace}
\renewcommand{\no}{\ensuremath{\bot}\xspace}
\newcommand{\pr}{\mathord\rightarrow}
\newcommand{\ppr}{\mathrel{\overset{*}{\Rightarrow}}}
\newcommand{\rR}{\mathrel R}

\newcommand{\reldom}{\ensuremath{\Sigma^*\times\Sigma^*}}
\newcommand{\fldom}{\ensuremath{\Sigma^*}}

\newcommand{\goal}{\ensuremath{\operatornamewithlimits{goal}}}
\newcommand{\argmax}{\ensuremath{\operatornamewithlimits{arg\,max}}}
\newcommand{\dom}{\ensuremath{\operatorname{dom}}}
\newcommand{\opt}{\ensuremath{\operatorname{opt}}}
\newcommand{\sol}{\ensuremath{\operatorname{sol}}}
\newcommand{\Ex}{\ensuremath{\operatorname{E}}}
\newcommand{\round}{\ensuremath{\operatorname{round}}}
\newcommand{\lcm}{\ensuremath{\operatorname{lcm}}}

\newcommand{\bool}{\{0,1\}}
\newcommand{\dop}{\operatorname{deg}}
\newcommand{\depth}{D}
\newcommand{\fL}{\textup{\textsf{msf}}}
\newcommand{\lin}[2]{#1_1, \ldots, #1_{#2}}

\newcommand{\perm}{\operatorname{perm}}
\newcommand{\pR}[2]{#1[#2]}
\newcommand{\pRn}[3]{\pR{#1}{\lin #2#3}}
\newcommand{\pRxn}{\pR{R}{\lin xn}}
\newcommand{\pbZxn}{\pR{\mathbf{Z}}{\lin xn}}
\newcommand{\sfR}[2]{#1_{\fL}[#2]}

\newcommand{\sfRxn}{\sfR{R}{\lin xn}}
\newcommand{\sfbZxn}{\sfR{\mathbf{Z}}{\lin xn}}
\newcommand{\size}{C}

\myprgruled

\renewcommand{\chaptermark}[1]{\markboth{#1}{}}

\fancypagestyle{plain}{\fancyhf{}}

\newcommand{\frontmatterpagestyle}{
\fancyhf{}
\fancyhead[LE,RO]{\thepage}
\fancyhead[CE,CO]{\scshape \leftmark}
}

\newcommand{\mainmatterpagestyle}{
\fancyhf{}
\fancyhead[LE,RO]{\thepage}
\fancyhead[LO,RE]{\S\ \thesection}
\fancyhead[CE]{\scshape \leftmark}
\fancyhead[CO]{\slshape \rightmark}
}

\newcommand{\backmatterpagestyle}{
\fancyhf{}
\fancyhead[LE,RO]{\thepage}
\fancyhead[CE,CO]{\scshape \leftmark}
}

\makeindex

\begin{document}
\raggedbottom
\shorthandoff{":!?;}
%
%
\frontmatter
\frontmatterpagestyle

\RCSfooter$Id: t.tex,v 2.3 1999/11/01 11:30:53 santini Exp $

\begin{titlepage}

\begin{center}{\scshape
    {\huge UNIVERSIT\`A DEGLI STUDI DI MILANO }\par
    {\Large 
      Dipartimento di Scienze dell'Informazione}
}\end{center}  
 
\vspace*{\stretch{1}}

\noindent\rule{\linewidth}{1mm}
\begin{flushright}{\huge
  Generazione casuale e\\ conteggio approssimato \\
  di strutture combinatorie \\
  \bigskip\slshape
  Random Generation and\\ Approximate Counting \\   
  of Combinatorial Structures \\
}\end{flushright}
\rule{\linewidth}{1mm}

\vspace{\stretch{0.5}}

\begin{center}
  \parbox[h]{1cm}{{\large\begin{tabbing} 
        \textsl{Tesi di Dottorato di}  \quad \= \kill
        \textsl{Tesi di Dottorato di}  \> Massimo Santini \\[2em]
        \textsl{Relatori}              \> Prof. Alberto Bertoni \\
                                       \> Prof. Massimiliano Goldwurm \\[1em]
        \textsl{Relatore esterno}      \> Prof. Bruno Codenotti
      \end{tabbing}}}
\end{center}

\vspace*{\stretch{2}}

\begin{center}{\Large\scshape 
    Dottorato in Scienze dell'Informazione --- XI Ciclo
}\end{center}

\end{titlepage}

\clearpage 

\thispagestyle{empty}
\mbox{}\vfill
\begin{minipage}[h]{7cm} 
  {\slshape Massimo Santini c/o\\[3pt]
  Dipartimento di Scienze dell'Informazione \\ 
  Universit\`{a} Statale degli Studi di Milano \\ 
  Via Comelico 39/41, 20135 Milano --- Italia \\[3pt]
  e-mail} \verb=santini@dsi.unimi.it=
\end{minipage}

\cleardoublepage 


\iffinal{\RCSfooter$Id: f.tex,v 2.2 1999/11/01 11:30:53 santini Exp $

\begin{otherlanguage}{italian}
    
    \thispagestyle{empty}
    \mbox{}
    \vspace{\stretch{1}}
    
    \begin{flushright}
      {\large

        Ad Ilenia, la persona che più di tutte\\
        ha cambiato la mia vita.

        }
    \end{flushright}
    
    \vspace{\stretch{3}} 
    \mbox{}
    
    \cleardoublepage 
    
    \thispagestyle{empty} 
    \mbox{} 
    \vspace{\stretch{1}}
    
    \begin{center}
      \begin{minipage}[c]{0.8\textwidth}
        
        Voglio ringraziare \textsl{Alberto Bertoni}, per il sostegno e
        l'ispirazione che non mi ha mai fatto mancare, dal periodo
        della mia laurea ad oggi. \textsl{Paola Campadelli}, che mi ha
        guidato in questi miei primi anni di ricerca, aiutandomi a
        maturare ed addolcire il mio carattere così bellicoso e
        critico.  \textsl{Massimiliano Goldwurm}, per la pazienza e
        l'attenzione infinita che ha dedicato a questo lavoro,
        mostrandomi sempre rispetto ed amicizia.

        \smallskip
        Ringrazio anche \textsl{Giuliano Grossi}, per le sue
        competenze ed il materiale che mi ha messo a disposizione in
        questi mesi di lavoro, in particolare riguardo al
        capitolo~\ref{cha:combopt}. \textsl{Giovanni Pighizzini}, per
        avermi offerto il suo tempo e per avermi reso partecipe di
        alcuni suoi interessanti risultati, in particolare riguardo
        alla sezione~\ref{sec:npda}.

        \smallskip
        Ringrazio infine \textsl{Paolo Boldi} e \textsl{Sebastiano
          Vigna}, per l'affetto che mi hanno dimostrato negli anni, i
        consigli ed il supporto che non mi hanno mai lesinato. Tutti i
        \textsl{colleghi del dottorato}, per aver costituito attorno a
        me, sin dall'inizio, un ambiente accogliente e stimolante per
        la ricerca. \textsl{Ilenia Epifani} (ultima, non certo per
        importanza), per essermi stata sempre a fianco, tollernado
        l'ansia di questi ultimi giorni di stesura e mai negandomi la
        sua revisione critica e obiettiva del mio lavoro.

      \end{minipage}
    \end{center}
    
    \vspace{\stretch{2}} 
    \mbox{}

\end{otherlanguage}

}

\RCSfooter$Id: c0.tex,v 2.9 1999/11/01 11:30:53 santini Exp $

\chapter*{Introduction}
\label{cha:preface}

Combinatorial counting problems have a long and distinguished history.
Apart from their intrinsic interest, they arise naturally from
investigations in numerous branches of mathematics and natural
sciences and have given rise to a rich and beautiful theory.  Ranking
problems, which consist in determining the position of a given element
in a well-ordered set, are closely related to counting.  Random
generation problems are less well studied but have a large number of
computational applications.

From the structural complexity viewpoint, the study of counting
problems was initiated by \citet{Val79}. A parallel approach to random
generation problems was proposed by \citet*{JVV86}; in particular,
they show how the standard reduction from generation to exact counting
can be modified to yield an almost uniform generator giving only
approximate counting estimates.  They also locate the almost uniform
generation and approximate counting problems for general \cNP
relations\footnote{by \cNP relations here we mean subsets
  $R\subseteq\reldom$ such that, for every $\alpha,\beta\in\fldom$,
  $|\beta|$ is polynomially related to $|\alpha|$ and the predicate
  $\alpha\rR\beta$ can be decided in time polynomial in $|\alpha|$
  (where $\Sigma$ is some finite alphabet).  For a more detailed
  discussion, see Section~\ref{sec:prel}.} within the second level of
the (probabilistic) polynomial time hierarchy~\citep{Sto77}.  Finally,
ranking has been studied by \citet{Huy90} and by \citet*{GS91} that
considered it as a special kind of optimal compression.

\bigskip

The aim of this thesis is to determine classes of \cNP relations for
which random generation and approximate counting problems admit an
efficient solution.  Since efficient rank implies efficient random
generation, we first investigate some classes of \cNP relations
admitting efficient \emph{ranking}.  On the other hand, there are
situations in which efficient random generation is possible even when
ranking is computationally infeasible.  We introduce the notion of
\emph{ambiguous description} as a tool for random generation and
approximate counting in such cases and show, in particular, some
applications to the case of formal languages.  Finally, we discuss a
limit of an heuristic for \emph{combinatorial optimization} problems
based on the random initialization of local search algorithms showing
that derandomizing such heuristic can be, in some cases, \cnP-hard.
More details follow.

\paragraph{Ranking.}

We extend some results about ranking for formal languages to the case
of \cNP relations, a fact that allows us to introduce two new classes
of relations admitting efficient (\ie polynomial time) random uniform
generation. In particular, we prove that the classes of \cNP relations
accepted by
\begin{enumerate}[(i)]
\item unambiguous \emph{auxiliary pushdown automata} working in polynomial
  time, and
\item nondeterministic \emph{Turing machines} using $s(n)$ space,
  $i(n)$ inversions and having ambiguity $d(n)$ with $s(n)\cdot
  i(n)\cdot d(n)=O(\log n)$,
\end{enumerate}
are such that their rank functions can be computed in polynomial time.
These results follow from the techniques used by \cite{Huy90} and,
respectively, \citet{BMP94} for the ranking of formal languages.
Hence, since we have also proved that efficient rank implies efficient
random generation, the above classes of relations both admit
polynomial time random uniform generation.

\paragraph{Ambiguous descriptions.}

To deal with the case in which ranking is computationally infeasible,
we introduce a simple notion of \emph{description} of a combinatorial
structure, together with a corresponding notion of \emph{ambiguity},
and study the problem of uniform random generation and approximate
counting for structures endowed with such descriptions. We prove a
general result stating that if a structure \bS has a description \bT
with polynomially bounded ambiguity and \bT admits a polynomial time
\emph{uniform random generator (\urg)}, then also \bS admits a \urg
working in polynomial time.  If, moreover, the counting problem for
\bT is solvable in polynomial time, then $\bS$ admits a fully
polynomial time \emph{randomized approximation scheme (\ras)} for its
counting problem.  Here, the proofs are based on the Karp-Luby
technique for sampling from a union of sets~\citep{KLM89} and on
Hoeffding's inequality~\citep{Hof63}, a classical tool for bounding
the tail probability of the sum of independent bounded random
variables.  Such general results can be applied to various classes of
languages:
\begin{enumerate}[(i)]
  
\item We show that, for \emph{finitely ambiguous context-free
    languages}, a word of length $n$ can be generated uniformly at
  random in $O(n^2\log n)$ time and $O(n^2)$ space, using $O(n^2\log
  n)$ random bits.  We observe that, in our model of computation, the
  same bounds for time and random bits are obtained for the uniform
  random generation of unambiguous \cf language~\citep{Gol95}.
  Similar bounds are obtained for the corresponding randomized
  approximation scheme. To prove these results we show in detail a
  multiplicity version of Earley's algorithm for context-free
  recognition~\citet{Ear70}. We prove that, for finitely ambiguous \cf
  languages, the number of derivation trees of an input word of size
  $n$ can be computed in $O(n^2\log n)$ time and $O(n^2)$ space;
  
\item We show how to generate, uniformly at random, words from
  languages accepted by \emph{one-way nondeterministic auxiliary
    push-down automata} working in polynomial time and using a
  logarithmic amount of work-space~\citep{Coo70,Coo71,Bra77}.  Also in
  this case, we obtain polynomial time \urg and \ras whenever the
  automaton has a polynomial number of accepting computations for each
  input word.  Notice that such results hold in particular for
  polynomially ambiguous context-free languages.
  
\item We consider the uniform random generation and approximate
  counting of \emph{rational trace languages}~\citep{DR95}.  Finitely
  ambiguous rational trace languages~\citep{BMS82, Sak87} admit \urg
  and \ras of the same time complexity of the algorithms for their
  recognition problem.  Analogously, we obtain polynomial time \urg
  and \ras for the rational trace languages that are polynomially
  ambiguous.

\end{enumerate}

\paragraph{Derandomization and combinatorial optimization.}

We focus our attention on an heuristic suggested by \citet{Gro99} for
improving \emph{local search} algorithms.  The basic idea is to use a
two phase (randomized) algorithm which in a first phase generates
uniformly at random some feasible solution and then, starting from the
best solution so obtained, performs a local search phase. For a large
class of problems, such heuristic is known to give (randomized)
approximation algorithm with a constant performance ratio.  We study
in detail the case in which the objective function of the
combinatorial optimization problem is represented by means of
arithmetic circuits: this is a quite common situation and applies to a
lot of natural problems.  Sufficient conditions are known in such a
case to prove that the first phase of the heuristic can be
``derandomized'' thus yielding to a deterministic approximation
algorithm. On the other hand, by investigating a problem of
transformation between arithmetic circuits, we bring to light a limit
of this approach and show that ``derandomizing'' the first phase of
the heuristic is, in some cases, a \cnP-hard problem.

\section*{Structure of the chapters.}

This thesis is organized as follows.  In Chapter~\ref{cha:intro} we
introduce some preliminary notions and definitions; in particular, in
Section~\ref{sec:compmodel} we present our model of computation: the
\mPrRAM, which is a probabilistic version of the standard random
access machine model endowed in addition with an unbiased coin tossing
device. In Section~\ref{sec:prel} we recall the definition of
$p$-relation \citep{JVV86}, a fundamental tool to describe formally
the problems we deal with in this work.  Section~\ref{sec:jvv}
concludes the chapter with a short survey on the known results about
approximate counting and random generation, as discussed by
\citet{JVV86,JS89}.

Chapter~\ref{cha:ranking} shows how the cited results on
\emph{ranking} for formal languages can be applied to the random
generation problem.  First of all, in Section~\ref{sec:rankrug} the
standard notion of ranking is recalled and a proposed extension to the
case of relations is given; hence, we prove
(Theorem~\ref{teo:rugprel}) that if a $p$-relation admits efficient
(\ie polynomial time) ranking then it also admits efficient random
uniform generation.  Hence, in Section~\ref{sec:sbtm} we extend the
result of \citet{BMP94} about the ranking of languages accepted by
\emph{Turing machines with simultaneous complexity bounds} to
$p$-relations (Theorem~\ref{teo:sbtm}); similarly, in
Section~\ref{ssec:pda} we extend the result of \citet{Huy90} about the
ranking of languages accepted by unambiguous \emph{one-way auxiliary
  pushdown automata} working in polynomial time to $p$-relations
(Theorem~\ref{teo:pda}).  Thanks to Theorem~\ref{teo:rugprel}, these
last results yield to two new classes of $p$-relations admitting
efficient uniform random generation (Corollaries~\ref{cor:prelsbtmurg}
and~\ref{cor:perlupdaurg}).

In Chapter~\ref{cha:ambparadigm}, for the sake of simplicity, we turn
our attention to combinatorial structures introducing a general
paradigm that, under suitable hypotheses, leads to polynomial time
algorithms both for random generation and approximate counting
problems.  We begin by recalling some useful definition in
Section~\ref{sec:combstrdef}, adapting in particular to the case of
combinatorial structures the notions of \emph{uniform random
  generator} (\urg) and \emph{randomized approximation scheme} (\ras)
given by \citet{JVV86} for the case of $p$-relations.  In
Section~\ref{sec:ambdescrdef} we introduce a simple notion of
\emph{description} for a combinatorial structure together with a
related definition of \emph{ambiguity} and prove the existence of a
polynomial time \urg (Theorem~\ref{teo:rugpa}) and of a fully
polynomial \ras (Theorem~\ref{teo:rasecpa}) for combinatorial
structures admitting suitable (ambiguous) descriptions.  To elucidate
such results, Section~\ref{sec:simpleapp} presents some simple
application of our general paradigm to the case of union and product
of combinatorial structures.

Chapter~\ref{cha:ambappl} gives less trivial applications of the
results of the previous chapter. In Section~\ref{sec:cfl} we show a
polynomial time \urg and a fully polynomial time \ras for polynomially
ambiguous \emph{context-free languages} (Theorem~\ref{teo:cf}).
Notice that even for finitely ambiguous context-free languages exact
counting is known to be $\cnP_1$-complete\footnote{the class $\cnP_1$
  is the restriction of $\cnP$ to functions having unary inputs.}
\citep{BGS91}). Our results depend on a multiplicity version of
Earley's algorithm for context-free recognition; in particular, for
finitely ambiguous \cf languages, we give an algorithm computing the
number of derivation trees of an input word of size $n$ in in
$O(n^2\log n)$ time and $O(n^2)$ space (Proposition~\ref{teo:cfd}).
Then, in Section~\ref{sec:npda}, we show that the class of languages
accepted by polynomially ambiguous \emph{one-way auxiliary pushdown
  automata} working in polynomial time admits a polynomial time \urg
and a fully polynomial time \ras (Theorem~\ref{teo:npda}).  Finally,
in Section~\ref{sec:rtl}, we give a polynomial time \urg and a fully
polynomial time \ras for polynomially ambiguous \emph{rational trace
  languages} (Theorem~\ref{teo:rtl}); in particular, in the case of
finitely ambiguous rational trace languages, such algorithms have the
same time complexity as the algorithms for their recognition problem.

We came back to the more general setting of $p$-relations in
Chapter~\ref{cha:boolean}, where we suggest how the results given for
the case of combinatorial structures can be generalized to the case of
$p$-relations. More precisely, in Section~\ref{sec:backtoprel} we
propose a generalization of Theorem~\ref{teo:rugprel} to the case of
$p$-relations (Corollary~\ref{cor:preldes}) which, in particular,
leads to a stronger version of Theorem~\ref{teo:pda}: this allows one
to define a broader class of $p$-relations admitting polynomial time
uniform random generations in terms of relations accepted by
polynomially ambiguous one-way pushdown automata working in polynomial
time (Corollary~\ref{cor:prelpdaurg}). Sections~\ref{sec:union}
and~\ref{sec:complement} conclude the chapter by discussing some
closure properties with respect to the union and complement of the
classes of $p$-relations admitting efficient ranking and/or random
generation.

As a very different applications of our results, in
Chapter~\ref{cha:combopt}, we consider the case of (\cNP)
\emph{combinatorial optimization problems}. After the very short
survey of basic notions and definitions of Section~\ref{sec:codefs},
we focus our attention to the \emph{local search} paradigm in
Section~\ref{sec:localsearch}, where we recall the approach of
\citet{Gro99} for improving local search algorithms. The determination
of new classes of $p$-relations admitting polynomial time uniform
random generators allows us to introduce a class of \cNP optimization
problems admitting a polynomial time randomized approximation
algorithm with a constant performance ratio
(Corollary~\ref{cor:combopt}).  On the other hand, in
Section~\ref{sec:negres}, we show some negative results about
``derandomization'' and local search. By studying a problem of
transformation between arithmetic circuits, we show that, in some
cases, derandomizing can be as hard as solving a \cnP-complete problem
(Theorems~\ref{teo:weak} and~\ref{teo:main}).

\clearpage 

\thispagestyle{empty}
\mbox{}\vfill
\mycite{italian}{%
  Per l'amore della bislunga ho tagliato dieci giovani alberi pioppi e
  glabri. Di tanti che erano ne ho fatto cinque cento fogli di carta
  bianchi e io su quelli ho scritto giorni e mesi per fare una storia.
  Ora che voi la leggete, sapete se vale o non vale quei pioppi padani
  e il tempo, la vita nei mesi, di un uomo.
}%
{Maurizio Maggiani}{Màuri màuri}


\tableofcontents

%
%
\mainmatter
\mainmatterpagestyle

\RCSfooter$Id: c1.tex,v 2.3 1999/10/29 10:23:58 santini Exp $

\chapter{Preliminary Notions}
\label{cha:intro}

\mycite{italian}{%
Bisognava concludere. Manifestai alla contessina Delrio ciò che
sentivo di non poterle dissimulare più a lungo. Si rassegnasse
all'idea: le diagonali del parallelogrammo si secano nel loro punto
mediano. E non è tutto: esse ne dividono l'area in quattro triangoli
equivalenti.}%
{Carlo Emilio Gadda}{La Madonna dei filosofi}

In this chapter we introduce some preliminary notion and definition. 
First of all, we discuss in some detail the model of computation we
refer to in the rest of this work. To this aim, we present the \mRAM
model together with a proposed probabilistic extension of such model:
the \mPrRAM; we also give an example of algorithm, discussing its
implementation on such model in order to elucidate some of the
concepts here introduced.

Then, we recall the definition of $p$-relation and discuss its
relevance as a unifying tool for the formal definition and analysis of
various versions of combinatorial problems, in particular with regard
to the notion of self-reducibility.

Finally, we present a brief survey of known results on random
generation of $p$-relations, discussing in particular the relationship
between (almost) uniform random generation and approximate counting.

\bigskip
\RCSfooter$Id: c1s1.tex,v 2.4 1999/11/01 11:30:53 santini Exp $

\section{The Model of Computation}
\label{sec:compmodel}

One of the fundamental issues arising in the design and analysis of
algorithms and in the investigation of the inherent computational
difficulty of various problems is the choice of the formal \idx{model
  of computation}.  Even though it is well known that all the
``reasonable'' computational models are polynomially related with
respect to the time and space complexity, nonetheless each model
enjoys some peculiar properties that can help us in the choice.

The Turing machine model, for instance, with its primitive instruction
repertoire, is suitable for the investigation of negative results and
lower bounds typically arising in the structural complexity
investigations. On the other hand, random access machines, or the
arithmetic machine model, lead to a more natural and easily
understandable notation for the description and design of high level
algorithms.

\subsection{The classic \mRAM model}
\label{ssec:rammodel}

Since the main aim of this work is to design efficient algorithms
rather than to provide lower bounds, we focus our attention on the
\mRAM model, that we now briefly recall (for a complete reference, and
for a discussion on the \idx{Turing machine} model, see
\citep{AHU74}).

\begin{figure}[ht]
  \begin{center}
    \input{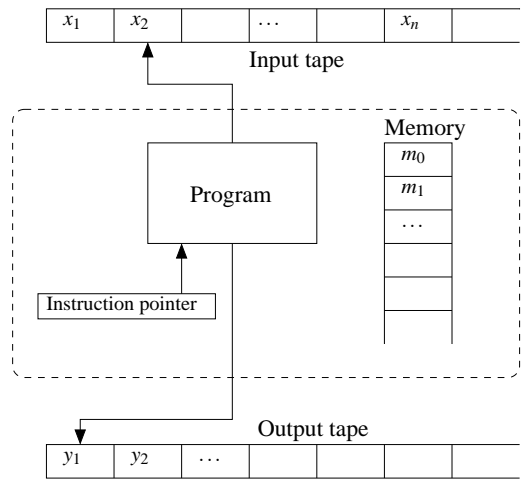}
    \caption{A random access machine.}
    \label{fig:ram}
  \end{center}
\end{figure}

A \eidx{random access machine} (\mRAM) consists of a read-only
\esidx{input tape}{\irRAM}, a write-only \esidx{output tape}{\irRAM},
a \esidx{memory}{\irRAM}, and a \esidx{program}{\irRAM}
(Figure~\ref{fig:ram}).  The input tape is a sequence of cells each of
which holds an integer; whenever an integer is read from the input
tape, the tape head moves one cell to the right. The output tape is a
(potentially infinite) sequence of cells that are initially all
blank; when a write instruction is executed, an integer is printed in
the cell of the output tape currently under the output tape head,
hence the head is moved one cell to the right.  The memory consists of
a sequence of registers $m_0,m_1,\ldots$ each of which is can hold an
integer; no upper bound is placed on the number of registers that can
be used.

The program is a fixed sequence of (optionally) labelled instructions
that are not modified during the execution. We assume that there are
input-output instructions manipulating the tapes (\texttt{read} and
\texttt{write}), instructions moving data across memory (\texttt{load}
and \texttt{store}), arithmetic instructions (\texttt{add},
\texttt{sub}, \texttt{mult}, and \texttt{div}), branching instructions
(\texttt{jump}, \texttt{jzero} and \texttt{jgtz}) and, finally,
\texttt{halt} to stop the computation.  Each instruction manipulating
data may operate in an \esidx{indirect addressing}{\irRAM} mode (for
example, for indexing arrays); the operand of such instructions can be
``${\mathord =}i$'' to indicate the integer $i\in\bZ$, ``$i$'' to
indicate the content of register $m_i$, or ``${\mathord *}i$'' to
denote the content of $m_j$ where $j$ is the content of $m_i$.

\subsubsection{Semantic}

The \esidx{semantic}{\irRAM} \mdidx{semantic}{\irRAM} of the
computation of a \mRAM can be easily defined with the help of two
elements: an \esidx{instruction pointer}{\irRAM} that determines the
next instruction of the program to be executed and a \esidx{memory
  map}{\irRAM} $M:\bN\to\bZ$ that gives, for every $i\in\bN$, the
contents $M(i)$ of the $i$-th register, that is the integer stored in
$m_i$. At the beginning of the computation, both tape heads scan the
respective first (leftmost) cell, the output tape cells are all blank,
all the registers are zero (\ie $M(i)=0$ for all $i\in\bN$) and the
instruction pointer points to the first instruction of the program.
During the execution the instruction pointer is modified so that the
instructions in the program are executed in sequential order unless
\texttt{jump} is executed, \texttt{jzero} is executed when $M(0)=0$,
or \texttt{jgtz} is executed when $M(0)\geq 0$: in these cases, the
instruction pointer is modified according to the label that follows
the jumping instruction.

\begin{table}[ht]
  \begin{center}
    \begin{mytabular}{Instruction}{Meaning}
      \texttt{read} $i$ & 
      $M(i)$ takes as value the integer in the current input cell
      \\
      \texttt{read} ${\mathord *}i$ & 
      $M(M(i))$ takes as value the integer in the current input cell
      \\
      \texttt{write} $a$ &
      $V(a)$ is written in the current output cell 
      \\
      \texttt{load} $a$ &
      $M(0)$ takes value $V(a)$ 
      \\
      \texttt{store} $i$ &
      $M(i)$ takes value $M(0)$
      \\
      \texttt{store} ${\mathord *}i$ &
      $M(M(i))$ takes value $M(0)$
      \\
      \texttt{add} $a$ & 
      $M(0)$ takes value $M(0)+V(a)$
      \\
      \texttt{sub} $a$ & 
      $M(0)$ takes value $M(0)-V(a)$
      \\
      \texttt{mult} $a$ & 
      $M(0)$ takes value $M(0)\times V(a)$
      \\
      \texttt{div} $a$ & 
      $M(0)$ takes value $\lfloor M(0)/V(a)\rfloor$
      \\
      \texttt{jump} $b$ & 
      the instruction pointer points to $b$\\
      \texttt{jzero} $b$ & 
      if $M(0)=0$, then the instruction pointer points to $b$, 
      else it points to the next instruction in the program\\
      \texttt{jgtz} $b$ & 
      if $M(0)\geq 0$, then the instruction pointer points to $b$, 
      else it points to the next instruction in the program\\
      \texttt{halt}  & 
      the execution stops
    \end{mytabular}
    \caption{Meaning of \mRAM instructions.}
    \label{tab:rammeaning}
  \end{center}
\end{table}

The meaning of the various instructions should be clear to anyone who
is familiar with some kind of assembly language and is briefly
recalled in Table~\ref{tab:rammeaning} where, for the sake of brevity,
we have denoted by $V(a)$ the \emph{value} of operand $a$ \msidx{value
  of an operand}{\irRAM} defined as follows: $V({\mathord =}i)=i$,
$V(i)=M(i)$ and $V({\mathord *}i)=M(M(i))$.

\smallskip

Given the semantic described so far, the computation of a \mRAM can be
essentially seen as a (partial) function from the content of its input
tape to the (non-blank) cells of its output tape (defined whenever the
computation halts). More formally, given a \mRAM $P$, with an abuse of
notation, we denote by $P$ also the (partial) function
$P:\bN^*\to\bN^*$ such that $P(x_1,\ldots,x_n)$ is the content of the
output tape of $P$ (omitting the trailing blank cells) whenever the
computation of $P$ starting with $x_1\ldots x_n$ on the input tape
halts; otherwise $P(x_1,\ldots,x_n)$ is undefined. The class of
functions so defined can be related to the class of \eidx{partial
  recursive functions} \citep{Dav58,Rog67}, it is in fact possible to
show that \citep{AHU74}
\begin{pro}
  The class of (partial) functions that can be computed in the \mRAM
  model coincides with the class of recursive (partial) functions.
\end{pro}

\subsubsection{Computational complexity}

To define the \esidx{computational complexity}{\irRAM}
\mdidx{computational complexity}{\irRAM} in the \mRAM model we need to
specify the time required to execute each instruction and the space
used by each register; to this end, two choices are usually made up in
the literature: \mdidx{cost criterion}{\irRAM} the \emph{uniform cost
  criterion} \mdidx{cost criterion}{uniform} and the \emph{logarithmic
  cost criterion} \mdidx{cost criterion}{logarithmic} \citep{AHU74}.
We briefly recall the latter since we shall use it for the analysis of
the algorithms that will be presented in this work. Let $l(i)$ be the
logarithmic function on the integers defined by\footnote{All the
  logarithms in this thesis are in base $2$, unless otherwise
  specified.}
\[
l(i)=
\begin{cases}
  \lfloor \log(i) \rfloor + 1 & \text{if $i>0$,} \\
  1 & \text{if $i=0$} 
\end{cases}
\]
and define the (logarithmic) cost $t(a)$ of accessing the operand $a$
as $t({\mathord =}i)=l(i)$, $t(i)=l(i)+l(M(i))$ and $t({\mathord
  *}i)=l(i)+l(M(i))+l(M(M(i)))$. Then the cost of the \mRAM
instruction is given in Table~\ref{tab:ramcost}.

\begin{table}[ht]
  \begin{center}
    \begin{mytabular}{Instruction}{Cost}
      \texttt{read} $i$ & 
      $l(\text{input})+l(i)$
      \\
      \texttt{read} ${\mathord *}i$ & 
      $l(\text{input})+l(i)+l(M(i))$
      \\
      \texttt{write} $a$ &
      $t(a)$
      \\
      \texttt{load} $a$ &
      $t(a)$
      \\
      \texttt{store} $i$ &
      $l(M(0))+l(i)$
      \\
      \texttt{store} ${\mathord *}i$ &
      $l(M(0))+l(i)+l(M(i))$
      \\
      \texttt{add} $a$ & 
      $l(M(0))+t(a)$
      \\
      \texttt{sub} $a$ & 
      $l(M(0))+t(a)$
      \\
      \texttt{mult} $a$ & 
      $l(M(0))+t(a)$
      \\
      \texttt{div} $a$ & 
      $l(M(0))+t(a)$
      \\
      \texttt{jump} $b$ & 
      $1$
      \\
      \texttt{jzero} $b$ & 
      $l(M(0))$
      \\
      \texttt{jgtz} $b$ & 
      $l(M(0))$
      \\
      \texttt{halt}  & 
      $1$
    \end{mytabular}
    \caption{Logarithmic cost of \mRAM instructions.}
    \label{tab:ramcost}
  \end{center}
\end{table}

\subsubsection{Some remarks}

We conclude this section observing that the choice of the \mRAM model
under logarithmic cost criterion enjoys two important properties we
look for in the design and analysis of the algorithms presented in
this work.  First of all, the high-level language and structure of the
\mRAM avoids the need of the boring specifications usually arising
when using the Turing machine model and moreover it allows us to
design algorithms that are easier to understand and more similar to
their implementation in today's programming languages.  On the other
and, by taking into reasonable account the size of the operands, the
logarithmic cost criterion will allow the analysis of our algorithms
to be realistic in the precise sense of the following proposition.
Recall that two functions $f,g$ are said to be \eidx{polynomially
  related} if two polynomials $p,q$ exist such that $f(n)\leq p(g(n))$
and $g(n)\leq q(f(n))$ for every $n>0$. Then, it is well known that
\citep{AHU74}
\begin{pro}
  The time (and space) cost in the \mRAM model under logarithmic cost
  criterion and in the (multi-tape) Turing machine model are
  (respectively) polynomially related.
\end{pro}
This means that every algorithm we design on a \mRAM can be
effectively implemented by some fixed hardware so that the logarithmic
time required by the algorithm on the \mRAM is polynomially related to
the time required by the hardware to actually carry on the
computation.

\subsection{The \mPrRAM model}
\label{ssec:prrammodel}

Since a large part of this work is devoted to randomized algorithms,
we need now to introduce some probabilistic model of computation.
Concerning the \idx{source of randomness}, two choices are usually
present in the literature. The first, which characterizes the
\idx{probabilistic Turing machine} model \citep{Gil77}, is to enrich
the deterministic model of computation with a device able to toss an
unbiased coin, that is to allow branches during the computation so
that each of the two branch can be taken with equal probability.  The
second choice, usually adopted when considering models of computation
similar to the \mRAM, is to enrich the deterministic model with one
(or more) coin of (different) rational, or even arbitrary, bias.  It
is clear that the two models have very different computational power:
if one has only a fair coin, then even the simple task of choosing
with equal probability between three numbers, that is rolling a
3-sided die, has no algorithm that always terminates!

This asymmetry leads to different approaches in the design of
probabilistic algorithms. The first possibility is to focus on
algorithms working in \emph{bounded time} \midx{bounded time
  probabilistic algorithms}, which always terminate but that,
specially when using the unbiased coin model, can fail to produce
their output exactly according to a desired distribution. If this
happens, there are two possibilities: a special symbol can be output
denoting the failure of the algorithm, or the algorithm may simply try
to approximate the desired distribution up to some specified precision
degree.  On the other hand, if one focuses on \emph{bounded expected
  time} \midx{bounded expected time probabilistic algorithms}
algorithms, then even in the unbiased coin model one can look for
(possibly very slow, or even nonterminating) algorithms whose output
exactly follows a desired distribution, or can choose to trade off the
computation time with the approximation precision.  Some examples of
the bounded time approach can be found for instance in the works of
\citet{JVV86,JS89} and of \citet{FINNRS93}, while examples of the
expect time approach can be found for instance in the works of
\citet{Alo94,BPS92}.

Without going further into details, one has to notice that a bounded
time algorithm using a special symbol to denote its failure can, in
principle, be iterated until success, leading to a bounded expected
time algorithm; on the other hand, a bounded expected time algorithm
whose output eventually follows exactly a desired distribution can, in
principle, be prematurely halted leading to a bounded time
approximating algorithm. As one can conclude even from this very short
presentation, the difference between these two approaches is very
subtle and often the choice between them is a matter of taste, or
opportunity, unless feasibility of the computational model is
considered.

As suggested by \citet{JVV86}, allowing the tossing of an arbitrary
(rationally) biased coin whose bias depends on the input violates the
philosophy of Turing machines, because the coin tossing steps of
different biases cannot be implemented with a fixed amount of hardware
and in a fixed time. \citet{FINNRS93} in fact show that at least two
coins of rational bias are necessary to simulate a $n$-sided die,
unless $n$ is a power of two, and that the simulation of every coin of
rational bias whose denominator is less then $n$ with $d$ coin flips
requires at least $\Omega(\log n/\log d)$ differently biased rational
coins.

Thus, to follow the principle of a \idx{feasible model of
  computation} we have stated in the previous section, enforced by
choosing the logarithmic cost criterion to allow a realistic analysis
of our algorithms, our choice of the source of randomness will be in
favor of the unbiased coin model.

\begin{figure}[ht]
  \begin{center}
    \input{fig/prram}
    \caption{A \mPrRAM.}
    \label{fig:prram}
  \end{center}
\end{figure}

Our model of \eidx{probabilistic random access machine} (\mPrRAM)
consists of a \mRAM machine endowed in addition with a read-only
\esidx{random tape}{\irPrRAM} (Figure~\ref{fig:prram}) and a further
instruction \texttt{rnd} to operate on it. The random tape is a
sequence of cells each containing $0$ or $1$; when the instruction
``\texttt{rnd} $a$'' is executed, the random tape head scans $V(a)$
cell and $M(0)$ takes as value the integer number given, in binary
notation, by the juxtaposition of the read cells. More formally, if
$R(j)$ denotes the content of the $j$-th random tape cell $r_j$, when
the random tape head is positioned on $r_i$ and ``\texttt{rnd} $a$''
is executed, then $M(0)=\sum_{j=0}^{V(a)-1} 2^j R(i+j)$ and the random
tape head is positioned on $r_{i+V(a)}$.

\begin{table}[ht]
  \begin{center}
    \begin{mytabular}{Instruction}{Meaning}
      \texttt{rnd} $a$ & 
      $M(0)$ takes value $\sum_{j=0}^{V(a)-1} 2^j R(i+j)$. 
    \end{mytabular}
    \caption{Meaning of \mPrRAM instructions.}
    \label{tab:prrammening}
  \end{center}
\end{table}

\subsubsection{Semantic}

To define the \esidx{semantic}{\irPrRAM}
\mdidx{semantic}{\irRAM!probabilistic} of a \mPrRAM computation, we
introduce the following notation: let $P(x_1,\ldots,x_n;
r_1,r_2,\ldots)$ be the (partial) function denoting the contents of
the output tape (omitting the trailing blank cells) at the end of the
computation of the \mPrRAM $P$ having $x_1\ldots x_n$ on the input
tape and $r_1 r_2\ldots$ on the random tape (whenever the computation
halts, otherwise we let $P(x_1,\ldots,x_n;r_1,r_2,\ldots)$ undefined).
Since $r_1,r_2,\ldots$ are fixed, the computation can be considered
deterministic and its semantic can be easily derived by the semantic
of the \mRAM model discussed above and from the given meaning of the
\texttt{rnd} instruction. Then $P(x_1,\ldots,x_n)$ is defined as the
random variable $P(x_1,\ldots,x_n;R_1,R_2,\ldots)$ where
$R_1,R_2,\ldots$ are independent and identically distributed Bernoulli
random variables such that $\Pr\{R_i=1\}=\Pr\{R_i=0\}=1/2$ for
$i\in\bN$.

\subsubsection{Computational complexity}

Again, to define the \sidx{computational complexity}{\irPrRAM}
\mdidx{computational complexity}{\irRAM!probabilistic} in the \mPrRAM
model we need to specify the time required to execute the \texttt{rnd}
instruction; to take into account the dimension of the operand also in
this case, we also assume that the cost of a ``\texttt{rnd} $a$''
instruction is $t(a)$ (where $t$ is defined as in the previous
section). Moreover, together with the usual time and space complexity
measures, for the \mPrRAM model we can also consider explicitly the
number of cells scanned on the random tape, a quantity which we call
the \esidx{number of random bits}{\irPrRAM} used by the computation.

\begin{table}[ht]
  \begin{center}
    \begin{mytabular}{Instruction}{Cost}
      \texttt{rnd} $a$ & 
      $t(a)$
    \end{mytabular}
    \caption{Cost of \mPrRAM instructions.}
    \label{tab:prramcost}
  \end{center}
\end{table}

\subsubsection{Some remarks} 

We want to highlight again that, thanks to the choice for the source
of randomness adopted in the \mPrRAM model discussed in this section,
we believe to have attained our purpose to suggest a model of
computation at the same time feasible and of sufficient high-level to
describe in a natural fashion the algorithms we will design in this
work. As for the \mRAM case, it is straightforward to check that the
following holds
\begin{pro}
  The time (and space) cost \mPrRAM model (under logarithmic cost
  criterion) and the (multi-tape) probabilistic Turing machine model
  are polynomially related.
\end{pro}

\subsection{An example of algorithm}
\label{ssec:algexampl}

All the algorithms presented in the following will be described using
a high-level language, usually known as ``Pidgin ALGOL'' (for its
informal description, see \citep{AHU74}); what is essential here, is
that such algorithms can be translated into a \mRAM program in a
straightforward manner, hence we shall never be concerned with the
details of this ``compilation'' process.  In our Pidgin ALGOL we allow
the use of every kind of usual mathematical statement and programming
language construct such as expressions, conditions, statements and
(recursive) procedures; moreover, the language has no fixed set of
data types: variables can represent integers, strings and arrays, or
even more complex objects such as sets, lists and graphs.  No attempt
will be made here to give a precise definition of all the constructs
used as long as their meaning will be clear and their translation into
\mPrRAM code is evident from the context.

\subsubsection{Uniform random generation of integers}

As an example of what stated in this section, we give an algorithm
(Algorithm~\ref{alg:inturg}) for the uniform random generation of an
integer in the range ${1,\ldots N}$. We discuss its implementation on
a \mPrRAM and the consequent analysis of its complexity and
correctness. We need to fix some $0<\delta<1$ as an upper bound to the
probability that the algorithm fails to give output according to the
uniform distribution, a fact that the algorithm will signal by use of
the special symbol \no (we recall that this depends on our choice of
an unbiased coin as a source of randomness).

\begin{alg}[ht]
  \caption{a uniform random integer numbers generator.}
  \label{alg:inturg}
  \begin{myprg*}
    \sinput $N$ \nl
    $r\stv\no$, $i\stv 0$ \nl
    \swhile $i < \lceil \log(1/\delta) \rceil$ and $r=\no$ \sdo \tnl
    $i\stv i+1$ \nl
    \textbf{generate} $u\in\{1, \ldots, 2^{\lceil\log N\rceil}\}$ 
       \emph{uniformly at random} \nl
    \sif $u\leq N$ \sthen $r\stv u$ \unl
    \soutput $r$.
  \end{myprg*}
\end{alg}

Consider first of all the implementation. The ``deterministic''
statements are all very simple to implement on a \mRAM, as it is easy
to verify; in particular, $\lceil\log\delta \rceil$ is a constant
(independent of the input), while $\lceil\log N\rceil$ can be computed
in $O(\log N\log\log N)$ time (see Lemma~\ref{lem:bit} for details).

On the other hand, for every $m\in\bN$, the ``probabilistic''
statement ``\textbf{generate} $u\in\{1, \ldots, 2^m\}$ \emph{uniformly
  at random}'', can be implemented on a \mPrRAM essentially using just
a ``\texttt{rnd} $m$'' instruction.

Concerning the correctness of the algorithm, if we denote by $R(N)$
the random variable giving the output of the \mPrRAM $R$ implementing
the algorithm, we are able to prove that, for every $N\in\bN$,
\begin{align}
  \Pr\{ R(N)=\no \} 
  & <  \delta, \;\text{and} \label{eq:rNgerr}\\ 
  \Pr\{ R(N)=n \mid R(N)\not=\no  \} 
  & =  1/N, \;\text{for every}\; n\in\{1,\ldots, N\}. \label{eq:rNgok}
\end{align}
Here, equation~(\ref{eq:rNgerr}) substantially tells that the
algorithm gives a ``correct'' output with probability $1-\delta$, and
equation~(\ref{eq:rNgok}) states that, if the algorithm terminates
``correctly'', then its output is actually uniformly distributed over
the desired range. The first equation follows by noting that the
output of the algorithm is \no iff $u$ is greater than $N$ for
$\lceil\log(1/\delta)\rceil$ times:
\begin{equation*}
  \begin{split}
    \Pr\{ R(N)=\no \}
    &= \Pr\{ u > N\}^{\lceil\log(1/\delta)\rceil} \\
    &= \left( 1 - \frac N{2^{\lceil\log N\rceil}} \right)^{\lceil\log(1/\delta)\rceil} \\
    &< 1/2^{\lceil\log(1/\delta)\rceil}\leq \delta,
  \end{split}
\end{equation*}
where the first inequality follows from the fact that $N/2^{\lceil\log
  N\rceil}>1/2$, for every $N>0$. Moreover, for every $n\in\{1,\ldots
, N\}$,
\[
\Pr\{ u=n \mid u\leq N \}=
\frac{\Pr\{ u=n \}}{\Pr\{ u\leq N\}}=
\frac 1{2^{\lceil\log N\rceil}}\left(\frac N{2^{\lceil\log N\rceil}}\right)^{-1}=
\frac 1N.
\]
Since the output of the algorithm, whenever it is different from \no,
is distributed as $u$ conditionally to the fact that $u\leq N$, by the
previous equality, we can finally derive equation~(\ref{eq:rNgerr}).
We can summarize the result of this section as
\begin{pro}
  \label{pro:rNg}
  For every $0<\delta<1$, there exists a \mPrRAM which, on input
  $N\in\bN$, gives in output an integer chosen uniformly at random in
  $\{1,\ldots, N\}$ with probability $1-\delta$, taking $O(\log
  N(\log\log N-\log \delta))$ time.
\end{pro}


\RCSfooter$Id: c1s2.tex,v 2.4 1999/11/01 11:30:53 santini Exp $

\section{The Notion of $p$-Relation}
\label{sec:prel}

We now recall a notion that allows us to formalize several kind of
combinatorial problems. The use of \emph{languages}, \ie subsets of
the free monoid \fldom (where $\Sigma$ is some finite alphabet), to
describe formally combinatorial problems is well established
\citep[see, for instance,][]{GJ79}. In a similar fashion, one can also
use \emph{relations} in \reldom to describe, with ``more richness'',
such problems. As an example, for a suitable encoding, consider the
relations
\begin{enumerate}[(1)]
\item \label{it:sat} $(F, \tau)$, where $F$ is a boolean formula and
  $\tau$ is one of its satisfying truth assignments;
\item \label{it:clique} $(G, K)$, where $G$ is a graph and $K$ is one
  of its cliques\footnote{$K$ is a subset of the vertices of $G$ such
    that every pair of vertices in $K$ are joined by an edge in $G$.};
\item \label{it:part} $(n, P)$, where $n\in\bN$ is a positive integer
  and $P$ is the set of partitions of $n$;
\item \label{it:gsub} $(G,K)$, where $K$ is a (induced)
  subgraph\footnote{$K$ is an induced subgraph of $G$ if it is
    isomorphic to a (proper) subgraph of $G$.}  of $G$.
\end{enumerate}

The use of relations instead of languages is more rich in the sense
that they enable us to define several ``versions'' of a combinatorial
problem \midx{versions of combinatorial problem} in a natural way.
Given some $R\subseteq\reldom$, define $R(\alpha)=\{\beta\in\fldom :
\alpha\rR \beta \}$, for every $\alpha\in\fldom$. Then, for some fixed
relation $R$, one have the following kinds of problems
\begin{enumerate}[(a)]
\item \label{it:decprob} \emph{decision}: given $\alpha\in\fldom$,
  decide whether there exists some $\beta\in R(\alpha)$;
\item \emph{construction}: given $\alpha\in\fldom$, construct, if it
  exists, some $\beta\in R(\alpha)$;
\item \label{it:cntprob} \emph{counting}: given $\alpha\in\fldom$,
  count the number of $\beta\in R(\alpha)$;
\item \emph{random generation}: given $\alpha\in\fldom$, generate an
  element of $R(\alpha)$ uniformly at random.
\end{enumerate}
Note on passing that, as we will see in Chapter~\ref{cha:combopt},
also combinatorial \emph{optimization} problems can be defined in a
natural way using relations.

Such a unifying view of combinatorial problems by means of relations
has appeared in the literature in the form of \eidx{string relations}
by \citet{GJ79} and of \eidx{search functions} by \citet{Val78}. It is
used in this thesis mainly to relate generation problems, which are
the main subject of this work, to more familiar combinatorial problems
such as existence and counting.

\subsection{$p$-Relations}
\label{ssec:prel}
 
A class of relations in $\reldom$ of particular interest is the
subclass of the relations that can be ``checked efficiently''; more
formally, \citet{JVV86} give the following:
\enlargethispage{3em} 
\begin{dfn}
  \label{dfn:prel}
  A relation $R\subseteq\reldom$ is a
  \emph{$p$-relation}\midx{p@$p$-relation} if
  \begin{enumerate}[(i)]
  \item \label{it:psize} there exists a polynomial $p$ for which
    $|\beta|=p(|\alpha|)$ for every $\alpha,\beta\in\fldom$ such that
    $\alpha\rR\beta$;
  \item \label{it:pdec} the relation $R$ can be decided\footnote{that
      is, the language $\{\alpha\sep\beta : \alpha,\beta\in\fldom,
      \alpha\rR\beta\}$, where $\sep$ is some symbol not belonging to
      $\Sigma$, belongs to \cP.} in polynomial time.
  \end{enumerate}
\end{dfn}
Two remarks are in order. First, we observe that, without loss of
generality, for the sake of simplicity we fix $|\beta|= p(|\alpha)$
while the original definition of \citet{JVV86} asks only that
$|\beta|\leq p(|\alpha|)$. Second, here and in the following, by an
abuse of notation, for every $p$-relation, by the same letter ``$p$''
we will always (implicitly) denote the polynomial whose existence is
required by the definition.

\bigskip

Note, for instance, that examples~(\ref{it:sat}) and~(\ref{it:clique})
above are obviously $p$-relations, whereas example~(\ref{it:part})
evidently violates condition~(\ref{it:psize}) of
Definition~\ref{dfn:prel} and example~(\ref{it:gsub}) violates
condition~(\ref{it:pdec}) of the same definition unless $\cP=\cNP$.

\bigskip

Finally, observe that it is not difficult to realize that the above
mentioned versions~(\ref{it:decprob}) and~(\ref{it:cntprob}) of
combinatorial problems give rise, in the case of $p$-relations, to the
well known classes \cNP, of decision problems \citep{GJ79} and,
respectively, \cnP, of counting problems \citep{Val79}.

\subsection{Self-reducible $p$-relations}
\label{ssec:selfred}

We now recall the definition of \emph{self-reducible} $p$-relation as
introduced by \citet{Sch76,JVV86}:
\begin{dfn}
  \label{dfn:selfred}
  A $p$-relation $R\subseteq\reldom$ is
  \sidx{self-reducible}{p@$p$-relation} if there exist two polynomial
  time computable functions $\psi:\reldom\to\fldom$
  and $\sigma:\fldom\to\bN$ such that, for every
  $\alpha,\omega,\beta=b_1\ldots b_n \in\fldom$,
  \begin{enumerate}[(i)]
  \item $\sigma(\alpha)=O(\log|\alpha|)$,
  \item $R(\alpha)\not=\emptyset$ implies $\sigma(\alpha)>0$,
  \item $|\psi(\alpha,\omega)|\leq |\alpha|$ and
  \item \label{it:regviol} $\alpha \rR \beta$ iff $\psi(\alpha,
    b_1\ldots b_{\sigma(\alpha)}) \rR b_{\sigma(\alpha)+1}\ldots b_n$.
  \end{enumerate}
\end{dfn}

Intuitively, a relation is self-reducible if, for every element
$\alpha$ of its domain, given a small (logarithmic) prefix $b_1\ldots
b_{\sigma(\alpha)}$ of a related $\beta$, it is possible to efficiently
compute a smaller (w.r.t.\ $\alpha$) element of its domain
$\psi(\alpha, b_1\ldots b_{\sigma(\alpha)})$ that is in relation with
the suffix $b_{\sigma(\alpha)+1}\ldots b_n$ of $\beta$.

\bigskip

Consider, for instance, the relation of example~(\ref{it:sat}) above.
Let $F$ be a boolean formula on the variables $v_1, \ldots, v_n$ and
$\tau:\{v_1, \ldots, v_n\}\to\{\text{true}, \text{false}\}$ be a
satisfying truth assignment for it. By substituting in $F$ the values
$\tau(v_1), \ldots, \tau(v_{\log n})$ in place of the variables $v_1,
\ldots, v_{\log n}$, one (efficiently) obtains a new formula $F'$
smaller (with less variables, and hence shorter) than $F$. It is also
evident that if $v'_1, \ldots, v'_{n-\log n}$ are the variables of
$F'$, then $\tau': \{v'_1, \ldots, v'_{n-\log n}\}\to\{\text{true},
\text{false}\}$ defined as $\tau'(v'_i)=\tau(v_{i+\log n})$ for $1\leq
i\leq n-\log n$ is a satisfying truth assignment for $F'$. Hence, such
a $p$-relation is self-reducible.

\subsubsection{Not every $p$-relation is self-reducible}

The notion of self-reducibility is very natural and applies to many of
the well known combinatorial problems to the extent that \citet{JVV86}
state that ``problems which cannot be formulated in a self-reducible
way seem to be the exception rather than the rule''.  Nonetheless,
here we want to note that there are some very natural cases of
relations which do not enjoy this property and this fact is very
relevant for this thesis, as we discuss in the following.

\medskip

A natural class of $p$-relations can be defined by means of formal
languages. For a fixed $L\subseteq\fldom$, we define the
corresponding \eidx{slice relation} $R_L$ as $a^n\rR_L \alpha$ iff
$\alpha\in\Sigma^n\cap L$ (for some fixed $a\in\Sigma$).

\smallskip

Take now as example the regular language $cd^*$ of words beginning
with a letter $c$ followed by any number of $d$'s.  It is very simple
to conclude that its corresponding slice relation is not
self-reducible.  For every function $\sigma,\psi$ (since for every
$n>0$ it must be $\sigma(a^n)>0$), every suffix
$b_{\sigma(a^n)+1}\ldots b_n$ of $cd^*$ will be made only of $d$'s,
but no word of this kind belongs to the original regular language.
Hence, whichever is the length $m$ such that $1^m=\psi(1^n, b_1\ldots
b_{\sigma(a^n)})$, condition~(\ref{it:regviol}) of
Definition~\ref{dfn:selfred} can never hold.


\RCSfooter$Id: c1s3.tex,v 2.3 1999/11/01 11:30:53 santini Exp $

\section{A Short Survey}
\label{sec:jvv}

In this section we want to briefly summarize some relationship between
uniform random generation, approximate counting and other ``versions''
of combinatorial problems as discussed by \citet{JVV86,JS89}.

\subsection{Basic definitions}
\label{ssec:basicdefsrgac}

First of all, we recall some basic definitions of \citet{JVV86} we
here adapt to our notation, for the sake of consistency with the
following chapters\footnote{however, it remains an easy task to verify
  that the all the definitions we give here, are, to our aim,
  equivalent to the original one.}. We begin with random generation:

\begin{dfn}
  \label{def:aug}
  An algorithm \sA is an \esidx{almost uniform generator}{\irpR}
  \mdidx{uniform generator}{\irpR~(almost)} for a $p$-relation
  $R\subseteq\reldom$ iff, for every \esidx{tolerance}{\irpR!almost
    uniform generator} $\epsilon\in(0,1)$ and $\alpha\in\fldom$ such
  that $R(\alpha)\not=\emptyset$,
  \begin{enumerate}[(i)]
  \item \sA on input $\alpha,\epsilon$ gives output $\sA(\alpha,\epsilon)\in
    R(\alpha)\cup\{\no\}$,
  \item for every\footnote{in the following, to avoid confusion with
      $|\cdot|$ which will be used to denote bote the size of an
      object in a combinatorial structure and the length of a word in
      a formal language, we denote by $\#A$ the cardinality of the set
      $A$.}  $\beta\in R(\alpha)$, $(1-\epsilon)/\#R(\alpha) \leq
  \Pr\{ \sA(\alpha,\epsilon)=\beta \mid \sA(\alpha,\epsilon)\not=\no
  \} \leq (1+\epsilon)/\#R(\alpha)$,
  \item $\Pr\{ \sA(\alpha,\epsilon)=\no \}<1/4$.
  \end{enumerate}
  Moreover, an almost uniform generator is said \emph{fully
    polynomial} iff it works in time polynomial in $|\alpha|$ and
  $\log( 1/\epsilon)$.
\end{dfn}
Observe that the inclusion of the logarithm means that a fully
polynomial almost uniform generator can, at a modest computational
expense, achieve an output distribution which is very close to
uniform.  If we let the tolerance $\epsilon=0$, we get the following
\begin{dfn}
  \label{def:eug}
  An algorithm \sA is an \esidx{exact uniform generator}{\irpR}
  \mdidx{uniform generator}{\irpR~(exact)} for a $p$-relation
  $R\subseteq\reldom$ iff, for every $\alpha\in\fldom$ such that
  $R(\alpha)\not=\emptyset$,
  \begin{enumerate}[(i)]
  \item \sA on input $\alpha$ gives output $\sA(\alpha)\in
    R(\alpha)\cup\{\no\}$,
  \item $\Pr\{ \sA(\alpha)=\beta \mid \sA(\alpha)\not=\no
    \}=1/\#R(\alpha)$, for every $\beta\in R(\alpha)$, and
  \item $\Pr\{ \sA(\alpha,\epsilon)=\no \}<1/4$.
  \end{enumerate}
\end{dfn}

We now turn to counting, more precisely, to approximate counting
algorithm, in the sense of \citet{Sto83,JVV86}.
\begin{dfn}
  \label{def:racr}
  An algorithm \sA is a \esidx{randomized approximate counter}{\irpR}
  \mdidx{randomized counter}{\irpR~(approximate)} within \emph{ratio}
  $\rho:\bN\to\bR^+$ for a $p$-relation $R\subseteq\reldom$ iff, for
  every $\alpha\in\fldom$ such that $R(\alpha)\not=\emptyset$,
  \begin{enumerate}[(i)]
  \item \sA on input $\alpha$ gives output
    $\sA(\alpha)\in\bQ\cup\{\no\}$,
  \item $\Pr\{ 1/\rho(|\alpha|) \leq \sA(\alpha)/\#R(\alpha) \leq
    \rho(|\alpha|) \mid \sA(\alpha)\not=\no \}>3/4$ and
  \item $\Pr\{ \sA(\alpha)=\no \}<1/4$.
  \end{enumerate}
\end{dfn}

\begin{dfn}
  \label{def:rasprel}
  An algorithm \sA is a \esidx{randomized approximation scheme}{\irpR}
  \mdidx{randomized approximation scheme}{\irpR} for a $p$-relation
  $R\subseteq\reldom$ iff, for every $\alpha\in\fldom$ such that
  $R(\alpha)\not=\emptyset$ and every $\epsilon\in(0,1)$,
  \begin{enumerate}[(i)]
  \item \sA on input $\alpha,\epsilon$ gives output
    $\sA(\alpha,\epsilon)\in\bQ\cup\{\no\}$,
  \item $\Pr\{ (1-\epsilon)\#R(\alpha)\leq \sA(\alpha,\epsilon) \leq
    (1+\epsilon)\#R(\alpha) \mid \sA(\alpha,\epsilon)\not=\no \}>3/4$
    and
  \item $\Pr\{ \sA(\alpha,\epsilon)=\no \}<1/4$.
  \end{enumerate}
  Moreover, a randomized approximation scheme is said to be
  \emph{fully polynomial time} whenever it works in time polynomial in
  $n$ and $1/\epsilon$.
\end{dfn}

\bigskip

We conclude this section noting that the constants $1/4$ and $3/4$
present in the definitions of (almost) uniform generator, randomized
approximate counter and approximation scheme can be replaced by other
suitably chosen constants, leaving such definitions substantially
unchanged by essentially the same arguments discussed in
Section~\ref{sec:combstrdef}.

\subsection{A theoretical solution to random generation}
\label{ssec:polyhyer}

A widely discussed topic in the literature is the strict relationship
existing between the problem of counting and generating combinatorial
objects.  This relationship is made precise in the work of
\citet{JVV86} by means of oracle (probabilistic) Turing machines with
oracles in the polynomial hierarchy \citep{Sto77}.

Here, we want only to briefly recall that a \emph{Turing machine with
  oracle $O\subset\fldom$} \mdidx{Turing machine}{with oracle} is a
Turing machine\footnote{see Section~\ref{sec:sbtm} for further details
  on Turing machines, and \citet{GJ79} for a formal definition of the
  oracle model.}  endowed in addition with an \esidx{oracle
  tape}{\irTM} and a distinguished \esidx{query state}{\irTM}. The
computation of such machine can be described as for the standard
Turing machine except for the fact that when the machine enters the
query state, with the string $\omega$ written on the oracle tape, the
``oracle'' answers the question ``$\omega\in O$'' by replacing, in one
single move, the contents of the oracle tape with (a suitable encoding
of) ``yes/no''.  The \eidx{polynomial hierarchy} of Stockmeyer is
defined building a hierarchy of classes of languages by means of such
oracle Turing machines (see \citep{Sto77} for more details and for a
definition of the classes of formal languages $\Sigma^P_n$).

For the case of exact uniform generation, \citet{JVV86} prove the
following
\begin{pro}
  Let $R\subseteq\reldom$ be a $p$-relation. Then there exists a
  polynomial time oracle (probabilistic) Turing machine that is an
  exact uniform generators for $R$ either taking a suitable oracle in
  \cnP, or in $\Sigma^P_2$.
\end{pro}

In the case of almost uniform generation, a weaker request for the
oracle is possible; again \citet{JVV86} show the following
\begin{pro}
  Let $R\subseteq\reldom$ be a $p$-relation. Then there exists an
  oracle (probabilistic) Turing machine, with a suitable oracle in
  \cNP ($\Sigma^P_1$), which is an almost uniform generator for $R$
  working in time $|\alpha|$ and $1/\epsilon$.
\end{pro}

\medskip

Observe that a usual computational complexity conjecture is that the
classes \cNP, \cnP and $\Sigma^P_2$ contain \emph{highly intractable}
problems \citep[see, for instance][]{BDG95}; for this reason, the
above mentioned results are to be considered mainly of theoretical
interest.

\subsection{Relationships with other versions of combinatorial problems}
\label{ssec:relvers}

In this subsection we discuss the relationships of the random
generation problem with other versions of combinatorial problems:
namely the decision and counting versions.

It is clear that a procedure for counting the elements of $R(\alpha)$
must in particular decide if $R(\alpha)=\emptyset$, and hence counting
is computationally at least as hard as decision. The first evidence
that the counting version can be harder than the decision one for
significant natural problems was proven by \citet{Val79}. Consider the
$p$-relation $(G,M)$ where $G=\langle V, E\rangle$ is a graph and
$M\subseteq E$ is a \emph{perfect matching}\footnote{a perfect
  matching of a graph is a set of edges such that every node of the
  graph is the endpoint of precisely one edge in the match.} of $G$;
Valiant showed that counting the number of perfect matching in a
bipartite graph (or, equivalently, to compute the permanent of a
$\{0,1\}$ valued matrix) is \cnP-complete and hence likely to be
computationally intractable, whereas deciding whether a bipartite
graph contains a perfect matching (or, equivalently, deciding if the
permanent of a $\{0,1\}$ valued matrix is non-zero) is in \cP, by
virtue of the classical ``augmenting path'' algorithm.

Now we give some evidence that uniform generation is, from the
computational complexity point of view, somewhere in between decision
and counting.

As an example of the fact that the uniform generation version of a
problem can be harder than the construction (and hence decision) one,
consider the $p$-relation $R_1=(G,c)$ where $G=\langle V, E\rangle$ is
a directed graph and $c$ is a directed simple cycle of $G$. As one can
verify, the decision version of the combinatorial problem given by the
relation $R_1$ is easily solvable; even the construction version:
given a graph $G$, to output one of its (directed, simple) cycles, is
clearly in \cP. Nonetheless, \citet{JVV86} proved the following
\begin{pro}
  If there exists a polynomial time uniform generator for $R_1$, then
  $\cNP=\cRP$.
\end{pro}

Hence, as an example of the fact that the uniform generation version
can be easier than the counting one, consider the $p$-relation
$R_2=(F,\tau)$ where $F$ is a boolean formula in disjunctive normal
form (DNF), and $\tau$ is one of its satisfying assignments. One can
verify \citep{JVV86} that the counting problem for $R_2$ is
\cnP-complete, a fact that comes from a reduction from the analogous
problem for conjunctive normal form (CNF) boolean formulas which has
been proven \cnP-complete by \citet{Sim77}. Nonetheless, \citet{JVV86}
prove the following
\begin{pro}
  \label{pro:dnfurg}
  There exists a polynomial time uniform generator for $R_2$.
\end{pro}

\subsection{Almost uniform generation and approximate counting}
\label{ssec:almost}

Going back to almost uniform generation, one finds again a strong
relationship with approximate counting. Let $R\subseteq\reldom$ be a
self-reducible $p$-relation and let $\psi$ and $\sigma$ be the
functions cited in Definition~\ref{dfn:selfred}. For some
$\alpha\in\fldom$ belonging to the domain of $R$, define the
\emph{tree of derivations} \citep[for more details, see][]{JS89}
$T_R(\alpha)$ of $R$ as the tree whose vertices are pairs
$v=(\delta,\gamma)\in\reldom$ such that
\begin{enumerate}[(a)]
\item the root is $(\alpha,\ew)$, where $\ew$ is the empty word;
\item for any other node $v=(\delta,\gamma)$
  \begin{itemize}
  \item[--] if $|\delta|=1$, then $v$ is a leaf, otherwise
  \item[--] if $|\delta|>1$, then $v$ has a child
    $v_\omega=(\psi(\delta,\omega),\gamma\cdot\omega)$ for every
    $\omega\in\Sigma^{\sigma(\gamma)}$ such that
    $R(\psi(\delta,\omega))\not=\emptyset$.
  \end{itemize}
\end{enumerate}
It should be clear that the second part of all the vertices are
distinct and that the second part of the leaves are precisely the
elements of $R(\alpha)$, without repetitions. The bounds on $\psi$ and
$\sigma$ in the definition of self-reducibility ensure that the depth
of $T_R(\alpha)$ and the number of children of every vertex of
$T_R(\alpha)$ are bounded by a polynomial in $|\alpha|$.

Most known uniform generation algorithms for combinatorial structures
(see, for example the book of \citet{NW78}, or even
Algorithm~\ref{alg:dturg} of Section~\ref{ssec:rnddt}) may be viewed
as instances of the following reduction to the corresponding counting
problem.  Given that the structures are described by a self-reducible
relation $R$, select a random path from the root of $T_R$ to a leaf,
at each stage choosing the next edge with probability proportional to
the number of solutions in the maximal subtree rooted at its lower
end.  This information may be obtained from a counter which evaluates
the function $\#R(\omega)$ for appropriate $\omega$ in the tree.
Moreover, by appending a correction process (based on the a posteriori
probability of the path), such a procedure can be made to work even if
the counter is slightly inaccurate.

More formally, let $R\subseteq\reldom$ be a self-reducible
$p$-relation such that $|\beta|=O(|\alpha|^{\kappa_R})$ if
$\alpha\rR\beta$, for some $\kappa_R>0$. Then, \citet{JVV86} have
proven that
\begin{enumerate}[(i)]
\item if there exists a polynomial time randomized approximate counter
  for $R$ within ratio $1+|\alpha|^{\kappa_R}$, then there exists a
  fully polynomial time almost uniform generator for $R$;
\item if there exists a polynomial time almost uniform generator for
  $R$ with tolerance $|\alpha|^{2\kappa_R}$, then there exists a fully
  polynomial time randomized approximate counter for $R$.
\end{enumerate}
This, in particular, implies that the following holds
\begin{pro}
  Let $R\subseteq\reldom$ be a self-reducible $p$-relation. Then there
  exists a fully polynomial time randomized approximate counter for
  $R$ iff there exists a fully polynomial time almost uniform
  generator for $R$.
\end{pro}

\bigskip

On the other hand, when a rather cruder counting information is
available (to within a constant factor, say) the above mentioned
scheme of random generation through (approximate) counting breaks down
owing to the accumulation of errors which are too large to be
corrected. In such a case a more flexible and self-correcting approach
is taken by \citet{JS89} in which a random process moves dynamically
around the tree $T_R$, with backtracking allowed. The generator hence
views the vertices of the tree as the states of a Markov chain in
which there is a non-zero transition probability between two states
iff they are adjacent in $T_R$. The transition probability themselves
are computed with the aid of the crude approximate counter. Clearly,
all states communicate, so that, leaving aside questions of
periodicity, if the chain is allowed to evolve for $t$ steps (from any
initial state), the distribution of its final state approaches a
unique stationary distribution as $t\to\infty$. If the transition
probabilities are such that the stationary distribution is uniform
over the leaves of $T_R$, then one gets an almost uniform generator by
simulating the Markov chain for sufficiently many steps.

The study of the \emph{rapid mixing} property of such a chain, \ie the
speed of the convergence of the chain to the stationary distribution,
is one of the main result of \citet{JS89}. Thanks to a very
sophisticated analysis, the authors are able to prove that, if
$R\subseteq\reldom$ is a self-reducible $p$-relation for which there
exists a polynomial randomized approximate counter within ratio
$1+O(|\alpha|^\kappa)$ for some \emph{arbitrary} $\kappa\in\bR$, then
\begin{enumerate}[(i)]
\item there exists a fully polynomial almost uniform generator for
  $R$, and
\item there exists a fully polynomial approximation scheme for $R$.
\end{enumerate}

This fact, in particular, proves that approximation counting problems,
in the case of $p$-relation, cannot give rise to an ``approximation
hierarchy'' of relations (as, for instance, in the case of
combinatorial optimization problems, as discussed in
Section~\ref{ssec:approxalgclass}). From the previous results, in
fact, one can obtain the following surprising
\begin{pro}
  Let $R\subseteq\reldom$ be a self-reducible
  $p$-relation. Let $A\subset\bR$ such that there exists a polynomial
  randomized approximate counter within ratio $1+O(|\alpha|^\kappa)$ for
  some $\kappa\in A$. Then, either $A=\emptyset$, or $A=\bR$.
\end{pro}

\bigskip



\RCSfooter$Id: c2.tex,v 2.4 1999/10/29 10:24:02 santini Exp $

\chapter{Uniform~Random~Generation Through~Ranking}
\label{cha:ranking}

\mycite{italian}{%
  Così si potesse dimezzare ogni cosa intera, \dots così ognuno
  potesse uscire dalla sua ottusa e ignorante interezza\dots Se mai
  diventerai metà di te stesso, e te l'auguro, ragazzo, capirai cose
  al di là della comune intelligenza dei cervelli interi. Avrai perso
  metà di te e del mondo, ma la metà rimasta sarà mille volte più
  profonda e preziosa.}%
{Italo Calvino}{Il visconte dimezzato}

In this chapter we discuss how to apply the concept of \emph{ranking}
to the problem of uniform random generation of $p$-relations. To this
end, we briefly recall the concept of ranking for formal languages and
suggest an extension of such concept to the case of relations. Then we
prove a theorem stating that if a $p$-relation is rankable in
polynomial time, then it admits a polynomial time uniform random
generator.  Finally, we extend some known results on ranking for
formal languages to the case of $p$-relations, a fact that allows us
to introduce two new classes of polynomial time rankable
$p$-relations.

\bigskip
\RCSfooter$Id: c2s1.tex,v 2.5 1999/09/27 07:48:45 santini Exp $

\section{Ranking $p$-Relations and Uniform Random Generation}
\label{sec:rankrug}

Suppose that, given a totally ordered set of objects, we are able to
tell how many they are and also to determine each of them once its
position in the order is given. Then, to generate such objects
uniformly at random, we can simply pick an integer uniformly at random
between $1$ and their total number, returning then the corresponding
object. We now show how this argument can be made precise in the case
of $p$-relations.

\subsection{The ranking for formal languages}  
\label{ssec:rankfl}

Suppose that \lexa is a total order relation over some finite alphabet
$\Sigma$, then this order can be extended to a total order relation
\lex over the elements of $\fldom$, called \eidx{lexicographic order}:
if $\omega,\omega'\in\fldom$, then $\omega\lex\omega'$ iff either
$\omega\sigma=\omega'$, or $\omega=\gamma a\sigma$, $\omega'=\gamma
b\sigma'$ with $a\lexa b$ and $a\not=b$ (where
$\gamma,\sigma,\sigma'\in\fldom$ and $a,b\in\Sigma$).  Hence, we have
the following
\begin{dfn}
  The \edidx{rank}{for formal languages} $r_L$ of a formal language
  $L$ is the function $r_L:\fldom\to \bN$ defined as
  $r_L(\omega)=\#\{\omega': \omega'\in L, |\omega'|<|\omega|
  \;\text{or}\; |\omega'|=|\omega| \;\text{and}\; \omega'\lex
  \omega\}$; conversely, the \edidx{unrank}{for formal languages}
  $u_L$ is a function $u_L:\bN\to\fldom$ such that $u_L(n)=\omega$,
  where $\omega$ satisfies $r_L(\omega)=n$ and for every $\omega'\lex
  \omega$, $r_L(\omega')<n$ (here we assume $|L|=\infty$).
\end{dfn}

The complexity of computing the rank function has been studied in
\citep{GS91} where it has been considered as a special kind of optimal
compression: ranking is in fact related to the more general problem of
storing and retrieving strings efficiently, a task very similar to the
one we sketched at the beginning of this section. As one can verify,
the ranking of every language in \cP belongs to \cnP; in particular,
several languages in \cP are very hard to rank.  This is the case, for
example, of languages accepted by nondeterministic log-time bounded
Turing machines, log-space bounded deterministic Turing machines,
one-way nondeterministic log-space bounded Turing machines, uniform
families of constant depth and polynomial size unbounded fan-in
circuits, CRCW P-RAM working in constant time with a polynomial number
of processor, two-way deterministic checking stack automata, two-way
deterministic pushdown automata and one-way two-head deterministic
finite state automata. For all these classes of languages, as proved
by \citet{Huy90}, the ranking is even \cnP-hard.

Nonetheless, classes of languages for which it is possible to
efficiently compute the rank function are known. For regular
languages, the ranking is $\cNC^1$-reducible to integer division
\citep{BBG91,Huy91}, and hence it can be solved by log-space uniform
boolean circuits of $O(\log n\log\log n)$ depth and polynomial size
\citep{Coo85}. Moreover, for every language accepted by one-way
unambiguous log-space bounded Turing machines, the rank function
belongs to the class DET \citep{GS91} of all functions
$\cNC^1$-reducible to computing the determinant of an integer matrix
\citep{Coo85}; this result has been further extended to one-way
log-space bounded Turing machines with bounded ambiguity degree by
\citet{BG93}, and similar results for Turing machines with
simultaneous complexity bounds are given by \citet{BMP94}.  Finally,
\citet{Huy90} showed that the rank function is in $\cNC^2$ for all
languages accepted in polynomial time by one-way unambiguous
(log-space) auxiliary pushdown automata hence, in particular, for
unambiguous context-free languages.

\subsection{The case of relations}
\label{ssec:rankrel}

We now extend the above-mentioned definition of ranking, usually given
for formal languages, to the case of relations.
\begin{dfn}
  The \edidx{rank}{for relations} $r_R$ of a relation
  $R\subseteq\reldom$ is the function $r_R:\reldom\to \bN$ defined as
  $r_R(\alpha,\beta)=\#\{\beta': \alpha\rR \beta', |\beta'|<|\beta|
  \;\text{or}\; |\beta'|=|\beta| \;\text{and}\; \beta'\lex \beta\}$;
  conversely, the \edidx{unrank}{for relations} $u_R$ is the function
  $u_R:\fldom\times\bN\to\fldom$ such that $u_R(\alpha,n)=\beta$,
  where $\beta$ satisfies $r_R(\alpha,\beta)=n$ and for every
  $\beta'\lex \beta$, $r_R(\alpha,\beta')<n$ whenever $n\leq
  \#\{\beta: \alpha\rR \beta\}$, else $u_R(\alpha,n)$ is undefined.
\end{dfn}

Observe that the notion of rank for languages and the one here given
for relations are strictly related. Given a relation
$R\subseteq\reldom$, define the ``corresponding'' language as
\[
L_R=\{\alpha\sep\beta: \alpha\rR\beta\},
\]
where $\sep$ is some distinct symbol not in $\Sigma$ such that
$\sigma\lexa\sep$ for every $\sigma\in\Sigma$; conversely, given a
language $L\subseteq\fldom$, define the ``corresponding'' \idx{slice
  relation}\footnote{see Section~\ref{sec:prel}.} as
\[
\alpha \mathord R_L\beta \;\text{iff}\;  \alpha=a^n, \beta\in L\cap\Sigma^n, n\in\bN,
\]
for some $a\in\Sigma$. It is then possible to check that if $R$ is a
$p$-relation, the following equalities hold
\begin{eqnarray}
    r_L(\alpha) & 
    = & r_{R_L}(\sep^{|\alpha|},\alpha)+\sum_{i=1}^{|\alpha|-1} r_{R_L}(\sep^i,\sep^i), \label{eq:rLRL}\\
    r_{R_L}(\sep^n,\alpha) & 
    = & r_L(\alpha) - \max_{|\gamma|<n} r_L(\gamma), \label{eq:rRLL} \\      
    r_R(\alpha,\beta) & 
    = & r_{L_R}(\alpha\sep\beta)- \max_{\gamma\lex \alpha}r_{L_R}(\gamma\sep\sep^{p(|\gamma|)}), \label{eq:rRLR} \\
    r_{L_R}(\alpha\sep\beta) & 
    = & r_R(\alpha,\beta) + \sum_{\gamma\lex \alpha} r_R(\gamma,\sep^{p(|\gamma|)}). \label{eq:rLRR} 
\end{eqnarray}
In particular, it is possible to prove that
\begin{lem}
  \label{lem:rlrr}
  If $R\subseteq\reldom$ is a $p$-relation, then $r_L\in\cFP$ iff
  $r_{R_L}\in\cFP$ and if $r_{L_R}\in\cFP$, then $r_R\in\cFP$.
\end{lem}
\begin{proof}
  The proof follows from equations~(\ref{eq:rLRL})--(\ref{eq:rRLR}),
  since the right hand side, in the given hypotheses, can be computed
  in polynomial time.  Observe that, for equation~(\ref{eq:rLRR}), the
  sum extends over an exponential number of items, so the lemma holds
  in the stated direction.
\end{proof}
Given these simple relationships, one can in principle directly apply
the known result about the ranking for formal languages to the case of
relations. Nonetheless, in the next two sections, we will give some
further independent result, strictly concerning the case of
$p$-relations.

\subsection{The unranking}
\label{ssec:unrank}

If the ranking is easy for some language $L$, in general it is not
also true that the unranking is similarly easy.  \mdidx{unranking}{for
  relations} \mdidx{unranking}{for formal languages} Let, for
instance, $L$ be the unary language $\{a^{2^n}:n\in\bN\}$: it is
evident that even if the ranking amounts simply to the computation of
the a logarithm in base $2$, the unranking function (even if its input
is encoded on a unary alphabet) has an exponentially long output, thus
it cannot at all be computed in polynomial time!

The situation, with $p$-relations, is different, since the codomain of
such a relation, for every element of the domain, has a cardinality
that is bouded above by a function the size of the element itself.
Hence
\begin{lem}
  If $R\subseteq\reldom$ is a $p$-relation and
  $r_R\in\cFP$, then $u_R\in\cFP$.
\end{lem}
\begin{proof}
  Since $\#\{\beta: \alpha\rR\beta\}\leq 2^{p(|\alpha|)}$, $u_L$ can
  be computed by binary search with at most
  $O\!\left(p(|\alpha|)\right)$ calls to $r_R$.
\end{proof}

\subsection{A  uniform random generator}
\label{ssec:rugprel}

We are now able to state formally the claim sketched at the beginning
of this section. Recall that in Section~\ref{ssec:algexampl} we have
designed an algorithm on \mPrRAM able to generate (with high
probability) a number in $\{1,\ldots, N\}$ uniformly at random in
time $O(\log N\log\log N)$ (see Proposition~\ref{pro:rNg}). It is
straightforward then to conclude that
\begin{teo}
  \label{teo:rugprel}
  If $R\subseteq\reldom$ is a $p$-relation and $r_R\in\cFP$, then $R$
  admits a polynomial time uniform random generator.
\end{teo}

\bigskip

In the next two sections of this chapter we show two classes of
$p$-relations, characterized in terms of acceptor devices, that admit
``efficient'' (\ie polynomial time) ranking. Thanks to the last
theorem of this section, for both such classes it will indeed be
possible to design ``efficient'' uniform random generators.


\RCSfooter$Id: c2s2.tex,v 2.4 1999/11/01 11:30:53 santini Exp $

\section{Turing Machines with Simultaneous Complexity Bounds}
\label{sec:sbtm}

In this section, we show how it is possible to extend to the case of
$p$-relations the result about the ranking of languages accepted by
nondeterministic Turing machines with simultaneous bounds on their
resources of \citet{BMP94}.

\subsection{The case of formal languages}
\label{ssec:sbtmfl}

We begin by briefly recalling the model of Turing machine we refer to
in this section (for a more detailed description, see~\citep{GJ79}).

\begin{figure}[ht]
  \begin{center}
    \input{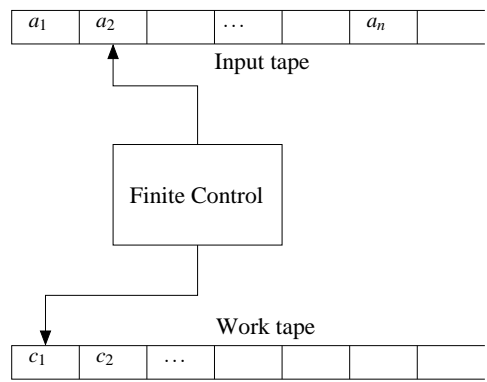}
    \caption{A Turing machine.}
    \label{fig:mdt}
  \end{center}
\end{figure}

A (deterministic) \edidx{Turing machine}{deterministic} (\mTM)
consists of a read-only \esidx{input tape}{\irTM}, a read-write
\esidx{output tape}{\irTM} and a \esidx{finite control}{\irTM}
(Figure~\ref{fig:mdt}). The tapes are sequences of cells each of which
can hold a \esidx{tape symbol}{\irTM} of some alphabet $\Sigma$; the
finite control can be in one of a finite set of \esidx{states}{\irTM},
two of which are distinguished and called the \esidx{initial
  state}{\irTM} and \esidx{accepting state}{\irTM}.  Given the symbols
under the input and work tape heads and the current state a
\esidx{next-move function}{\irTM} determines the next state, the head
movements on both tapes and possibly a symbol to be written on the
work tape. The \esidx{semantic}{\irTM} \mdidx{semantic}{\irTM} of the
computation of the machine on input $\omega=a_1\ldots a_n$ can be
defined as follows: initially the input tape contains $a_1 \ldots a_n$
all work tape cells are blank and both tape heads scan the first
(leftmost) cell of the tapes.  The input $\omega$ is said to be
\emph{accepted} iff the \mTM, started in the initial state, makes a
sequence of moves which eventually enters the accepting state.  The
\esidx{language accepted}{\irTM} \mdidx{language accepted}{\irTM} by a
\mTM is the set accepted inputs.

A \emph{nondeterministic} Turing machine
\msidx{nondeterministic}{\irTM} \mridx{nondeterministic Turing
  machine}{\irTM} (\mNTM) is simply a Turing machine in which the
next-move function, instead of being a (partial) map from the tuple
``state, input tape symbol, work tape symbol'' to the tuple ``state,
input tape move, work tape move and symbol to write'' as in the
deterministic case, becomes a map from the same domain to subsets of
the codomain. The \msidx{semantic}{\irTM}
\mdidx{semantic}{\irTM!nondeterministic} computation can then take
``nondeterministically'', at every step, one of the possible next
moves given by such mapping. In this case a string is said to be
accepted if at least one of such nondeterministic computations ends in
an accepting state.

Various computational resources for such models can be defined as
follows, according to the usual worst case criterion; observe that we
assume without loss of generality that all the computations eventually
halt.  We say that a (deterministic) Turing machine
\mdidx{computational complexity}{\irTM} works in \esidx{time}{\irTM}
$t(n)$ (respectively, uses \esidx{space}{\irTM} $s(n)$, or makes
$i(n)$ \esidx{inversions}{\irTM}) iff, for every input
$a_1,\dots,a_n\in\Sigma$, the computation of the machine makes at most
$t(n)$ moves (respectively, consumes at most $s(n)$ cells of the work
tape, or has the input tape head change its direction at most $i(n)$
times). For the nondeterministic case, \mdidx{computational
  complexity}{\irTM!nondeterministic} the time, space and inversion
resources are similarly defined by taking the \emph{smallest} amount
of each resource consumed on every input by one of the possible
nondeterministic computations; moreover, the \esidx{ambiguity}{\irTM}
\mdidx{ambiguity}{\irTM} $d(n)$ is defined as the maximum number of
accepting computations, taken over all the inputs
$a_1,\dots,a_n\in\Sigma$.

\bigskip

We can now define a class of languages accepted by such machines with
simultaneous bounds on the computation:
\enlargethispage{2em}
\begin{dfn}
  \label{dfn:bmdtl}
  A language $L\subseteq\fldom$ belongs to $\cB(s(n),i(n),d(n))$,
  whenever it is accepted by a nondeterministic Turing machine that
  simultaneously uses space $s(n)$, makes $i(n)$ inversions and has
  ambiguity $d(n)$.
\end{dfn}


Finally, we are able to formally state the following result obtained
by \citet{BMP94}.
\begin{pro}
  If a language $L\subseteq\fldom$ belongs to $\cB(s(n),i(n),d(n))$
  and $s(n)\cdot i(n)\cdot d(n)=O(\log n)$, then $r_L\in\cFP$.
\end{pro}

%
%

\subsection{An extension to $p$-relations}
\label{sec:sbtmprel}

In order to apply the previous result to the case of $p$-relations,
first of all we need to extend the model of computation. The more
natural way of doing so, as depicted in Figure~\ref{fig:rmdt}, is to
add to the standard Turing machine an additional (two-way read-only) input
tape and to reserve each of the two input tapes to, respectively, the
domain and codomain part of the relation.

\begin{figure}[ht]
  \begin{center}
    \input{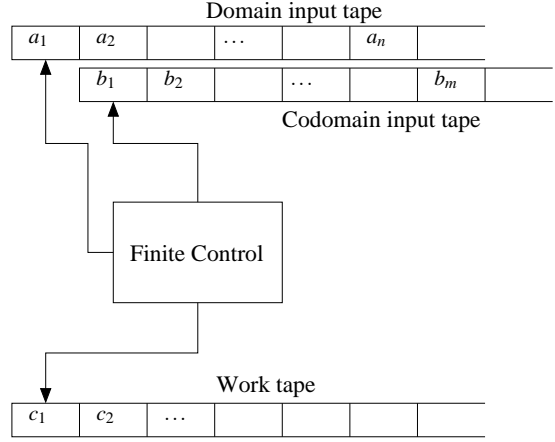}
    \caption{An extended Turing machine for recognizing relations.}
    \label{fig:rmdt}
  \end{center}
\end{figure}

The \emph{semantic} \msidx{semantic}{\ireTM}
\mdidx{semantic}{\irTM!extended} of the computation of the
\emph{extended} Turing machine \midx{Turing
  machine!extended|see{extended Turing machine}} \midx{extended Turing
  machine} on input $(\alpha,\beta)$ can be defined as follows:
initially the domain input tape contains $\alpha=a_1\dots a_n$, the
codomain input tape contains $\beta=b_1\ldots b_m$, all work tape
cells are blank and both the tape heads scan the first (leftmost) cell
of the tapes.  The input $(\alpha,\beta)$ is said to be
\emph{accepted} iff the extended \mTM, started in the initial state,
makes a sequence of moves which eventually enters the accepting state.
The \esidx{relation accepted}{\ireTM} \mdidx{relation
  accepted}{\ireTM} by an extended \mTM is the set of ordered pairs of
strings of input symbols so accepted.  The semantic for the
nondeterministic version can be defined similarly.

The resources of \esidx{time}{\ireTM}, \esidx{space}{\ireTM} and
\esidx{ambiguity}{\ireTM} \mdidx{ambiguity}{\irTM!extended} for the
extended model are defined as before, \mdidx{computational
  complexity}{\irTM!extended} but \emph{the number of input tape
  inversions are counted only for the codomain input tape head}
\msidx{inversions}{\ireTM} (the domain input tape head has always an
unrestricted number of moves). So we can finally extend
Definition~\ref{dfn:bmdtl} to the case of relations, with a little
abuse of notation.
\begin{dfn}
  \label{dfn:bmdtr}
  A relation $R\subseteq\reldom$ belongs to
  $\cB(s(n),i(n),d(n))$, whenever it is accepted by an extended
  nondeterministic Turing machine which simultaneously uses space
  $s(n)$, makes $i(n)$ inversions and has ambiguity $d(n)$.
\end{dfn}

We can finally give our first result on ``efficiently'' rankable
relations. By adapting a proof of \citet{BMP94}, we can indeed state
the following
\begin{teo}
  \label{teo:sbtm}
  If a $p$-relation $R\subseteq\reldom$ belongs to
  $\cB(s(n),i(n),d(n))$ and $s(n)\cdot i(n)\cdot d(n)=O(\log n)$, then
  $r_R\in\cFP$.
\end{teo}
\begin{proof}
  Note that every \mTM accepting a language in $\cB(s(n),i(n),d(n))$
  can be efficiently simulated by one-way (\ie with no input tape head
  inversions) \mTM using space $O(s(n)\cdot i(n))$ of the same order
  of ambiguity \citep{BMP94}. It is then easy to check that the same
  holds for the extended model.
  
  Moreover, if $M$ is an extended \mTM accepting a $p$-relation $R$,
  for every $\alpha\in\fldom$ and $\beta\in\Sigma^{p(|\alpha|)}$ it
  is immediate to construct a one-way extended \mTM $M^{(\beta)}$, of
  the same order of space and ambiguity, accepting the $p$-relation
  $R^{(\beta)}$ satisfying $\alpha\rR^{(\beta)}\gamma$ iff
  $\alpha\rR\gamma$ and $\gamma\lex\beta$.
  
  Hence, to prove the statement of the theorem, it is enough to show
  that if $R$ is a $p$-relation accepted by some one-way extended \mTM
  using space $s(n)$ and having ambiguity $d(n)$ such that $s(n)\cdot
  d(n)=O(\log n)$, then the cardinality of $\{\gamma :
  \alpha\rR^{(\beta)}\gamma\}=r_R(\alpha,\beta)$ can be computed in a
  time polynomial in $|\alpha|$.
  
  It is not hard to see that $r_R(\alpha,\beta)$ is exactly the number
  of $\gamma\in \Sigma^{p(|\alpha|)}$ for which $M^{(\beta)}$, having
  $\alpha$ on the first input tape and $\gamma$ on the second, halts
  in an accepting state.
  
  Since both input tapes are read-only, for every $\alpha\in\Sigma^n$,
  $\beta\in\Sigma^{p(n)}$ and $n\in\bN$, the configurations of
  $M^{(\beta)}$ having $\alpha$ on the first input tape are at most
  $2^{O(s(n))}=n^{O(1)}$.  Then (for instance by simulating
  $M^{(\beta)}$ with $\alpha$ on the first input tape) one can
  efficiently build two matrices $C^{(\alpha,\beta)}_a$, for
  $a\in\Sigma$, of polynomial size and values in $\Sigma$, such that
  the $(i,j)$th entry of $C^{(\alpha,\beta)}_a$ is $1$ iff
  $M^{(\beta)}$ moves from the $i$th to $j$th configuration having
  $\alpha$ on the first input tape and reading $a$ on the second input
  tape.
  
  If $\pi$ and $\eta$ are, respectively, the characteristic vector of
  then initial and accepting configurations of $M^{(\beta)}$, then one
  can verify that $\pi C^{(\alpha,\beta)}_{c_1}
  C^{(\alpha,\beta)}_{c_2} \cdots
  C^{(\alpha,\beta)}_{c_{p(n)}}\eta\leq d(n)$ is the number of
  accepting computations of $M^{(\beta)}$ on input $(\alpha,\gamma)$
  where $\gamma=c_1c_2\cdots c_{p(n)}$. Here we have implicitly
  assumed that $M^{(\beta)}$ always moves the head on its second input
  tape, but this is not a restriction since stationary moves can be
  eliminated \citep{BG93} by computing the transitive closure of
  suitable transition matrices representing stationary moves of
  $M^{(\beta)}$.
  
  Let now $q$ be an integer polynomial of degree $d(n)$ such
  that $q(0)=0$ and $q(\ell)=1$ for $1\leq \ell\leq d(n)$;
  then, it is possible to check that
  \[
  r_R(\alpha,\beta)=
    \sum_{i=1}^{p(n)} 
    \sum_{c_1,c_2,\ldots,c_{p(n)}\in\Sigma}
    q(%
    \pi C^{(\alpha,\beta)}_{c_1} C^{(\alpha,\beta)}_{c_2} \cdots C^{(\alpha,\beta)}%
    _{c_{p(n)}}\eta ). 
  \]  
  The above computation is essentially the same performed as the final
  step of Proposition~3 in \citep{BG93} and, as shown in the same
  paper, it can be transformed, using the matrix algebra given by
  direct sum and Kronecker product, in a form suitable to be computed
  by boolean circuits of polynomial size and depth $O(\log^2 n)$,
  hence, in overall polynomial time.
\end{proof}

\clearpage 

From this result, by immediate application of
Theorem~\ref{teo:rugprel}, we can introduce, via extended Turing
machines with simultaneous complexity bounds, a new class of
$p$-relations admitting a polynomial time uniform random generation
\begin{cor}
  \label{cor:prelsbtmurg}
  If a $p$-relation $R\subseteq\reldom$ belongs to
  $\cB(s(n),i(n),d(n))$ and $s(n)\cdot i(n)\cdot d(n)=O(\log n)$, then
  it admits a polynomial time uniform random generator.
\end{cor}


\RCSfooter$Id: c2s3.tex,v 2.3 1999/09/23 14:43:06 santini Exp $

\section{One-way Auxiliary Pushdown Automata}
\label{ssec:pda}

In this section, we show how to extend to the case of $p-$relation the
result about the ranking of languages accepted by auxiliary pushdown
automata by \citet{Huy90}.

\subsection{The case of formal languages}
\label{ssec:pdafl}

Also in this case, we begin by briefly recalling the model of
auxiliary pushdown automata we refer to in this section (for a more
detailed description, see \citep{HU79}).

\begin{figure}[ht]
  \begin{center}
    \input{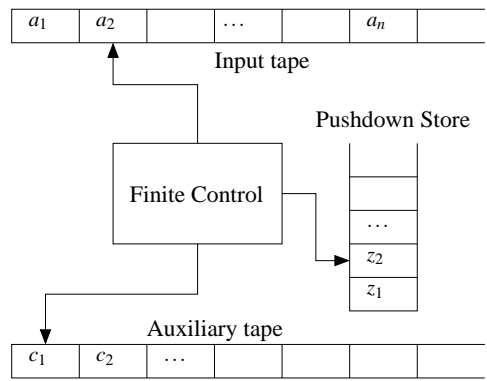}
    \caption{An auxiliary pushdown automaton.}
    \label{fig:apda}
  \end{center}
\end{figure}

A (deterministic) \eidx{one-way auxiliary pushdown automaton}
(\moAuxPDA) consists of a one-way read-only \esidx{input
  tape}{\irAuxPDA}, a two-way read-write \esidx{auxiliary
  tape}{\irAuxPDA}\footnote{as will be stated in the following, the
  auxiliary tape has a logarithmic bound on the number of usable
  cells.}, a \esidx{pushdown store}{\irAuxPDA} and a \esidx{finite
  control}{\irAuxPDA} (Figure~\ref{fig:apda}). The tapes and the store
are sequences of cells each of which can hold a
\esidx{symbol}{\irAuxPDA} of some alphabet $\Sigma$; one of these
symbols is distinguished and called the \esidx{initial pushdown
  symbol}{\irAuxPDA}.  The finite control can be in one of a finite
set of \esidx{states}{\irAuxPDA}, two of which are distinguished and
called the \esidx{initial state}{\irAuxPDA} and \esidx{accepting
  state}{\irAuxPDA}.  Given the symbols under the input and auxiliary
tape heads, the symbol on top of the pushdown store and the current
state, a \esidx{next-move function}{\irAuxPDA} determines the next
state, the auxiliary tape head movement with possibly a symbol to be
written on it and an action regarding the store which can either be a
push of a symbol on top, or a pop of a symbol from the top of the
store.  The \esidx{semantic}{\irAuxPDA} \mdidx{semantic}{\irAuxPDA} of
the computation of the automaton on input $\omega=a_1\dots a_n$ can be
defined as follows: initially the input tape contains $a_1\dots a_n$,
the pushdown store contains the initial symbol and all auxiliary tape
cells are blank and both tape heads scan the first (leftmost) cell of
the tapes.  The input $\omega$ is said to be \emph{accepted} iff the
automaton, started in the initial state, makes a sequence of moves
which eventually enters the accepting state, with the work tape empty
and the pushdown store containing only the initial symbol.  The
\esidx{language accepted}{\irAuxPDA} \mdidx{language
  accepted}{\irAuxPDA} by such an automaton is the set of accepted
strings.

A \emph{nondeterministic} one-way auxiliary pushdown automaton
\mridx{nondeterministic one-way auxiliary pushdown
  automata}{\irAuxPDA} \msidx{nondeterministic}{\irAuxPDA}
(\moNAuxPDA) is simply a one-way auxiliary pushdown automaton in which
the next-move function, instead of being a (partial) map from the
tuple ``state, input tape symbol, work tape symbol and top of the
pushdown store symbol'' to the tuple ``state, auxiliary tape move and
symbol to write on, action on the pushdown store'' as in the
deterministic case, becomes a map from the same domain to subsets of
the codomain. The computation can then take ``nondeterministically'',
at every step, one of the possible next-moves given by such mapping.
In this case a string is said to be accepted if at least one of such
nondeterministic computations ends in an accepting state.

Similarly to what happens with \mTM, various computational resources
for such models can be defined, according to the usual worst case
criterion; \msidx{computational complexity}{\irAuxPDA}
\mdidx{computational complexity}{\irAuxPDA!nondeterministic} observe
that we assume without loss of generality that all the computations
eventually halt.  We say that a (deterministic) one-way auxiliary
pushdown automaton works in \esidx{time}{\irAuxPDA} $t(n)$
(respectively, uses \esidx{space}{\irAuxPDA} $s(n)$) iff, for every
input $a_1\ldots a_n\in\fldom$, the computation of the automaton makes
at most $t(n)$ moves (respectively, consumes at most $s(n)$ cells of
the auxiliary tape).  For the nondeterministic case, the time and
space resources are similarly defined by taking the \emph{smallest}
amount of each resource consumed on every input by one of the possible
nondeterministic computations; moreover, the
\mdidx{ambiguity}{\irAuxPDA} \esidx{ambiguity}{\irAuxPDA} $d(n)$ is
defined as the maximum number of accepting computations, taken over
all the inputs $a_1\ldots a_n\in\fldom$.

Once we have defined the space resource, which measures only the space
on the auxiliary tape, not of the pushdown store, we \emph{restrict
  the above definitions of \moAuxPDA and \moNAuxPDA to automata which
  use at most logarithmic space}\footnote{without such restriction, as
  one can verify, such devices become in fact (computationally)
  equivalent to standard Turing machines.}, \ie for which $s(n)=O(\log
n)$.  Moreover, the one-way \emph{unambiguous} auxiliary pushdown
automata \mridx{one-way unambiguous auxiliary pushdown
  automata}{\irAuxPDA} \msidx{unambiguous}{\irAuxPDA} (\moUAuxPDA) is
a \moNAuxPDA for which $d(n)=1$, \ie an automaton which has at most
one accepting computation for every accepted string.

\bigskip

Finally, we can formally state the following result of \citet{Huy90}.
\begin{pro}
  If a language $L\subseteq\fldom$ is accepted by a polynomial time
  one-way unambiguous auxiliary pushdown automaton, then $r_L\in\cFP$.
\end{pro}

Notice that the class of languages accepted by polynomial time one-way
nondeterministic auxiliary pushdown automata is exactly the class of
languages that are reducible to context-free languages via one-way
log-space reductions \citep{BL92}.

\subsection{An extension to $p$-relations}
\label{ssec:pdaprel}

Again, in order to apply the previous result to the case of
$p$-relations, we need to extend the model of computation. As in the
case of Turing machines, the more natural way of doing so is to add to
the standard auxiliary pushdown automaton an additional two-way
read-only input tape for the \emph{domain} part of the relation,
leaving the usual one-way input tape for the \emph{codomain} part
(Figure~\ref{fig:rmdt}).

\begin{figure}[ht]
  \begin{center}
    \input{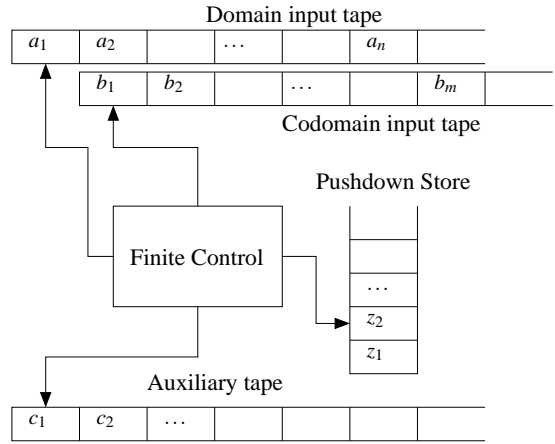}
    \caption{An extended auxiliary pushdown automaton for recognizing relations.}
    \label{fig:rapda}
  \end{center}
\end{figure}

The \emph{semantic} of the computation of the \emph{extended} one-way
auxiliary pushdown automaton \midx{extended one-way auxiliary pushdown
  automata} \midx{one-way auxiliary pushdown
  automata!extended|see{\ireAuxPDA}}
\mdidx{semantic}{\irAuxPDA!extended} on input $(\alpha,\beta)$ can be
defined as follows: initially the domain input tape contains
$\alpha=a_1\dots a_n$, the codomain input tape contains
$\beta=b_1\ldots b_m$, the pushdown store contains the initial symbol,
all auxiliary tape cells are blank and all tape heads scan the first
(leftmost) cell of the tapes.  The input $(\alpha,\beta)$ is said to
be \emph{accepted} iff if the extended \moAuxPDA, started in the
initial state, makes a sequence of moves which eventually enters the
accepting state.  The \esidx{relation accepted}{\ireAuxPDA}
\mdidx{relation accepted}{\ireAuxPDA} by an extended \moAuxPDA is the
set of ordered pairs of strings of input symbols so accepted.  The
semantic for the nondeterministic version can be similarly defined.

The resources of \esidx{time}{\ireAuxPDA}, \esidx{space}{\ireAuxPDA}
and \esidx{ambiguity}{\ireAuxPDA}
\mdidx{ambiguity}{\irAuxPDA!extended} for the extended model are
defined as before, \mdidx{computational
  complexity}{\irAuxPDA!extended} but \emph{only the codomain input
  tape is restricted to be one-way} (while the domain input tape is an
unrestricted two-way tape).

\bigskip

We are ready to give our second result on ``efficiently'' rankable
relations. By adapting a proof of \citet{BMP94}, we can prove the
following
\begin{teo}
  \label{teo:pda}
  If a $p$-relation $R$ is accepted by a polynomial time extended
  one-way unambiguous auxiliary pushdown automaton, then $r_R\in\cFP$.
\end{teo}
\begin{proof}
  Given an extended \moUAuxPDA $A$, with state set $Q$, and some
  $\alpha\in\fldom$, it is always possible to build (in time
  polynomial in $|\alpha|$) a \moUAuxPDA $A^{(\alpha)}$, with state set
  $Q\times\{1,\ldots,|\alpha|\} \times\Sigma$, such that the
  computation of $A^{(\alpha)}$ on input $\beta$ efficiently simulates
  the computation of $A$ on input $(\alpha,\beta)$ for every
  $\beta\in\fldom$. This is made possible by using the state
  $(q,i,a_i)$ of $A^{(\alpha)}$ to represent the automaton $A$ being in
  state $q$ having the the two-way head on position $i$ reading
  symbol $a_i$ (where $\alpha=a_1\ldots a_n$ for some $n\in\bN$).  It
  is then clear that if $L^{(\alpha)}$ is the language accepted by
  $A^{(\alpha)}$, $r_{L^{(\alpha)}}(\beta)=r_R(\alpha,\beta)$.
  
  As it is shown in \citep[Theorem 1.1]{Huy90}, for every language $L$
  accepted by a polynomial time \moUAuxPDA, $r_L(\beta)$ can be
  computed in $|\beta|^{O(1)}$ time for every $\beta\in\fldom$ by
  means of a sequential algorithm working on a suitably defined
  \emph{path system} depending on the \emph{surface configurations} of
  the automaton \citep{Coo70}.  In order to use this result on
  $A^{(\alpha)}$ we need a more accurate analysis of the constants
  hidden in the asymptotics.
  
  As one can verify, the (sequential) algorithm of \citep[Theorem
  1.1]{Huy90}, applied on a \moUAuxPDA $A'$, works on input
  $\beta\in\fldom$ in $k N^4 q(|\beta|)$ time for some constant $k$
  and a polynomial $q$, depending on the running time of $A'$, where
  $N$ is the number of nodes of the path system built by the
  algorithm. As an immediate consequence of the particular
  construction of the path system, $N=S^2 q(|\beta|)$, where $S$ is
  the number of surface configurations of $A'$.  Recall that a surface
  configuration \citep{Coo70} of a \moUAuxPDA is a tuple
  $(z,a,i,\gamma q\delta)$ where: $z$ is the topmost symbol of the
  stack, $a$ is the symbol on the input tape under the head that is in
  position $i$, $\gamma\delta$ is the content of the working tape and
  $q$ is the status of the automaton.  Thus, $S=h |Q'||\beta|$, where
  $h$ is a constant depending on the cardinality of the alphabets
  (which here are assumed to be constant) and $Q'$ is the set of
  states of $A'$.
  
  We are then able to conclude that, since $A^{(\alpha)}$ has $2|Q|\log
  |\alpha|$ states and $R$ is a $p$-relation, $r_{L^{(\alpha)}}(\beta)$ can be
  computed in time
  $k\bigl(2h|Q|p(|\alpha|)\log |\alpha|\bigr)^8 q^5\bigl(p(|\alpha|)\bigr)=|\alpha|^{O(1)}$.
\end{proof}

\bigskip

Also in this case, by immediate application of
Theorem~\ref{teo:rugprel}, we can introduce, via extended one-way
auxiliary pushdown automata, a new class of $p$-relations admitting a
polynomial time uniform random generation.
\begin{cor}
  \label{cor:perlupdaurg}
  If a $p$-relation $R\subseteq\reldom$ is accepted by
  a polynomial time extended one-way unambiguous auxiliary pushdown
  automaton, then it admits a polynomial time uniform random generator.
\end{cor}



\RCSfooter$Id: c3.tex,v 2.2 1999/10/29 10:24:06 santini Exp $

\chapter{Ambiguous Descriptions}
\label{cha:ambparadigm}

\mycite{french}{%
  Vous vous étonnez comme cette matière, brouillée pêle-mêle, au gré
  du hasard, peut avoir constitué un homme, vu qui'il avait tant de
  choses nécessaires à la construction de son être, mais vous ne savez
  pas que cent millons de fois cette matière, s'acheminant au dessein
  d'un homme, s'est arrêtée à former tantôt une pierre, tantôt du
  plomb, tantôt du corail, tantôt une fleur, tantôt une comète, pur le
  trop ou le trop peu de certains figures qu'il fallait ou ne fallait
  pas à désigner un homme?}%
{Cyrano de Bergerac}{Voyage dans la lune}

In this chapter we discuss a new approach to uniform random generation
based on the notion of ambiguous description. For the sake of
simplicity, here we restrict our attention to the case of
combinatorial structures instead of $p$-relations. After a brief
restatement of the notions of uniform random generator and
approximation scheme for this case, we give a formal definition of a
description of a combinatorial structure, together with its ambiguity
function. Then we give some general results relating combinatorial
structures admitting (possibly ambiguous) description with uniform
random generation, approximate and exact counting. Finally, we discuss
two very simple applications of this paradigm.

\bigskip
\RCSfooter$Id: c3s1.tex,v 2.4 1999/11/01 11:30:53 santini Exp $

\section{Combinatorial Structures}
\label{sec:combstrdef}

We start this section by recalling some notions widely used in
literature \citep{Flaj88}.  A \eidx{combinatorial structure} is a pair
\combs{S}, where the \esidx{domain}{\irCS} \bS is a finite or
denumerable set and the \esidx{size}{\irCS} $|\cdot|:\bS\to\bN$ is a
function such that $\#\{ s\in\bS : |s|=n\}$ is finite for every
$n\in\bN$. Here, we implicitly assume the elements $s\in\bS$ admit a
(recursive) binary representation such that each $|s|$ is polynomially
related to the length of its binary representation; this allows our
model of computation to manipulate the elements of combinatorial
structures.  The \esidx{census function}{\irCS} $C_{\bS}$ of a
combinatorial structure \combs{S} is the function $C_{\bS}:\bN\to\bN$
such that $C_{\bS}(n)=\#\{ s\in\bS : |s|=n\}$ for every $n\in\bN$. In
the following, for the sake of brevity, we denote by $\bS_n$ the
subset of the domain of a combinatorial structure \combs{S} defined as
$\bS_n=\{s\in\bS : |s|=n\}\subseteq\bS$ (so that
$C_{\bS}(n)=\#\bS_n$).

\bigskip

Different kinds of combinatorial structures are usually studied in
literature, for instance: families of graphs, or trees, where the size
can be considered as the number of vertices, or edges; formal
languages, where the length can be taken as the size of words;
discrete geometrical objects, such as polyominoes \citep{DV84}, or
tilings, where the size can be considered as the number of faces, or
elementary objects constituting the whole figure.

\bigskip

By adapting to our case analogous definitions of \citet{JVV86}
recalled in Section~\ref{sec:jvv}, we now introduce the concepts of
\emph{uniform random generator} and \emph{randomized exact counter}
for a combinatorial structure \combs{S} together with the notion of
\emph{randomized approximation scheme} for the census function
$C_{\bS}$. From now onwards we assume \no to be a distinguished
element not belonging to the domain of any combinatorial structure.

\subsection{Uniform random generator}
\label{ssec:urgdef}

\begin{dfn}
  \label{def:urg}
  An algorithm \sA is a \esidx{uniform random generator}{\irCS}
  \mdidx{uniform generator}{\irCS} (\urg) for a combinatorial
  structure \combs{S} iff, for every $n>0$ such that $C_{\bS}(n)>0$,
  \begin{enumerate}[(i)]
  \item \label{def:urg1} \sA on input $n$ gives output
    $\sA(n)\in\bS_n\cup\{\no\}$,
  \item \label{def:urg2} $\Pr\{ \sA(n)=s \mid \sA(n)\not=\no
    \}=1/C_{\bS}(n)$, for every $s\in\bS_n$, and
  \item \label{def:urg3} $\Pr\{ \sA(n)=\no \}<1/4$.
  \end{enumerate}
\end{dfn}
Observe that the constant $1/4$ in the previous definition can be
replaced by any positive number strictly less than $1$ leaving the
definition substantially unchanged in the sense of the following
\begin{lem}
  \label{lem:powurg}
  Given a combinatorial structure \combs{S}, let \sA be an algorithm
  for which~(\ref{def:urg1}) and~(\ref{def:urg2}) of
  Definition~\ref{def:urg} hold and such that $\Pr\{ \sA(n)=\no
  \}<\delta$, for some $0<\delta<1$.  Then, for every $0<\delta'<1$
  there exists an algorithm $\sA'$ for which~(\ref{def:urg1})
  and~(\ref{def:urg2}) of Definition~\ref{def:urg} hold and such that
  $\Pr\{ \sA'(n)=\no \}<\delta'$. Moreover, if \sA works in
  $T_{\sA}(n)$ time and uses $R_{\sA}(n)$ random bits, then $\sA'$
  works in $O(T_{\sA}(n))$ time and uses $O(R_{\sA}(n))$ random bits.
\end{lem}
\begin{proof}
  If $\delta\leq\delta'$ the statement is trivial. Otherwise, let
  $\sA'$ be algorithm~\ref{alg:powurg}.

  \begin{alg}[ht]
    \caption{An algorithm to increase the success probability of a \urg.}
    \label{alg:powurg}
    \begin{myprg*}
      \sinput $n$ \nl
      $s\stv\no$, $i\stv 0$ \nl
      \swhile $i < \lceil \log\delta'/\log\delta \rceil$ and $s=\no$ \sdo \tnl
      $i\stv i+1$ \nl
      $s\stv \sA(n)$ \unl
      \soutput $s$.
    \end{myprg*}
  \end{alg}
  
  To prove the correctness\footnote{for a detailed discussion of the
    probabilistic aspects of such proof, see
    Appendix~\ref{assec:impr}.} of $\sA'$, let $n$ be such that
  $C_{\bS}(n)>0$, then
  \[
  \Pr\{ \sA'(n)=\no \}=\Pr\{ \sA(n)=\no\}^{\lceil \log\delta'/\log\delta \rceil}
  <\delta^{\lceil\log\delta'/\log\delta \rceil}
  <\delta';
  \] 
  on the other hand, as it is easy to verify, $\Pr\{ \sA'(n)=s \mid
  \sA'(n)\not=\no \}=\Pr\{ \sA(n)=s \mid \sA(n)\not=\no \}$ for every
  $s\in\bS_n$. Finally, the statements about the computation time and
  number of random bits used by $\sA'$ follow immediately from its
  definition.
\end{proof}

\subsection{Randomized approximation scheme}
\label{ssec:rasdef}

Following \citet{Sto83,KLM89}, we now introduce the notion of
approximate counting for combinatorial structures.
\begin{dfn}
  \label{def:ras}
  An algorithm \sA is a \esidx{randomized approximation scheme}{\irCS}
  \mdidx{randomized approximation scheme}{\irCS} (\ras) for the census
  function $C_{\bS}$ of a combinatorial structure \combs{S} iff, for
  every $n>0$ such that $C_{\bS}(n)>0$ and every $\epsilon\in(0,1)$,
  \begin{enumerate}[(i)]
  \item \label{def:ras1} \sA on input $n,\epsilon$ gives output
    $\sA(n,\epsilon)\in\bQ\cup\{\no\}$,
  \item \label{def:ras2} $\Pr\{ (1-\epsilon)C_{\bS}(n)\leq
    \sA(n,\epsilon) \leq (1+\epsilon)C_{\bS}(n) \mid
    \sA(n,\epsilon)\not=\no \}>3/4$ and
  \item \label{def:ras3} $\Pr\{ \sA(n,\epsilon)=\no \}<1/4$.
  \end{enumerate}
  Moreover, a \ras is said to be a \emph{fully polynomial time} \ras
  whenever it works in time polynomial in $n$ and $1/\epsilon$.
\end{dfn}
Observe that even in this case the constant $1/4$ of
point~\emph{(\ref{def:ras3})} can be replaced by any positive number
strictly less than $1$ by essentially the same argument of
Lemma~\ref{lem:powurg}; moreover, following an idea of \citet{JVV86},
the constant $3/4$ of point~\emph{(\ref{def:ras2})} can be replaced by
any number strictly between $1/2$ and $1$ again leaving the definition
substantially unchanged in the sense of the following
\begin{lem}
  \label{lem:powras}
  Given a combinatorial structure \combs{S}, let \sA be an algorithm
  for which~(\ref{def:ras1}) and~(\ref{def:ras3}) of
  Definition~\ref{def:ras} hold and such that $\Pr\{
  (1-\epsilon)C_{\bS}(n)\leq \sA(n,\epsilon) \leq
  (1+\epsilon)C_{\bS}(n) \mid \sA(n,\epsilon)\not=\no \}>1/2+\delta$,
  for some $0<\delta<1/2$.  Then, for every $0<\delta'<1$, there
  exists an algorithm $\sA'$ for which~(\ref{def:ras1})
  and~(\ref{def:ras3}) of Definition~\ref{def:ras} hold and such that
  $\Pr\{ (1-\epsilon)C_{\bS}(n)\leq \sA'(n,\epsilon) \leq
  (1+\epsilon)C_{\bS}(n) \mid \sA'(n,\epsilon)\not=\no \}>1-\delta'$.
  Moreover, if \sA works in $T_{\sA}(n,\epsilon)$ time and uses
  $R_{\sA}(n,\epsilon)$ random bits, then $\sA'$ works in
  $O(T_{\sA}(n,\epsilon))$ time and uses $O(R_{\sA}(n,\epsilon))$
  random bits.
\end{lem}
\begin{proof}
  If $\delta'\geq 1/2-\delta$ the statement is trivial. Otherwise, let
  $\sA'$ be Algorithm~\ref{alg:powras}, where $\kappa>0$ is an integer
  constant whose value will be determined in the following.

  \begin{alg}[ht]
    \caption{An algorithm to increase the probability of correct approximation of a \ras.}
    \label{alg:powras}
    \begin{myprg*}
      \sinput $n,\epsilon$ \nl
      $i\stv 0$, $j\stv 0$ \nl
      \swhile $i < \kappa\lceil \delta^{-2}\log{\delta'}^{-1}\rceil$ \sdo \tnl
      $i\stv i+1$ \nl
      $c\stv \sA(n,\epsilon)$ \nl
      \sif $c\not=\no$ \sthen \tnl
      $j\stv j+1$\nl
      $c_j\stv c$ \untab\unl
      \sif $j>0$ \sthen \tnl
      \soutput the median of $c_1, \ldots, c_j $ \unl
      \selse \tnl
      \soutput $\no$.
    \end{myprg*}
  \end{alg}
  
  The statements about computation time and number of random bits of
  $\sA'$ follow immediately from its definition; we prove in detail
  only the correctness\footnote{for a detailed discussion of the
    probabilistic aspects of such proof, see
    Appendix~\ref{assec:impr}.} of $\sA'$. Fix $n$ such that
  $C_{\bS}(n)>0$ and let $J$ be the random variable representing the
  value of $j$ at the end of the execution. It is easy to observe that
  \[
  \Pr\{ \sA'(n,\epsilon)=\no \}= \Pr\{ J=0 \}
  = \Pr\{ \sA(n,\epsilon)=\no\}^{\kappa\lceil \delta^{-2}\log{\delta'}^{-1}\rceil } <1/4.
  \] 
  Consider now the case $J>0$.  Let $C_1, \ldots, C_J$ be the random
  variables representing the values assumed by $c_1, \ldots, c_J$ at
  the end of the execution and let $M$ be the random variable
  corresponding to the median of $C_1, \ldots, C_J$.  It is then
  evident that if $M\not\in I(n,\epsilon)= [(1-\epsilon)C_{\bS}(n),
  (1+\epsilon)C_{\bS}(n)]$, then at most one half of the $C_i$'s with
  $1\leq i\leq J$ are such that $C_i\in I(n, \epsilon)$.  Hence, by
  Lemma~\ref{lem:hoefbinom} and for every $k>0$, $\Pr\{ M\not\in
  I(n,\epsilon) \mid J=k \} \leq e^{-2k\delta^2}$ where $\Pr\{ C_i\in
  I(n,\epsilon) \}=\Pr\{ \sA(n,\epsilon) \in I(n, \epsilon) \mid
  \sA(n,\epsilon)\not=\no \}>1/2+\delta$ by the definition of \sA.
  Hence, if $N=\kappa\lceil \delta^{-2}\log{\delta'}^{-1}\rceil$,
  \begin{equation*}
    \begin{split}
      \Pr\{ \sA'(n,\epsilon)\not\in I(n,\epsilon) \mid \sA'(n,\epsilon)\not=\no \} 
      &= \frac{\Pr\{ M\not\in I(n,\epsilon)\cap J>0 \}}
      {\Pr\{ J>0 \}} \\
      &= \frac{\sum_{k=1}^N \Pr\{ M\not\in I(n,\epsilon) \mid J=k \}\Pr\{J=k\}} 
      {\Pr\{ J>0 \}} \\
      &\leq \frac 43\sum_{k=1}^N e^{-2k\delta^2} {N \choose k} q^k (1-q)^{N-k} \\
      &\leq \frac 43\sum_{k=1}^N {N \choose k} \left( \frac q{e^{2\delta^2}}\right)^k (1-q)^{N-k} \\
      &< \frac 43\left( 1 - q\left(1-\frac 1{e^{2\delta^2}}\right) \right)^N \\
      &\leq \frac 43\left( 1 - \frac 34 \delta^2 \right)^{\kappa\lceil
        \delta^{-2}\log{\delta'}^{-1}\rceil}
    \end{split}
  \end{equation*}
  where $q=\Pr\{ \sA(n,\epsilon)\not=\no \}>3/4$ and the last
  inequality follows from the fact that $1-e^{-x}\geq x/2$ for
  $x\in[0,1]$. Finally, by Lemma~\ref{lem:expineq}, the integer
  $\kappa>0$ can be chosen such that
  \[
  \frac 43\left( 1 - \frac 34 \delta^2 \right)^{\kappa \lceil
    \delta^{-2}\log{\delta'}^{-1}\rceil}<\delta'. \qed
  \] 
\end{proof}

\subsection{Randomized exact counter}
\label{ssec:recdef}

\begin{dfn}
  \label{def:rca}
  An algorithm \sA is a \esidx{randomized exact counter}{\irCS}
  \mdidx{randomized counter}{\irCS~(exact)} (\rec) for a combinatorial
  structure \combs{S} iff, for every $n>0$ such that $C_{\bS}(n)>0$,
  \begin{enumerate}[(i)]
  \item \sA on input $n$ gives output $\sA(n)\in\bN\cup\{\no\}$,
  \item  $\Pr\{ \sA(n)=C_{\bS}(n) \mid \sA(n)\not=\no
    \}>3/4$ and
  \item  $\Pr\{ \sA(n)=\no \}<1/4$.
  \end{enumerate}
\end{dfn}
Also for this definition, one can show that the choice of the
constants $1/4$ and $3/4$ is not restrictive by reasoning as in
Lemma~\ref{lem:powurg} and~\ref{lem:powras}.

\subsection{An example: regular languages}
\label{ssec:examplereg}

We now give a very simple example of the definitions given in this
section.  Let us consider a \eidx{deterministic finite automaton}
$\cA=(Q,q_0,\delta,F)$ over a finite alphabet $\Sigma$, where $Q$ is
the set of \emph{states}, $q_0\in Q$ is the \emph{initial} state,
$\delta:Q\times\Sigma\to Q$ is the \emph{transition function} and
$F\subseteq Q$ is the family of \emph{final} states \citep{HU79}. Let
$L_{\cA}\subseteq\fldom$ be the language recognized by $\cA$.  Hence,
the combinatorial structure we consider is $\langle L_{\cA}, |\cdot|
\rangle$, where $|\cdot|$ denotes the length. Our aim is to design a
\urg for such a structure.

The uniform random generation algorithm first computes, for every
$q\in Q$ and $1\leq\ell\leq n$, the number $C_q(\ell)$ of all words of
length $\ell$ accepted by $(Q,q,\delta,F)$. This computation takes
$O(n^2)$ time on a \mRAM under logarithmic cost criterion since each
$\{C_q(\ell)\}_{\ell\geq 1}$ is the sequence of coefficients of a
rational generating function \citep{CS63}.  For every of these terms,
the algorithm also computes the number $b_q(\ell)=\lceil \log
C_q(\ell) \rceil$ of bits required to represent $\{1,\ldots,
C_q(\ell)\}$ which can be obtained in $O(\ell\log \ell)$ time (see
Lemma~\ref{lem:bit}).  Hence, such a precomputation phase requires
$O(n^2\log n)$ time.

Then, the algorithm executes the procedure Generate$(q_0,n)$ described
by Algorithm~\ref{alg:rlurg}. Here, $\kappa\in\bN$ is a global
parameter whose value will be determined in the following, while
$\Sigma$ and all the sets $\{ (s,\sigma)\in Q\times\Sigma :
\delta(q,\sigma)=s \}$, for $q\in Q$, are endowed with some total
order relation $\preccurlyeq$.

\begin{alg}[ht]
  \caption{a uniform random generator for regular languages.}
  \label{alg:rlurg}
  \begin{myprg*}
    \sprocedure Generate$(q, \ell)$ \nl 
    $i\stv 0$, $r\stv\no$, $w\stv\no$ \nl
    \swhile $i < \kappa$ and $r=\no$ \sdo \tnl
    $i\stv i+1$ \nl
    \textbf{generate} $u\in\{1, \ldots, 2^{b_q(\ell)}\}$ 
    \emph{uniformly at random} \nl
    \sif $u\leq C_q(\ell)$ \sthen $r\stv u$ \unl
    \sif $r\not=\no$ \sthen \tnl
    \sif $n=1$ \sthen \tnl
    
    let $a$ be the $r$-th symbol in the set $\{\sigma\in\Sigma :
    \delta(q,\sigma)\in F \}$ \nl
    $w\stv a$ \unl
    
    \selse \tnl
    
    choose the smallest element $(p,a)$ in the set $\{ (s,\sigma)\in
    Q\times\Sigma : \delta(q,\sigma)=s \}$ \tab\tnl such that $\displaystyle 
    \sum_{(s,\sigma)\preccurlyeq (p,a)} C_s(\ell-1) \geq r$ \untab\unl
    
    $w_p\stv$Generate$(p,\ell-1)$\nl
    \sif $w_p\not=\no$ \sthen $w\stv a w_p$ \untab\unl
    
    \sreturn $w$.
  \end{myprg*}
\end{alg}

Reasoning by induction on $\ell$, it is easy to verify that if
Generate$(q,\ell)$ returns an output $w$ different from $\no$, then
$w$ is uniformly distributed in the set of words of length $\ell$
accepted by $(Q,q,\delta,F)$. An upper bound to the probability that
Generate$(q,\ell)$ gives output $\no$ depends on the global parameter
$\kappa$.  Denoting by $e(\ell)$ the maximum of all probabilities that
Generate$(q,\ell)$ gives output $\no$ for $q\in Q$, one can easily
show that $e(1)\leq (1/2)^\kappa$ and $e(\ell)\leq (1/2)^\kappa +
e(\ell-1)$ for every $\ell>1$.  A simple induction proves $e(n)\leq
n/2^\kappa$ and hence, for every constant $t>0$, by fixing
$\kappa=t+\lceil\log n\rceil$ we obtain $e(n)\leq 1/2^t$.

\enlargethispage{2em} 

As far as the time complexity is concerned, if we denote by $T(\ell)$
the maximum time cost of procedure Generate$(q,\ell)$ for $q\in Q$,
for every $1\leq \ell\leq n$, one can write the recursion
$T(\ell)=O(\kappa\cdot\ell)+T(\ell-1)$ which proves that
$T(n)=O(\kappa n^2)=O((t+\log n)n^2)$.

We summarize our discussion by the following
\begin{pro}
  \label{pro:raturg}
  Given a deterministic finite automaton $\cA$, there exists a \mPrRAM
  which, on input $n\in\bN$, generates a word uniformly at random in
  $L_\cA\cap\Sigma^n$ with probability $1-1/2^t$ working in
  $O(n^2(t+\log n))$ time. Hence the combinatorial structure $\langle
  L_{\cA}, |\cdot| \rangle$ admits a \urg working in $O(n^2\log
  n)$ time.
\end{pro}


\RCSfooter$Id: c3s2.tex,v 2.4 1999/11/01 11:30:53 santini Exp $

\section{Ambiguous Description}
\label{sec:ambdescrdef}

We formally introduce the concept of \emph{(ambiguous) description}
\msidx{description}{\irCS} \mridx{description}{\irCS} \mridx{ambiguous
  description}{\irCS} (see Figure~\ref{fig:ambdes}):
\begin{dfn}
  \combs{T} is a \emph{description} of \combs{S} via the function
  $f:\bT\to\bS$, if $f$ is a surjective function preserving $|\cdot|$,
  \ie~$|f(t)|=|t|$ for every $t\in\bT$. The \emph{ambiguity} of the
  description is the function $d:\bS\to\bN$ defined by
  $d(s)=\#\{t\in\bT : f(t)=s \}$, for every $s\in\bS$. We say that the
  description is \emph{ambiguous} \mdidx{ambiguity}{description, of a
    combinatorial structure} if $d(s)>1$ for some $s\in\bS$.
  Moreover, the description is said to be \emph{polynomial} whenever
  $f$ and $d$ are computable in polynomial time and there exists some
  $D\in\bN$ such that $d(s)=O(|s|^D)$.
\end{dfn}

\begin{figure}[ht]
  \begin{center}
    \input{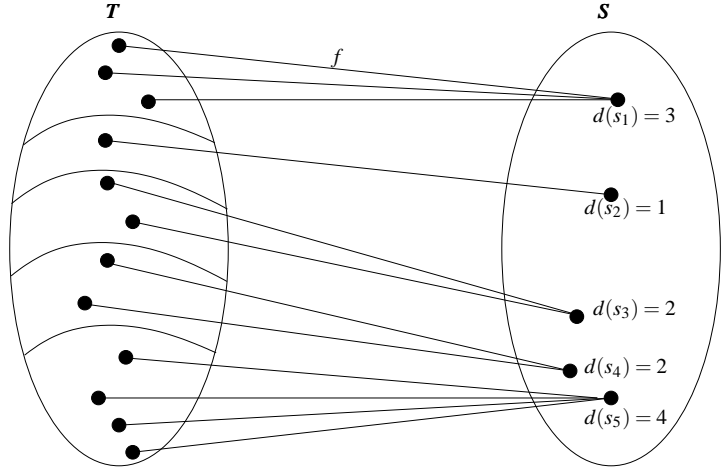}
    \caption{An Example of Ambiguous description.}
    \label{fig:ambdes}
  \end{center}
\end{figure}

In this thesis, we will give examples of (ambiguous) descriptions,
such as the family of derivation trees of a grammar as a (possibly
ambiguous) description of the strings derived in the grammar (see
Section~\ref{sec:cfl}) and the family of computations of a
nondeterministic device as a (possibly ambiguous) description of the
strings accepted by the device (see Section~\ref{sec:npda}).

\clearpage 

\subsection{Uniform random generation}
\label{ssec:urgteo}

We start by considering the uniform random generation problem.
\begin{teo}
  \label{teo:rugpa}
  If a combinatorial structure \combs{S} admits a polynomial
  (ambiguous) description \combs{T} and there exists a polynomial time
  \urg for \combs{T}, then there exists a polynomial time \urg for
  \combs{S}.
\end{teo}
To prove this theorem, we show a stronger result derived by applying
the \citet*{KLM89} technique for sampling from a union of sets.
\begin{lem}
  \label{lem:rug}
  Let \combs{S} admit a polynomial (ambiguous) description \combs{T}
  via the function $f$ and let $T_f(n)$ and $T_d(n)$ be respectively
  the computation time of $f$ and of its ambiguity $d$; moreover,
  assume $d(s)=O(|s|^D)$ for some $D\in\bN$. If \sB is a \urg for
  \combs{T} working in $T_{\sB}(n)$ time and using $R_{\sB}(n)$ random
  bits, then there exists a \urg for \combs{S} working in $O\bigl(
  n^{2D}+ n^D (T_{\sB}(n)+T_f(n)+T_d(n)) \bigr)$ time and using $O(
  n^{2D} + n^D R_{\sB}(n))$ random bits.
\end{lem}
\begin{proof}
  Define, for the sake of brevity, $D(n)=\kappa_1 n^D+\kappa_2$ where
  $\kappa_1,\kappa_2$ are two integer constants such that $d(s)\leq
  D(|s|)$; let $\sA$ be Algorithm~\ref{alg:adurg}, where $\kappa>0$ is
  an integer constant whose value will be determined in the following.

  \begin{alg}[ht]
    \caption{An \urg for (ambiguously) described combinatorial structures.}
    \label{alg:adurg}
    \begin{myprg*}
      \sinput $n$ \nl
      $m\stv \lcm\{1, \ldots, D(n)\}$, $\ell\stv \lceil\log m\rceil$ \nl
      $i\stv 0$, $s\stv \no$\nl
      \swhile $i < \kappa D(n)$ and $s=\no$ \sdo \tnl
      $i\stv i+1$ \nl
      $t\stv \sB(n)$ \nl
      \sif $t\not=\no$ \sthen \tnl
      $s\stv f(t)$ \nl
      \textbf{generate} $r\in\{1, \ldots, 2^\ell\}$ 
      \emph{uniformly at random} \nl
      \sif $r> m/d(s)$ \sthen \tnl
      $s\stv\no$ \cnl 
      \soutput $s$
    \end{myprg*}
  \end{alg}
  
  First of all, we focus on the computation time of $\sA$ on input
  $n$; the time to precompute $D(n)$, $m$ and $\ell$, by
  Lemma~\ref{lem:lcm} and~\ref{lem:bit}, equals $O(n^{2D})$ and the
  time of each iteration equals $O(T_{\sB}(n)+T_f(n)+T_d(n))$ plus the
  time $O(n^D)$ required for the generation of $r$.  Moreover, each
  iteration uses at most $R_{\sB}(n)$ and $O(n^D)$ random bits to
  compute $t$ and $r$ respectively, hence the total amount of random
  bits used by \sA adds up to $O(n^{2D}+n^DR_{\sB}(n))$.
  
  Now, we prove the correctness of $\sA$. Assume that $C_{\bS}(n)>0$
  so that, since $f$ is surjective, $C_{\bT}(n)>0$ and let $S$ and $T$
  be the random variables representing respectively the value of $s$
  and $t$ at the end of a \textbf{while} iteration (observe that $S$
  and $T$ are well defined since their value is independent of the
  outcomes of the preceding iterations). Then, $S$ takes value $s\in
  \bS_n$ whenever $T\in f^{-1}(s)\subseteq \bT_n$ and $r\leq
  m/d(f(T))$.  Moreover, by definition of \sB, there exists some
  $0<\delta<1/4$ such that $\Pr\{ T=t \}=(1-\delta)C_{\bT}(n)^{-1}$
  for every $t\in \bT_n$ and, as it is easy to verify since $r$ is
  independent of $S$ and $T$, $\Pr\{ r\leq m/d(f(T)) \mid T=t \}=
  d(f(t))^{-1} m 2^{-\lceil\log m\rceil}$. Hence, for every $s\in
  \bS_n$,
  \begin{equation*}
    \begin{split}
      \Pr\{ S=s \}
      &= \sum_{t\in f^{-1}(s)} \Pr\{ T=t, r\leq m/d(f(T)) \} \\
      &= \sum_{t\in f^{-1}(s)} \Pr\{ r\leq m/d(f(T)) \mid T=t \}\Pr\{ T=t \} \\
      &= d(s) \bigl(d(s)^{-1} m 2^{-\lceil\log m\rceil} (1-\delta)C_{\bT}(n)^{-1} \bigr) \\
      &= (1-\delta) m 2^{-\lceil\log m\rceil}C_{\bT}(n)^{-1}
    \end{split}
  \end{equation*}
  which is independent of $s$. On the other hand, at the end of each
  \textbf{while} iteration, $\Pr\{ S=\no \} = 1 - \Pr\{ S\in \bS_n \}
  = 1 - C_{\bS}(n)\left( (1-\delta) m 2^{-\lceil\log
      m\rceil}C_{\bT}(n)^{-1} \right)$.  Then, since $m 2^{-\lceil\log
    m\rceil}>1/2$ and $C_{\bT}(n)\leq C_{\bS}(n)D(n)$,
  \[
  \Pr\{ S=\no \}\leq 1 - \frac 38 \frac 1{D(n)}.
  \]
  
  Finally, since $\Pr\{ \sA(n)=\no \}=\Pr\{ S=\no \}^{\kappa D(n)}$,
  the integer $\kappa>0$ can be chosen such that
  \[
  \left(1 - \frac 38 \frac 1{D(n)} \right)^{\kappa D(n)} < 1/4
  \]
  and hence $\Pr\{ \sA(n)=\no \}\leq 1/4$.

  Moreover, as it is easy to verify, for every $s\in\bS_n$, it holds
  $\Pr\{ \sA(n)=s \mid \sA(n)=\no\}=\Pr\{ S=s\mid S\not=\no \}$ and,
  since $\Pr\{S=s\}$ is constant, it is immediate to conclude that
  $\Pr\{ \sA(n)=s \mid \sA(n)\not=\no \}=C_{\bS}(n)^{-1}$ .
\end{proof}
In particular, in the case of finite ambiguity, it holds that
\begin{cor}
  \label{cor:rugfa}
  Under the same hypotheses as in Lemma~\ref{lem:rug}, with the
  stronger condition $d(s)\leq D'$ for every $s\in\bS$ and some
  $D'\in\bN$, there exists a \urg for \combs{S} working in time
  $O(T_{\sB}(n)+T_f(n)+T_d(n))$ and using $O(R_{\sB}(n))$ random bits.
\end{cor}

\subsection{Randomized counting}
\label{ssec:rasteo}

Now, let us consider the approximate and exact counting problem.
\begin{teo}
  \label{teo:rasecpa}
  Let \combs{S} be a combinatorial structure admitting a polynomial
  (ambiguous) description \combs{T} and assume there exists a
  polynomial time \urg for \combs{T}. If $C_{\bT}(n)$ is computable in
  polynomial time, then there exists a fully polynomial time \ras for
  $C_{\bS}$. If moreover $C_{\bT}(n)$ is polynomially bounded, then
  there exists a polynomial time \rec for \combs{S}.
\end{teo}
To prove this theorem, we show two stronger results essentially
obtained by an application of Hoeffding's \citeyearpar{Hof63}
inequality.
\begin{lem}
  \label{lem:ras}
  Under the same hypotheses as in Lemma~\ref{lem:rug}, if $C_{\bT}$ is
  computable in $T_{C_{\bT}}(n)$ time, then there exists a \ras for
  $C_{\bS}$ working in $O\bigl( T_{C_{\bT}}(n)+ n^{2D}\epsilon^{-2}
  (T_{\sB}(n)+T_f(n)+T_d(n)) \bigr)$ time and using
  $O(n^{2D}\epsilon^{-2}R_{\sB}(n) )$ random bits.
\end{lem}
\begin{proof}
  Define, for the sake of brevity, $D(n)=\kappa_1 n^D+\kappa_2$ where
  $\kappa_1,\kappa_2$ are two integer constants such that $d(s)\leq
  D(|s|)$; let $\sA$ be Algorithm~\ref{alg:adras}, where $\kappa>0$ is
  an integer constant whose value will be determined in the following.

  \begin{alg}[ht]
    \caption{A \ras for ambiguously described combinatorial structures.}
    \label{alg:adras}
    \begin{myprg*}
      \sinput $n$ \nl
      $i\stv 0$, $j\stv 0$, $m\stv 0$ \nl
      \swhile $i < \kappa \lceil D(n)/\epsilon \rceil^2$ \sdo \tnl
      $i\stv i+1$ \nl
      $t\stv \sB(n)$ \nl
      \sif $t\not=\no$ \sthen \tnl
      $j\stv j+1$\nl
      $m\stv m+1/d(f(t))$ \untab\unl
      \sif $j>0$ \sthen \tnl
      \soutput $mC_{\bT}(n)/j$ \unl
      \selse \tnl
      \soutput $\no$.
    \end{myprg*}
  \end{alg}
  
  The statements about computation time and number of random bits of
  \sA follow immediately from its definition; we prove in detail only
  the correctness\footnote{for a detailed discussion of the
    probabilistic aspects of such proof, see
    Appendix~\ref{assec:impr}.} of \sA. Fix $n$ such that
  $C_{\bS}(n)>0$ and let $J$ be the random variable representing the
  value of $j$ at the end of the execution. It is easy to observe
  that,
  \[
  \Pr\{ \sA(n)=\no \}= \Pr\{ J=0 \}=
  \Pr\{ \sB(n)=\no \}^{\kappa \lceil D(n)/\epsilon \rceil^2} <1/4.
  \]
  Consider now the case $J>0$ and define for the sake of brevity
  \begin{equation*}
    \begin{split}
      \mu=E(1/d(f(\sB(n)))\mid \sB(n)\not=\no )
      &=\sum_{t\in\bT_n} 1/d(f(t)) C_{\bT}(n)^{-1} \\
      &=\sum_{s\in\bS_n} C_{\bT}(n)^{-1} \sum_{t\in f^{-1}(s)} 1/d(s) \\
      &=C_{\bS}(n)C_{\bT}(n)^{-1}.
    \end{split}
  \end{equation*}
  If $M$ is the random variable representing the values assumed
  by $m$ at the end of the execution and $I(n,\epsilon)=
  [(1-\epsilon)\mu, (1+\epsilon)\mu]$, by Lemma~\ref{lem:hoefrel},
  $\Pr\{ M/J\not\in I(n,\epsilon) \mid J=k \}\leq
  2e^{-2k(\mu\epsilon)^2}$, for every $k>0$. Hence, if $N=\kappa
  \lceil D(n)/\epsilon \rceil^2$,
  \begin{equation*}
    \begin{split}
      \Pr\{ M/J\not\in I(n,\epsilon) \mid J>0 \} 
      &= \frac{\Pr\{ M/J\not\in I(n,\epsilon)\cap J>0 \}}
      {\Pr\{ J>0 \}} \\
      &= \frac{\sum_{k=1}^N \Pr\{ M/J\not\in I(n,\epsilon) \mid J=k \}\Pr\{J=k\}} 
      {\Pr\{ J>0 \}} \\
      &\leq \frac 83 
        \sum_{k=1}^N e^{-2k(\mu\epsilon)^2} {N \choose k} q^k (1-q)^{N-k} \\
      &\leq \frac 83 
        \sum_{k=1}^N {N \choose k} \left( \frac q{e^{2(\mu\epsilon)^2}}\right)^k (1-q)^{N-k} \\
      &< \frac 83\left( 1 - q\left(1-\frac 1{e^{2(\mu\epsilon)^2}}\right) \right)^N \\
      &\leq \frac 83\left( 
        1 -\frac 34 \left(\frac \epsilon{D(n)}\right)^2 
      \right)^{\kappa \lceil D(n)/\epsilon \rceil^2} 
    \end{split}
  \end{equation*}
  where $q=\Pr\{ \sB(n)\not=\no \}>3/4$ and the last inequality
  follows from the fact that $1-e^{-x}\geq x/2$ for $x\in[0,1]$ and
  that $C_{\bT}(n)\leq C_{\bS}(n)D(n)$.
  
  Finally, by Lemma~\ref{lem:expineq} the integer $\kappa>0$ can be
  chosen such that
  \[     
  \frac 83\left( 1 -\frac 34 \left(\frac \epsilon{D(n)}\right)^2 
  \right)^{\kappa \lceil D(n)/\epsilon \rceil^2} < \frac 14.
  \]
  Then, since $\sA(n,\epsilon)= M C_{\bT}(n)/J$, it holds
  \[
  \Pr\{ M/J\in I(n,\epsilon) \mid J>0 \} 
  =  \Pr\{(1-\epsilon)C_{\bS}(n)\leq \sA(n,\epsilon) \leq (1+\epsilon)C_{\bS}(n)
  \mid \sA(n,\epsilon)\not=\no \},
  \]
  hence, $\Pr\{(1-\epsilon)C_{\bS}(n)\leq \sA(n,\epsilon) \leq
  (1+\epsilon)C_{\bS}(n) \mid \sA(n,\epsilon)\not=\no \}\geq 3/4$.
\end{proof}
We now consider the problem of obtaining a \rec
\begin{lem}
  \label{lem:rec}
  Under the same hypotheses as in Lemma~\ref{lem:ras}, there exists a
  \rec for \combs{S} working in $O\bigl(T_{C_{\bT}}(n)+
  n^{2D}C_{\bT}(n)^2(T_{\sB}(n)+T_f(n)+T_d(n)) \bigr)$ time and using
  $O(n^{2D}C_{\bT}(n)^2R_{\sB}(n) )$ random bits.
\end{lem}
\begin{proof}
  Using the same notation of Lemma~\ref{lem:ras}, define the algorithm
  $\sA'$ as
  \[
  \sA'(n)=\round\!\left(\sA\biggl(n, \frac 1{3C_{\bT}(n)}\biggr)\right).
  \] 
  Since $f$ is surjective $C_{\bT}(n)\geq C_{\bS}(n)$, hence $\sA'$ is
  correct since $|\sA(n,\epsilon(n))/C_{\bS}(n)-1| < 1/(3C_{\bT}(n)) $
  implies $|\sA(n,\epsilon(n))-C_{\bS}(n)| < 1/3$.  Finally, the
  statement about computation time and number of random bits used
  follows immediately from Lemma~\ref{lem:ras}.
\end{proof}

\clearpage 

In particular, in the case of finite ambiguity, it holds that
\begin{cor}
  \label{cor:rasecfa}
  Under the same hypotheses as in Lemma~\ref{lem:ras}, with the
  stronger condition $d(s)\leq D'$ for every $s\in\bS$ and some
  $D'\in\bN$, there exists a \ras for $C_{\bS}$ working in $O\bigl(
  T_{C_{\bT}}(n)+ \epsilon^{-2} (T_{\sB}(n)+ T_f(n)+T_d(n)) \bigr)$
  time and using $O(\epsilon^{-2}R_{\sB}(n))$ random bits and a \rec
  for \combs{S} working in $O\bigl(T_{C_{\bT}}(n)+
  C_{\bT}(n)^2(T_{\sB}(n)+T_f(n)+T_d(n)) \bigr)$ time and using
  $O(C_{\bT}(n)^2R_{\sB}(n) )$ random bits.
\end{cor}


\RCSfooter$Id: c3s3.tex,v 2.3 1999/09/23 14:43:06 santini Exp $

\section{Simple Applications}
\label{sec:simpleapp}

In the remaining part of this chapter, we give two simple applications
of the above results to the case of formal languages over a given
alphabet $\Sigma$, \ie for combinatorial structures \combs{S} such
that $\bS\subseteq\fldom$ and $|\cdot|$ is simply the word length.
Given two such combinatorial structures \combs{S} and \combs{T}, it is
natural to define the \esidx{union}{\irCS} as $\langle \bS\cup\bT,
|\cdot| \rangle$ and the \esidx{product}{\irCS} as $\langle
\bS\cdot\bT, |\cdot| \rangle$.

We start from the uniform random generation of the product:
\begin{teo}
  \label{teo:produrg}
  Let \combs{S} and \combs{T} be two combinatorial structures such
  that $\bS\subseteq\fldom$ and $\bT\subseteq\fldom$ are
  recognizable in polynomial time. If both structures admit a
  polynomial time \urg, then there exists a polynomial time \urg for
  $\langle \bS\cdot\bT, |\cdot| \rangle$.  
\end{teo}
\begin{proof}
  If $\bP=\bS\times\bT$ denotes the Cartesian product of $\bS$ and
  $\bT$, then \combs{P} is a polynomial (possibly ambiguous)
  description of $\langle \bS\cdot\bT, |\cdot| \rangle$ via the
  function $f((s,t))=s\cdot t$, for every $(s,t)\in\bP$; the ambiguity
  is defined as
  \[
  d(q)=\sum_{s\in\bS,t\in\bT: s\cdot t=q} \chi_{\bS}(s) \chi_{\bT}(t)
  \]
  for every $q\in\bS\cdot\bT$, where $\chi_A$ denotes the
  characteristic function of the set $A$. Then, if $\sA$ and $\sB$ are
  the polynomial time \urg respectively of \combs{S} and \combs{T},
  the integer $\kappa>0$ can be chosen such that
  Algorithm~\ref{alg:produrg} is a polynomial time \urg for \combs{P}.

  \begin{alg}[ht]
    \caption{A \urg for the product of combinatorial structures.}
    \label{alg:produrg}
    \begin{myprg*}
      \sinput $n$ \nl
      $i\stv 0$, $p\stv\no$ \nl
      \swhile $i < \kappa$ and $p=\no$ \sdo \tnl
      $i\stv i+1$ \nl
      $s\stv\sA(n)$, $t\stv\sB(n)$ \nl
      \sif $s\not=\no$ and $t\not=\no$ \sthen \tnl
      $p\stv (s,t)$ \untab \unl
      \soutput $p$. 
    \end{myprg*}
  \end{alg}

  Therefore, by Theorem~\ref{teo:rugpa}, there exists a polynomial
  time \urg for $\langle \bS\cdot\bT, |\cdot| \rangle$.
\end{proof}

Now we consider the case of the union:
\begin{teo}
\label{teo:unionurg}  
Under the same hypotheses as in Theorem~\ref{teo:produrg}, if
$C_{\bS}$ and $C_{\bT}$ are computable in polynomial time, then there
exist a polynomial time \urg for $\langle \bS\cup\bT, |\cdot|
\rangle$.
\end{teo}
\begin{proof}
  If $\bU=\bS\oplus\bT$ denotes the disjoint union of $\bS$ and $\bT$,
  then \combs{U} is a polynomial (possibly ambiguous) description of
  $\langle \bS\cup\bT, |\cdot| \rangle$ via the identity function (we
  denote here by $f$) and the ambiguity is $d=\chi_{\bS}+\chi_{\bT}$.
  As before, the integer $\kappa>0$ can be chosen such that
  Algorithm~\ref{alg:unionurg} is a polynomial time \urg for
  \combs{U}, where $\sA$ and $\sB$ are the polynomial time \urg
  respectively of \combs{S} and \combs{T}.

  \begin{alg}[ht]
    \caption{A \urg for the union of combinatorial structures.}
    \label{alg:unionurg}
    \begin{myprg*}
      \sinput $n$ \nl $i\stv 0$, $\ell\stv {\lceil\log
        C_{\bS}(n)+C_{\bT}(n)\rceil}$, $u\stv\no$ \nl \swhile $i <
      \kappa$ and $u=\no$ \sdo \tnl $i\stv i+1$ \nl \textbf{generate}
      $r\in\{1, \ldots, 2^\ell\}$ \emph{uniformly at random} \nl \sif
      $r\leq C_{\bS}(n)$ \sthen \tnl $u\stv\sA(n)$ \unl \selse \sif
      $r\leq C_{\bS}(n)+C_{\bT}(n)$ \sthen \tnl $u\stv\sB(n)$
      \untab\unl \soutput $u$.
    \end{myprg*}
  \end{alg}
  
  Again, by Theorem~\ref{teo:rugpa}, there exists a polynomial time
  \urg for $\langle \bS\cup\bT, |\cdot| \rangle$.
\end{proof}
    
Finally, since $C_{\bP}(n)=C_{\bS}(n)C_{\bT}(n)$ and
$C_{\bU}(n)=C_{\bS}(n)+C_{\bT}(n)$, from Theorem~\ref{teo:rasecpa}
simply follows
\begin{teo}
  Under the same hypotheses as in Theorem~\ref{teo:unionurg}, there
  exist and a fully polynomial time \ras for the census functions
  $C_{\bS\cup\bT}$ and $C_{\bT\cdot\bS}$.
\end{teo}



\RCSfooter$Id: c4.tex,v 2.3 1999/10/29 10:24:09 santini Exp $

\chapter{Applications to Formal~Languages}
\label{cha:ambappl}

\mycite{italian}{%
  A questo piacere contribuisce la varietà, l'incertezza, il non veder
  tutto, il potersi perciò spaziare coll'immaginazione, riguardo a ciò
  che non si vede\dots È piacevolissima ancora \dots la vista di una
  moltitudine innumerabile, come delle stelle, o di persone ec.~un
  moto molteplice, incerto, confuso, irregolare, disordinato, un
  ondeggiamento vago ec., che l'animo non possa determinare, né
  concepire definitamente e distintamente \dots}%
{Giacomo Leopardi}{Zibaldone di pensieri}

In this chapter we present some nontrivial application of the general
paradigm introduced in the previous chapter to some class of formal
languages. In particular, we show how to obtain, under suitable
hypotheses, a polynomial time uniform random generator and a fully
polynomial time randomized approximation schemes for polynomially
ambiguous context-free languages, languages accepted in polynomial
time by nondeterministic one-way auxiliary pushdown automata of
polynomial ambiguity and, finally, polynomially ambiguous rational
trace languages.

\bigskip
\RCSfooter$Id: c4s1.tex,v 2.5 1999/11/01 11:30:53 santini Exp $

\section{Context-Free Languages}
\label{sec:cfl}

A natural example of our general paradigm is given by the family of
derivation trees of a context-free grammar, considered as (possibly
ambiguous) description of the corresponding language.  Thus, we can
design simple \urg and \ras of census functions for inherently
ambiguous context-free languages.  These procedures work in polynomial
time whenever the degree of ambiguity of the associated grammar is
polynomially bounded.  Moreover, if such a degree is bounded by a
constant, both procedures require $O(n^2\log n)$ time on input $n$
under logarithmic cost criterion, the same order of growth required by
the best known algorithms \citep{FZV94,Gol95} for the analogous
problems in the unambiguous case (once their time cost is adapted to
the \mPrRAM model of computation).

\medskip

To show this application in detail, consider a \idx{context-free
  grammar} (\cf) $G=\langle V,\Sigma,S, P\rangle$, where $V$ is the
set of nonterminal symbols (also called variables), $\Sigma$ is the
alphabet of terminals, $S\in V$ is the initial variable and $P$ is the
family of productions.  We assume $G$ in \sidx{Chomsky normal
  form}{\irCFg} \citep{HU79} without useless variables, \ie every
nonterminal appears in a derivation of some terminal word\footnote{it
  is well known that every \cf language not containing the empty word
  $\ew$ can be generated by such a grammar.}.  Moreover, for every
$A\in V$, let $T_A$ be the family of derivation trees with root
labelled by $A$ deriving a word in $\Sigma^+$.  It is easy to see that
there are finitely many $t\in T_S$ deriving a given $x\in\Sigma^+$; in
the following, we denote by $\hat{d}_G(x)$ the number of such trees
and call \esidx{ambiguity}{\irCFg}
\mdidx{ambiguity}{context-free!grammar} of $G$ the function
$d_G:\bN\to\bN$ defined by $d_G(n)=\max_{x\in\Sigma^n} \hat{d}_G(x)$,
for every $n\in\bN$.  The grammar $G$ is said \esidx{finitely
  ambiguous}{\irCFg} if there exists a $k\in\bN$ such that $d_G(n)\leq
k$ for every $n>0$; in particular, $G$ is said
\esidx{unambiguous}{\irCFg} if $k=1$.  On the other hand, $G$ is said
\esidx{polynomially ambiguous}{\irCFg} if, for some polynomial $p(n)$,
we have $d_G(n)\leq p(n)$ for every $n>0$.

Clearly, our idea is to use the structure $\langle T_S,|\cdot|
\rangle$ as a (possibly ambiguous) description of the language $L$
generated by $G$, where, for every $t\in T_S$, $|t|$ is the length of
the derived word.  However, to get a \urg for $L$ and a \ras for its
census function, we first need two preliminary procedures: one for
generating a tree of given size in $T_S$ uniformly at random, the
other for computing the degree of ambiguity $\hat{d}_G(x)$ for the
words $x\in \Sigma^+$.

\subsection{Uniform random generation of derivation trees}
\label{ssec:rnddt}

A \urg for derivation trees can be designed by following a general
approach to the uniform random generation of combinatorial structures
proposed by \citet{FZV94}.  The algorithm we obtain is similar to the
procedure given by \citet{Gol95} for generating words uniformly at
random in unambiguous \cf languages.  Here, we essentially adapt that
routine to our case and evaluate its time complexity with respect to
our model of computation.

\medskip

To fix the notation, for every $A\in V$ and $\ell\in\bN$, let
$T^\ell_A$ be the subset $\{t\in T_A : |t|=\ell\}$, and let
$C_A(\ell)$ be its cardinality.  It is well known that every sequence
$\{C_A(\ell)\}_{\ell\geq 1}$ has an algebraic generating function
\citep{CS63} and hence, applying Comtet's recurrence equation
\citep{Com64}, its first $n$ terms $C_A(1), \ldots, C_A(n)$ can be
computed in $O(n^2)$ time on a \mRAM under logarithmic cost criterion.
For each $C_A(\ell)$ with $1\leq\ell\leq n$, the integers
$b_A(\ell)=\lceil\log C_A(\ell)\rceil$ are then computed, by
Lemma~\ref{lem:bit}, in $O(\ell\log \ell)$ time, for an overall time
cost of $O(n^2\log n)$. Observe that this precomputations has to be
executed only once even if we have to generate several derivation
trees in $T^\ell_A$.

Then, to generate a tree uniformly at random in $T^n_S$, we apply
Algorithm~\ref{alg:dturg} which works on input $(A,\ell)\in
V\times\bN$ and uses a global parameter $\kappa$ (depending on $n$ but
not on $\ell$), whose value will be determined in the following. In
the procedure, $P_A$ denotes the subset of productions in $P$ of the
form $A\pr BC$ with $B,C\in V$, and $P^1_A$ denotes the set of
productions in $P$ of the form $A\pr a$ with $a\in\Sigma$.

The procedure chooses an element uniformly at random either in $P^1_A$
(if $\ell=1$), or in the set $P_A\times\{1, \ldots, \ell-1\}$ (if
$\ell>1$), the choice depending on a total order relation $\leq$ among
the elements of these sets. To define $\leq$, we assume a
lexicographic order $\lex$ in both alphabets $\Sigma$ and $V$ and set
$A\pr a\leq A\pr b$ in $P^1_A$ if $a\lex b$, while $(A\pr BC, h)\leq
(A\pr DE, k)$ in $P_A\times\{1, \ldots, \ell-1\}$ if either $h<k$, or
$h=k$ and $BC\lex DE$.

\begin{alg}[ht]
  \caption{a uniform random generator of derivation trees.}
  \label{alg:dturg}
  \begin{myprg*}
    \sprocedure RandomTree$(A, \ell)$ \nl
    $i\stv 0$, $r\stv\no$, $t\stv\no$ \nl
    \swhile $i < \kappa$ and $r=\no$ \sdo \tnl
    $i\stv i+1$ \nl
    \textbf{generate} $u\in\{1, \ldots, 2^{b_A(\ell)}\}$ 
    \emph{uniformly at random} \nl
    \sif $u\leq C_A(\ell)$ \sthen $r\stv u$ \unl
    \sif $r\not=\no$ \sthen \tnl
    \sif $\ell=1$ \sthen \tnl
    let $A\pr a$ be the $r$-th element of $P^1_A$ \nl
    $t\stv (A,a)$ \unl
    \selse \tnl
    compute the smallest element $(A\pr BC, k)$ in $P_A\times\{1,
    \ldots,\ell-1\}$ \tab\tnl such that $\displaystyle \sum_{(A\pr DE, h)\leq (A\pr BC, k)}
    C_D(h)C_E(\ell-h)\geq r$ \` $(*)$\untab\unl
    $t_B\stv $RandomTree$(B,k)$ \nl
    $t_C\stv $RandomTree$(C,\ell-k)$ \nl
    \sif $t_B\not=\no$ and $t_C\not=\no$ \sthen $t\stv (A,t_B,t_C)$ \untab\unl
    \sreturn $t$.
  \end{myprg*}
\end{alg}

Reasoning by induction on $\ell$, it is easy to verify that, if the
output $A(\ell)$ of RandomTree$(A,\ell)$ is different from $\no$, then
$A(\ell)$ is uniformly distributed in $T^\ell_A$. More precisely, for
every $t\in T^n_A$, we have
\[
\Pr\{ A(n)=t \mid A(n)\not=\no \} = 1/{C_A(n)}.
\]

Now, let us give an upper bound to $\Pr\{ A(n)=\no \}$.  Clearly, this
value depends on the global parameter $\kappa$ used by the procedure.
Denoting by $e(\ell)$ the maximum of all values $\Pr\{ A(\ell)=\no \}$
for $A\in V$, one can easily show that $e(1)\leq (1/2)^\kappa$ and
$e(\ell)\leq (1/2)^\kappa +\max_{1\leq k\leq \ell-1}
\{e(k)+e(\ell-k)\}$ for every $1\leq\ell\leq n$. A simple induction
proves $e(n)\leq (2n-1)/2^\kappa$ and hence fixing
$\kappa=3+\lceil\log n\rceil$ we have
\[
\Pr\{ A(n)=\no \}\leq e(n)< 1/4 \qquad\text{for every $A\in V$.}
\]

\clearpage 

As far as the time complexity is concerned, we assume to search the
element $(A\pr BC, k)$ of step $(*)$ by a boustrophedonic\footnote{\ie
  turning like oxen in ploughing (Webster).} routine \citep{FZV94}.
Again, let $N(\ell)$ be the maximum number of \mPrRAM instructions
executed by RandomTree$(A,\ell)$ for $A\in V$. Then, for a suitable
constant $c>0$, one can write the following recursion, for every
$1\leq \ell\leq n$,
\[
N(\ell)=
\begin{cases}
  O(\kappa) & \text{if $\ell=1$} \\
  O(\kappa) 
    + \max_{1\leq j\leq \ell} \{ c\min\{ j, \ell-j \} + N(j) + N(\ell-j) \} 
    & \text{if $\ell>1$}.
\end{cases}
\]
This proves that $N(n)=O(\kappa n)+c f(n)$, $f(n)$ being the solution
of the minimax equation $f(n)=\max\{ f(k) + f(n-k) + \min\{ k, n-k \}
\}$ usually arising in the evaluation of the cost of boustrophedonic
search \citep{GK81}. Since it is known that $f(n)=O(n\log n)$,
assuming $\kappa=O(\log n)$, we get $N(n)=O(n\log n)$ which, under our
model of computation, gives a total time cost $O(n^2\log n)$, since
all integers involved in the calls of this routine have $O(n)$ bits.
Finally, as it is straightforward to check, the number of random bits
used by the procedure is $O(n^2\log n)$.

We observe that, following an idea described by \citet{Gol95}, the
computation of the coefficients $\{ (C_A(\ell), b_A(\ell)) : A\in V,
1\leq \ell\leq n\}$ and the actual process of generating a tree
uniformly at random in $T^n_S$ can be mixed together into a unique
procedure which only requires space $O(n)$ under logarithmic cost
criterion (leaving unchanged the order of growth of the time
complexity).

\medskip

\noindent We summarize the result of this section by the following
\begin{pro}
  \label{pro:cfrug}
  Given a context-free grammar $G=\langle V,\Sigma,S, P\rangle$ in
  Chomsky normal form, there exists a \urg for $\langle T_S, |\cdot|
  \rangle$ working in $O(n^2\log n)$ time and using $O(n^2\log n)$
  random bits.
\end{pro}

\subsection{Earley's algorithm for counting derivations}
\label{ssec:earley}

The number of derivation trees of a terminal string in a \cf grammar
can be computed by adapting Earley's algorithm \citep{Ear70} for
context-free recognition. The main advantage of this procedure, with
respect to the well known CYK algorithm \citep{Har78}, is that in the
case of a grammar with bounded ambiguity, the computation only
requires quadratic time on a \mRAM under unit cost criterion
\citep{Ear70,AU72}.

\medskip

Let again $G=\langle V,\Sigma,S, P\rangle$ be a \cf grammar in Chomsky
normal form without useless variables. In passing, we note that the
algorithm can be easily modified to work on every \cf grammar without
$\ew$ and unit productions.  Our algorithm manipulates a
\midx{weighted dotted productions} \mridx{dotted productions}{weighted
  dotted productions} \emph{weighted} version of the so called
\emph{dotted productions} of $G$, \ie expressions of the form
$A\pr\alpha\cdot\beta$, where $A\in V$, $\alpha,\beta\in (\Sigma\cup
V)^*$ and $A\pr \alpha\beta\in P$.

Given an input string $x=a_1a_2\ldots a_n$, the algorithm computes a
table of entries $S_{i,j}$, for $0\leq i\leq j\leq n$, each of which
is a list of term of the form $[A\pr\alpha\cdot\beta, t]$, where
$A\pr\alpha\cdot\beta$ is a dotted production in $G$ and $t$ is a
positive integer. Each pair $[A\pr\alpha\cdot\beta, t]$ is called
\emph{state} and $t$ is the \emph{weight} of the state.

The table of lists $S_{i,j}$ computed by the algorithm has the
following properties for every pair of indices $0\leq i\leq j\leq n$:
\begin{enumerate}
\item \label{en:er1} $S_{i,j}$ contains at most one state
  $[A\pr\alpha\cdot\beta, t]$ for every dotted production
  $A\pr\alpha\cdot\beta$ in $G$;
\item \label{en:er2} a state $[A\pr\alpha\cdot\beta, t]$ belongs to
  $S_{i,j}$ iff there exists $\delta\in V^*$ such that $S\ppr a_1\dots
  a_i A \delta$ and $\alpha\ppr a_{i+1}\ldots a_j$;
\item \label{en:er3} if $[A\pr\alpha\cdot\beta, t]$ belongs to
  $S_{i,j}$, then $t=\#\{ \alpha\ppr a_{i+1}\ldots a_j \}$,
  \ie~the number of leftmost derivations $\alpha\ppr a_{i+1}\ldots
  a_j$.
\end{enumerate}
Note that, since there are no $\ew$-productions,
$[A\pr\alpha\cdot\beta, t]\in S_{i,i}$ implies $\alpha=\ew$ for every
$0\leq i\leq n$. Furthermore, once the lists $S_{i,j}$ are completed
for every $0\leq i\leq j\leq n$, the number of parse trees deriving $x$
can be obtained by the sum $\sum_{[S\pr AB\cdot, t]\in S_{0,n}} t$.

\medskip

The algorithm first computes the list $S_{0,0}$ of all the states
$[A\pr\cdot\alpha,1]$ such that $S\ppr A\delta$ for some $\delta\in
V^*$. Then, it executes, for $1\leq j\leq n$, the cycle of
\emph{Scanner}, \emph{Predictor} and \emph{Completer} loops given in
Algorithm~\ref{alg:earley} computing, at the $j$-th loop, the lists
$S_{i,j}$ for $0\leq i\leq j$.  To this end the procedure maintains a
family of sets $L_{B,i}$ for $B\in V$ and $1\leq i\leq j$; each
$L_{B,i}$ contains all indices $k\leq i$ such that a state of the form
$[A\pr\alpha\cdot B\beta, t]$ belongs to $S_{k,i}$ for some $A\in V$,
$\alpha,\beta\in V\cup\{\ew\}$, $t\in\bN$.  Moreover, during the
computation every state in $S_{i,j}$ can be unmarked or marked
according to whether it can still be used to add new states in the
table.

The statement ``\textsc{add} $D$ \textsc{to} $S_{i,j}$'' simply
appends the state $D$ as unmarked to $S_{i,j}$ and updates $L_{B,i}$
whenever $D$ is of the form $[A\pr\alpha\cdot B\beta,t]$; the
statement ``\textsc{update} $[A\pr\alpha\cdot\beta, t]$ \textsc{in}
$S_{i,j}$'' replaces the state $[A\pr\alpha\cdot\beta, u]$ in
$S_{i,j}$ for some $u\in\bN$ by $[A\pr\alpha\cdot\beta, t]$, finally,
``\textsc{mark} $D$ \textsc{in} $S_{i,j}$'' transforms the state $D$
in $S_{i,j}$ into a marked state.

\begin{alg}[ht]
  \caption{Part of Earley's algorithm modified to count derivation trees.}
  \label{alg:earley}
  \begin{myprg}
  
  \sfor $j=1\ldots n$ \sdo

  \\\\

  \emph{Scanner}: \nl
  \sfor $i=j-1\ldots 0$ \sdo \tnl
     \sfor $[A\pr \cdot a, t]\in S_{i,j-1}$ \sdo \tnl
        \textsc{mark} $[A\pr \cdot a, t]$ \textsc{in} $S_{i,j-1}$ \nl
        \sif $a=a_j$ \sthen \label{ln:add1}%
             \textsc{add} $[A\pr a\cdot, t]$ \textsc{to} $S_{i,j}$ 

  \cleartab \\ \\ 
  \emph{Completer}: \nl

  \label{ln:comp} 
  \sfor $i=j-1\ldots 0$ \sdo \tnl
     \sfor $[B\pr \gamma\cdot, t]\in S_{i,j}$ \sdo \tnl
        \textsc{mark} $[B\pr \gamma\cdot, t]$ \textsc{in} $S_{i,j}$ \nl
        \sfor $k\in L_{B,i}$ \sdo \tnl
           \sfor $[A\pr \alpha\cdot B\beta, u]\in S_{k,i}$ \sdo \tnl
              \sif $[A\pr \alpha B\cdot\beta, v]\in S_{k,j}$ \tnl 
              \sthen \label{ln:upd}
                     \textsc{update} $[A\pr \alpha B\cdot\beta,v+tu]$ \textsc{in} $S_{k,j}$ \nl
              \selse \label{ln:add2}%
                     \textsc{add} $[A\pr \alpha B\cdot\beta,tu]$ \textsc{to} $S_{k,j}$ 

  \cleartab \\ \\ \emph{Predictor}: \nl

  \sfor $i=0\ldots j-1$ \sdo \tnl
     \sfor $[A\pr \alpha\cdot B\beta, t]\in S_{i,j}$ \sdo \tnl
        \label{ln:mark3}%
        \textsc{mark} $[A\pr \alpha\cdot B\beta, t]$ \textsc{in} $S_{i,j}$ \nl
        \sfor $B\pr\gamma\in P$ \sdo \tnl
           \sif $[B\pr \cdot\gamma,1]\not\in S_{j,j}$ \sthen \label{ln:add3}%
              \textsc{add} $[B\pr \cdot\gamma,1]$ \textsc{to} $S_{j,j}$ \untab\untab \unl 

  \swhile $\exists$ \textsc{unmarked} $[A\pr \cdot B\beta, t]\in S_{j,j}$ \sdo \tnl
     \label{ln:mark4}%
     \textsc{mark} $[A\pr \cdot B\beta, t]$ \textsc{in} $S_{j,j}$ \nl
     \sfor $B\pr\gamma\in P$ \sdo \tnl
        \sif $[B\pr \cdot\gamma,1]\not\in S_{j,j}$ \sthen \label{ln:add4}%
           \textsc{add} $[B\pr \cdot\gamma,1]$ \textsc{to} $S_{j,j}$

\end{myprg}
\end{alg}

We are then able to prove the following
\begin{lem}
  The table of lists $S_{i,j}$, $0\leq i\leq j\leq n$, computed by the
  procedure described above satisfies the properties~\ref{en:er1}),
  \ref{en:er2}) and~\ref{en:er3}).
\end{lem}
\begin{proof}
  First, observe that statement~\ref{en:er1}) is easily verified: at
  line~\ref{ln:add1} distinct states are added to an initially empty
  list $S_{i,j}$; at lines~\ref{ln:add2}, \ref{ln:add3}, \ref{ln:add4}
  a state is added to a list provided no state with the same dotted
  production already appears in the list.
  
  Statement~\ref{en:er2}) only refers to dotted productions appearing
  in the lists and does not concern the weight of the states.
  Moreover, disregarding the computations on the weight of the states,
  the procedure works on the dotted production exactly like Earley's
  algorithm; hence statement~\ref{en:er2}) is a consequence of its
  correctness (for a detailed proof see \citep[Theorem~4.9]{AU72}).
  
  Now, let us prove statement~\ref{en:er3}). First note that all
  states in $S_{i,j}$, for $1\leq i\leq j\leq n$, are marked during
  the computation. Hence, we can reason by induction on the order of
  marking states. The initial condition is satisfied because all
  states in each $S_{j,j}$, $0\leq j\leq n$, are of the form
  $[A\pr\cdot\alpha,t]$ and have weight $t=1$. Also the states of the
  form $[A\pr a\cdot,t]$ have weight $t=1$ and again
  statement~\ref{en:er3}) is satisfied.
  
  Then, consider a state $D=[A\pr\alpha\cdot B\beta,w]\in S_{k,j}$
  with $k<j$. We claim that $w$ is the number of leftmost derivations
  $\alpha B\ppr a_{k+1}\ldots a_j$. A state of this form is first
  added by the \emph{Completer} at line~\ref{ln:add2}. This means that
  there exists a set of indices $I_k$ such that for every $i\in I_k$
  there is $[A\pr\alpha\cdot B\beta, u_i]\in S_{k,i}$ with
  $u_i\in\bN$, and a family $U_i$ of sates $[B\pr \gamma\cdot]\in
  S_{i,k}$ such that
  \[
  w=\sum_{i\in I_k}\sum_{[B\pr\gamma\cdot,t]\in U_i} tu_i.
  \]
  Observe that $k\leq i< j$ for every $i\in I_k$ and $U_i$ is the
  subset of all states in $S_{i,j}$ with a dotted production of the
  form $B\pr\gamma\cdot$.  Moreover, each state $[A\pr\alpha\cdot
  B\beta, u_i]\in S_{k,i}$ is marked at line~\ref{ln:mark3}
  or~\ref{ln:mark4} before $D$ is added to $S_{k,j}$. Also all states
  in $U_i$, for all $i\in I_k$, are marked during the computation of
  the weight $w$. Observe that, due to the form of the grammar,
  updating such a weight $w$ cannot modify the weight of any state in
  $U_i$. As a consequence all the sates in $U_i$ are marked before
  $D$. Hence, by inductive hypothesis, we have for every $i\in I_k$
  \begin{equation}
    \label{eq:erui}
    u_i = \#\{ \alpha\ppr a_{k+1}\ldots a_i \}
  \end{equation}
  and, for each $[B\pr\gamma\cdot,t]\in U_i$,
  \begin{equation}
    \label{eq:ert}
    t =  \#\{ \gamma\ppr a_{i+1}\ldots  a_j \}.
  \end{equation}
  Now the number of leftmost derivations $\alpha B\ppr a_{k+1}\ldots
  a_j$ is clearly given by
  \[
  \sum_{k\leq i <j} \#\{ \alpha\ppr a_{k+1}\ldots a_i\} 
  \sum_{B\pr\gamma \in P} \#\{ \gamma\ppr a_{i+1}\ldots a_j \}
  \]
  and the claim follows from statement~\ref{en:er1}) and
  equalities~(\ref{eq:erui}) and~(\ref{eq:ert}).
\end{proof}

\begin{teo}
  \label{teo:cfd}
  Given a context-free grammar $G$ in Chomsky normal form and assuming
  the \mRAM model under logarithmic cost criterion, the algorithm
  described above computes the number of derivation trees of an input
  string of length $n$ in $O(n^4)$ time, consuming $O(n^3)$ space.  If
  the grammar $G$ is finitely ambiguous, then the algorithm has time
  complexity $O(n^2\log n)$ and space complexity $O(n^2)$.
\end{teo}
\begin{proof}
  We first observe that every list $S_{i,j}$ for $0\leq i\leq j\leq n$
  contains at most $O(1)$ states, each of which can be represented by
  one integer of size respectively $O(1)$ or $O(n)$ according to
  wheather $G$ is finitely ambiguous or not. Since the space taken by
  the algorithm is essentially due to the memory required by the table
  $S_{i,j}$, we obtain a space complexity $O(n^2)$ for finitely
  ambiguous grammars and $O(n^3)$ in the general case.
  
  As far as the time complexity is concerned, note that in each loop,
  for a fixed $1\leq j\leq n$, the \emph{Scanner} and \emph{Predictor}
  phase execute $O(j)$ statements, while the \emph{Completer} phase
  requires $O(j\ell)$ unit steps, where $\ell$ is the maximum size of the
  sets $L_{B,i}$, $B\in V$, $0\leq i\leq j-1$. Now, for a general
  grammar, we have $\ell=O(j)$ which implies a total number of unit steps
  $O(n^3)$: each of them requires logarithmic time $O(n)$ leading the
  overall time complexity to $O(n^4)$.
  
  On the contrary, in the case of a finitely ambiguous grammar $G$, we
  have $\ell=O(1)$ and hence we only need $O(\log n)$ logarithmic time
  to locate each state in the table yielding a total time complexity
  $O(n^2\log n)$.
\end{proof}

\subsection{Inherently ambiguous context-free languages}
\label{ssec:inherentamb}

Now, let us go back to our original problem and let $L\subseteq
\fldom$ be a \cf language. We recall that $L$ is
\mdidx{ambiguity}{context-free!language} \sidx{unambiguous}{\irCFl} if
it is generated by an unambiguous \cf grammar, while it is
\esidx{inherently ambiguous}{\irCFl} whenever every \cf grammar $G$
generating $L$ is ambiguous. We also say that $L$ is \emph{finitely}
\mdidx{finitely ambiguous}{\irCFl} (respectively, \emph{polynomially})
\mdidx{polynomially ambiguous}{\irCFl} \msidx{polynomially
  ambiguous}{\irCFl} \emph{ambiguous} if it is generated by a finitely
(respectively polynomially) ambiguous \cf grammar.

Observe that there are natural examples of polynomially ambiguous \cf
languages which are not finitely ambiguous. For instance, if $L=\{
ww^R : w\in\{a,b\} \}$, then it turns out that $L^2$ is inherently
ambiguous, but not finitely ambiguous \citep[Section 7.3]{Har78}:
however, it is easy to verify that $L^2$ is generated by a grammar of
ambiguity $O(n)$ given by the set of productions $S\pr AB$, $A\pr
aAa|bAb|\ew$ and $B\pr aBa|bBb|\ew$. Similarly, for every $k\in\bN$,
the language $(L^2\cdot\{\sep\})^k$ is generated by a grammar of
ambiguity $O(n^k)$.

Then, applying Propositions\ref{pro:cfrug} and Theorem~\ref{teo:cfd}
to Theorems~\ref{teo:rugpa} and~\ref{teo:rasecpa}, we obtain
\begin{teo}
  \label{teo:cf}
  If $L$ is a polynomially ambiguous context-free language, then there
  exists a polynomial time \urg for $L$ and a fully polynomial time
  \ras for its census function $C_L$. If moreover $C_L$ is polynomially
  bounded, then there exists a polynomial time \rec for $L$.
\end{teo}
In particular, in the case of finite ambiguity, one gets the following
\begin{cor}
  \label{cor:cf}
  If $L$ is a finitely ambiguous context-free language, then there
  exists a \urg for $L$ working in $O(n^2\log n)$ time on a \mPrRAM
  under logarithmic cost criterion and a \ras for its census function
  $C_L$ of the same time complexity.
\end{cor}

\subsection{The grammar as a part of the input}
\label{ssec:ug}

Observe that the results of Sections~\ref{ssec:rnddt}
and~\ref{ssec:earley} assume as fixed a \cf grammar $G$. In view of an
application presented in the next section, here we study the
complexity of similar algorithms taking $G$ as a part of the input.

To this end, we need a definition of the dimension of a \cf grammar;
if $G=\langle V,\Sigma,S, P \rangle$, we define its dimension $|G|$ as
the number of bits needed to encode it, which is $O(\#P\ell_G
\log(\#V+\#\Sigma))$, where $\ell_G$ is the maximum number of symbols
appearing in the right hand side of a production in $P$. Also in this
case, we restrict to \cf grammars in Chomsky normal form, since it is
known \citep{Har78} that every \cf grammar $G$ can be transformed in a
\cf grammar $G'$ in Chomsky normal form in time polynomial in $|G|$
(and hence $|G'|$ is polynomially related to $|G|$).

Given a \cf grammar $G=\langle V,\Sigma,S, P \rangle$ (in Chomsky
normal form) and $n>0$, what is essentially needed to generate
(uniformly at random) a derivation tree of a word of length $n$, are
the coefficients $C_A(\ell)$, for $1\leq\ell\leq n$ and $A\in V$ ( see
Section~\ref{ssec:rnddt}). Observe that the method used in that
section, fundamentally based on Comtet's recurrence equation, can, in
general, require time exponential in $|G|$. Nonetheless, by a modified
version of the CYK algorithm \citep{Har78}, we obtain the following
\begin{lem}
  \label{lem:treecnt}
  There exists an algorithm that, having as input a \cf grammar
  $G=\langle V,\Sigma,S, P \rangle$ (in Chomsky normal form), $A\in V$
  and $n>0$, computes $C_A(n)$ in $O(n^3|G|)$ time on a \mRAM under
  logarithmic cost criterion.
\end{lem}
\begin{proof}
  We design such an algorithm by an application of dynamic programming
  and of the well known convolution property
  \[
  C_A(\ell)=\sum_{A\pr BC\in P}\sum_{i=1}^{\ell-1} C_B(i)C_C(\ell-i)
  \]
  which holds for every $A\in V$ and $\ell>0$. Consider
  Algorithm~\ref{alg:treecnt}, where $c(A;k)$, for $A\in V$ and $1\leq
  k\leq n$, is an array of integer.

  \begin{alg}[ht]
    \caption{Counting derivation trees.}
    \label{alg:treecnt}
    \begin{myprg}
      \sinput $G=\langle V,\Sigma,S, P \rangle, A, n$ \nl
      \sforall $B\in V$  \sdo $c(B;1)\stv 0$ \label{ln:forall1} \nl 
      \sforall $B\pr a\in P$  \sdo $c(B;1)\stv c(B;1)+1$ \label{ln:forall2} \nl
      \sfor $k=2\ldots n$ \sdo\label{ln:fork} \tnl
      \sforall $B\pr CD\in P$ \sdo \label{ln:coll} \tnl
      $c(B;k)\stv c(B;k)+\sum_{0<i<k} c(C;i) c(D;k-i)$ \untab \label{ln:sum} \unl
      \soutput $c(A;n)$.
    \end{myprg}
  \end{alg}
  
  To prove the correctness of the algorithm, we prove a somehow
  stronger result: at the beginning of each iteration of the loop at
  line~\ref{ln:fork}, $c(B;\ell)=C_B(\ell)$ for every $B\in V$ and
  $1\leq\ell<k$.  By induction on $k$: if $k=2$, then the loop at
  lines~\ref{ln:forall1} and~\ref{ln:forall2} compute exactly $C_A(1)$
  for every $B\in V$, moreover, the loop at line~\ref{ln:fork} makes
  no iteration, hence the statement is true.  If $k>2$, then
  $\sum_{0<i<k} c(C;i) c(D;k-i)$, by inductive hypothesis, is equal to
  the number of parse trees in $T^k_B$ with the first derivation
  corresponding to $B\pr CD$.  Then the loop at line~\ref{ln:coll}
  collects in $c(B;k)$ the value $C_B(k)$ by iteratively summing over
  all the production in $P$ with $B$ on the left, for every $B\in V$.
  To obtain an upper bound on the computation time, observe that the
  sum of line~\ref{ln:sum} has at most $n$ summands and is in a nested
  loop which is executed $O(n\#P)$ times, and that all the involved
  integers have size $O(n)$ bits.
\end{proof}

Then, as discussed in Section~\ref{ssec:rnddt}, we can apply
Algorithms~\ref{alg:treecnt} and~~\ref{alg:dturg} to obtain the
following
\begin{pro}
  There exists a \urg of derivation trees that, given a \cf grammar
  $G$ in Chomsky normal form and $n>0$ as input, works on a \mPrRAM in
  time polynomial in $|G|$ and $n$.
\end{pro}

Furthermore, the algorithm of Section~\ref{ssec:earley} to compute the
number of derivation trees of a terminal string, can be used also
assuming the grammar as part of the input. A simple analysis leads to
the following
\begin{pro}
  The number of derivation trees of a terminal string $x$ in a
  context-free grammar $G$ in Chomsky normal form can be computed in
  time polynomial in $|G|$ and $|x|$ on a \mRAM under logarithmic cost
  criterion.
\end{pro}


\RCSfooter$Id: c4s2.tex,v 2.4 1999/11/01 11:30:53 santini Exp $

\section{One-way Nondeterministic Auxiliary Pushdown Automata}
\label{sec:npda}

In this section we show how to apply the general technique of
Chapter~\ref{cha:ambparadigm} to the case of languages accepted by
polynomial time \emph{one-way nondeterministic auxiliary pushdown
  automata} (\moNAuxPDA, see Section~\ref{ssec:pda} for a definition of
the model).

We recall that, given a \moNAuxPDA $M$, the \emph{ambiguity} of $M$
\mdidx{ambiguity}{\irAuxPDA} is the function $d_M:\bN\to\bN$ defined
as the maximum number of accepting computations for every input string
of length $n\in\bN$.  Hence, we say that $M$ is \esidx{polynomially
  ambiguous}{\irAuxPDA} \mdidx{polynomially ambiguous}{\irAuxPDA} if,
for some polynomial $p(n)$, we have $d_M(n)\leq p(n)$ for every $n>0$.
We also recall that it is well known that, given an integer input
$n>0$, a context-free grammar $G_n$ can be built, in time polynomial
in $n$, such that $L(G_n)\cap\Sigma^n=L(M)\cap\Sigma^n$, where
$L(G_n)\subseteq\fldom$ is the language generated by $G_n$ and
$L(M)\subseteq\fldom$ is the language accepted by $M$.  This allows us
to apply the results of the previous section to the languages accepted
by \moNAuxPDA.

\bigskip

Here, we describe a modified version of the usual construction of
$G_n$ given by \citet{Coo71,ABP93} which allows to bound the ambiguity
of $G_n$ with respect to the ambiguity of $M$.  First of all, we
assume without loss of generality that the automaton cannot
simultaneously consume input and modify the content of the stack and
that at most one symbol can be pushed or popped for each single move.

A \esidx{surface configuration}{\irAuxPDA} \citep{Coo70} of a
\moNAuxPDA $M$ on input $x\in\Sigma^+$ of length $n$ is a $5$-tuple
$(q,w,i,\Gamma,j)$ where $q$ is the state of $M$, $w$ the content of
its work tape, $1\leq i\leq |w|$ the work tape head position, $\Gamma$
the symbol on top of the stack and $1\leq j\leq n+1$ the input tape
head position. Observe that there are $n^{O(1)}$ surface
configurations on any input of length $n\geq 0$.  Two surface
configurations $C_1,C_2$ form a \msidx{realizable pair}{\irAuxPDA}
\emph{realizable pair $(C_1,C_2)$ (on a word $y\in\Sigma^+$)} iff $M$
can move (consuming input $y$) from $C_1$ to $C_2$, ending with its
stack at the same height as in $C_1$, without popping below this
height at any step of the computation. If $(C_1,D)$ is a realizable
pair on $y'$ and $(D,C_2)$ is a realizable pair on $y''$, then it is
straightforward to check that $(C_1,C_2)$ is a realizable pair on
$y=y'y''$.  Let $\cS_n$ be the set of surface configurations of $M$ on
inputs of length $n$ and define the \cf grammar $G_n(M)$ by the
following statements:
\begin{enumerate}[(a)]

\item the set of nonterminal symbol $V$ contains each $(C_1,C_2,\ell)$
  such that $C_1,C_2\in\cS_n$ and $\ell\in\{0,1\}$;

\item the set of terminal symbols is $\Sigma$;
  
\item the initial variable is $(C_{\text{\textsf{in}}},
  C_{\text{\textsf{fin}}})$, where $C_{\text{\textsf{in}}}$ and
  $C_{\text{\textsf{fin}}}$ represent respectively the initial and
  final surface configuration of $M$;

\item the set $P$ of productions is given as follows:

  \begin{enumerate}[(1)]
    
  \item \label{rl:term} $(C_1,C_2,0)\pr\sigma \in P$ iff $(C_1,C_2)$
    is a realizable pair on $\sigma\in\Sigma\cup \{\ew\}$ via a
    single move computation;
  
  \item \label{rl:anyd} $(C_1,C_2,0)\pr(C_1,D,1)(D,C_2,\ell)\in P$,
    for $\ell\in\{0,1\}$, iff $C_1,C_2,D\in\cS_n$;
  
  \item \label{rl:unamb} $(C_1,C_2,1)\pr(D_1,D_2,\ell)\in P$, for
    $\ell\in\{0,1\}$, iff $C_1,D_1,D_2,C_2\in\cS_n$, $D_1$ can be
    reached from $C_1$ by a single move pushing a symbol on top of the
    stack and $C_2$ can be reached from $D_2$ by a single move popping
    the same symbol from the top of the stack.

  \end{enumerate}

\end{enumerate}

For the sake of brevity define, for each realizable pair $(C_1,C_2)$,
the set $\cC(C_1,C_2)$ of all computations of $M$ starting from $C_1$
and ending in $C_2$ with the stack at the same height as in $C_1$,
without popping below this height at any point of the computation;
define also the subset $\cC_1(C_1,C_2)\subseteq \cC(C_1,C_2)$ of such
computations (of at least two steps) during which the stack height
never equals the stack height at the extremes $C_1,C_2$ and the subset
$\cC_o(C_1,C_2)= \cC(C_1,C_2)\setminus \cC_1(C_1,C_2)$ of computations
during which the stack height equals, at least once, the stack height
at the extremes $C_1,C_2$. 

Fist of all, observe that, following \citet{Coo71} and \citet{ABP93},
it is possible to prove the
\begin{pro}
  Given a \moNAuxPDA $M$ working in polynomial time, there exists an
  algorithm computing, on input $n>0$, the \cf grammar $G_n(M)$ in
  polynomial time on a \mRAM under logarithmic cost criterion.
\end{pro}

We now turn to show that the \cf grammar $G_n(M)$ defined above
exactly generates the set of strings of length $n$ accepted by $M$:
\begin{lem}
  \label{lem:gmcorr}
  Given a \moNAuxPDA $M$, the \cf grammar $G_n(M)$ is such that
  \begin{enumerate}[(i)]
  \item \label{st:com2der} if $(C_1,C_2)$ is a realizable pair on $y$
    via a computation in $\cC_\ell(C_1,C_2)$, then $(C_1,C_2,\ell)\ppr
    y$;
  \item \label{st:der2com} if $(C_1,C_2,\ell)\ppr y$, then $(C_1,C_2)$
    is a realizable pair on $y$ via a computation in
    $\cC_\ell(C_1,C_2)$.
  \end{enumerate}
\end{lem}
\begin{proof}
  We prove statement~(\ref{st:com2der}) by induction on the number $m$
  of computation steps from $C_1$ to $C_2$.  If $m=1$ then the
  computation can only be in $\cC_0(C_1,C_2)$ and consumes at most one
  symbol $\sigma\in\Sigma\cup\{\ew\}$; then, by~(\ref{rl:term}),
  $(C_1,C_2,0)\pr\sigma\ppr\sigma$.
  
  If $m>1$ and the computation is in $\cC_0(C_1,C_2)$ let $D$ be the
  first surface configuration after $C_1$ in the computation for which
  the stack height equals that of $C_1,C_2$; it is then
  straightforward to check that $(C_1,D)$ is a realizable pair on some
  $y'$ via a computation in $\cC_1(C_1,D)$ and $(C_2,D)$ is a
  realizable pair on some $y''$ via a computation in
  $\cC_\ell(D,C_2)$, for some $\ell\in\{0,1\}$, where $y=y'y''$;
  moreover these computations have both fewer steps then the one
  between $C_1$ and $C_2$. Hence, by inductive hypothesis,
  $(C_1,D,1)\ppr y'$ and $(D,C_2,\ell)\ppr y''$ and,
  by~(\ref{rl:anyd}), $(C_1,C_2,0) \pr(C_1,D,1)(D,C_2,\ell)\ppr
  y'y''=y$.
  
  Finally, if $m>1$ and the computation is in $\cC_1(C_1,C_2)$, then
  it is straightforward to check that the first move after $C_1$ (say,
  to a surface configuration $D_1$) has to be a push and the last move
  before $C_2$ (say, from a surface configuration $D_2$), has to be a
  pop of the same symbol; moreover, since the moves from $C_1$ to
  $D_1$ and from $D_2$ to $C_2$ alter the stack and hence consume no
  input, $(D_1,D_2)$ is a realizable pair on $y$ via a computation in
  $\cC_\ell(D_1,D_2)$, for some $\ell\in\{0,1\}$, have fewer steps
  than the computation between $C_1$ and $C_2$.  Then, by inductive
  hypothesis, $(D_1,D_2,\ell)\ppr y$ and, by~(\ref{rl:unamb}),
  $(C_1,C_2,1)\pr (D_1,D_2,\ell)\ppr y$.

  \medskip
  
  Now we prove statement~(\ref{st:der2com}) by induction on the number
  $k$ of derivation steps. If $k=1$ the only production leading to a
  terminal symbol is the one in~(\ref{rl:term}), hence the statement
  is obvious.
  
  On the other hand, if $k>1$ and $\ell=0$, then, by~(\ref{rl:anyd}),
  the derivation can only be of the form $(C_1,C_2,0)\pr
  (C_1,D,1)(D,C_2,\ell') \ppr y'y''=y$, for some $\ell'\in\{0,1\}$,
  where $(C_1,D,1)\ppr y'$ and $(D,C_2,\ell')\ppr y''$ and both
  derivations have less then $k$ steps; then, by inductive hypothesis,
  $(C_1,D)$ is a realizable pair on $y'$ and $(D,C_2)$ is a realizable
  pair on $y''$, hence $(C_1,C_2)$ is a realizable pair on $y$ via a
  computation in $\cC_0(C_1,C_2)$.
  
  If $k>1$ and $\ell=1$, then, by~(\ref{rl:unamb}), the derivation can
  only be of the form $(C_1,C_2,1)\pr$ $(D_1,D_2,\ell')\ppr y$, for some
  $\ell'\in\{0,1\}$, where $(D_1,D_2,\ell')\ppr y$ in $k-1$ steps; by
  inductive hypothesis, $(D_1,D_2)$ is a realizable pair on $y$;
  moreover, by~(\ref{rl:unamb}), $D_1$ can be reached from $C_1$ with
  a single push move and $C_2$ can be reached from $D_2$ with a single
  pop of the same symbol, hence it is straightforward to verify that
  $(C_1,C_2)$ is a realizable pair on $y$ via a computation in
  $\cC_1(C_1,C_2)$.
\end{proof}

In order to apply the result on \cf grammars of Section~\ref{sec:cfl}
in this case, we have to bound the ambiguity of $G_n(M)$. To this end,
we have the following
\begin{lem}
  \label{lem:pdaamb}
  Given a \moNAuxPDA $M$, the number of leftmost derivations
  $(C_1,C_2,\ell)\ppr y$ in $G_n(M)$ is less than or equal to the
  number of computations in $\cC_\ell(C_1,C_2)$ consuming $y$.
\end{lem}
\begin{proof}
  We want to prove that, for every $C_1,C_2\in\cS_n$ and
  $y\in\fldom$, the proofs of Lemma~\ref{lem:gmcorr} define a
  bijective map from the leftmost derivations $(C_1,C_2,\ell)\ppr y$
  to the computations in $\cC_\ell(C_1,C_2)$ consuming $y$.
  
  By the construction, it is clear that such a map is surjective.
  Hence, we simply have to show that it is injective, \ie if the
  computations associated with two leftmost derivations are equal,
  then the two derivations themselves are the same. We work by
  induction on the number $m$ of steps of the computation. If $m=1$
  the statement is obvious.
  
  If $m>1$ and $\ell=0$, let, for $i\in\{1,2\}$, $(C_1,C_2,0)\pr
  (C_1,D_i,1)(D_i,C_2,\ell_i)\ppr y$ be two leftmost derivations (for
  some $\ell_i\in\{0,1\}$ and $D_i\in\cS_n$). By construction of the
  grammar, if the associated computations are equal, then $D_1=D_2$,
  $\ell_1=\ell_2$ and there exist two words $y',y''$ such that
  $y=y'y''$, $(C_1,D_i,1)\ppr y'$ and $(D_i,C_2,\ell_i)\ppr y''$, for
  $i\in\{1,2\}$.  Then, by inductive hypothesis, $(C_1,D_i,1)\ppr y'$
  are the same leftmost derivation for $i\in\{1,2\}$ and
  $(D_i,C_2,\ell_i)\ppr y''$ are the same leftmost derivation for
  $i\in\{1,2\}$. Hence $(C_1,C_2,0)\pr (C_1,D_i,1)(D_i,C_2,\ell_i)\ppr
  y$ are the same leftmost derivation for $i\in\{1,2\}$.
  
  If $m>1$ and $\ell=1$, let, for $i\in\{1,2\}$, $(C_1,C_2,1)\pr
  (D_{1,i},D_{2,i},\ell_i)\ppr y$ be two leftmost derivations (for
  some $\ell_i\in\{0,1\}$ and $D_{1,i},D_{2,i}\in\cS_n$).  If the
  associated computations are equal, then again by the construction of
  the grammar, $D_{1,1}=D_{1,2}$, $D_{2,1}=D_{2,2}$ and
  $\ell_1=\ell_2$, hence, for $i\in\{1,2\}$,
  $(D_{1,i},D_{2,i},\ell_i)\ppr y$. Moreover, by inductive hypothesis,
  the two derivations are the same leftmost derivation for
  $i\in\{1,2\}$, hence $(C_1,C_2,1)\pr (D_{1,i},D_{2,i},\ell_i) \ppr
  y$ are the same leftmost derivation for $i\in\{1,2\}$. \qed
\end{proof}

\bigskip

We can now conclude our discussion: by Theorem~\ref{teo:rugpa}
and~\ref{teo:rasecpa}, an application of the results of
Section~\ref{ssec:ug} and of this section, leads to the following
\begin{teo}
  \label{teo:npda}
  Let $L$ be the language accepted by a polynomial time \moNAuxPDA with
  polynomial ambiguity. Then there exists a polynomial time \urg for
  $L$ and a fully polynomial time \ras for its census function $C_L$.
  If moreover $C_L$ is polynomially bounded, then there exists a
  polynomial time \rec for $L$.
\end{teo}


\RCSfooter$Id: c4s3.tex,v 2.4 1999/11/01 11:30:53 santini Exp $

\section{Rational Trace Languages}
\label{sec:rtl}

Another application concerns the uniform random generation and the
census function of trace languages.  To study this case we refer to
\citet{DM97} for basic definitions.  We only recall that, given a
trace monoid $\bbM(\Sigma,I)$ over the independence alphabet
$(\Sigma,I)$, a \eidx{\irTl}, \ie~a subset of $\bbM(\Sigma,I)$, is
usually specified by considering a string language $L\subseteq\fldom$
and taking the closure $[L]=\{t\in\bbM(\Sigma,I): t=[x] \mbox{ for
  some } x\in L\}$.  In particular, a trace language
$T\subseteq\bbM(\Sigma,I)$ is called \esidx {rational}{\irTl} if
$T=[L]$ for some regular language $L\subseteq\fldom$.  In this case we
say that $T$ is \emph{represented} by $L$ and the \emph{ambiguity} of
this representation is the function $d_L:\bN\to\bN$, defined by
$d_L(n)=\max_{x\in\Sigma^n} \#(L\cap [x])$. We say that a rational
trace language $T$ is \esidx{finitely ambiguous}{\irTl}
\mdidx{finitely ambiguous}{\irTl} if it is represented by a regular
language $L$ such that, for some $k\in\bN$, $d_L(n)\leq k$ for every
$n>0$; in this case we say that $T$ has \mdidx{ambiguity}{trace
  language} \esidx{ambiguity}{\irTl} $k$.

In the following, assuming a given independence alphabet $(\Sigma,I)$,
we denote by $\cR$ the set of all rational trace languages in
$\bbM(\Sigma,I)$ and, by $\cR_k$, the subset of trace languages in
$\cR$ of ambiguity $k$.  Clearly, for every independence alphabet
$(\Sigma,I)$, we have $\cR_1\subseteq \cR_2\subseteq\cdots \subseteq
\bigcup_{k=1}^\infty\cR_k \subseteq \cR$.

The properties of these families of languages have been studied in the
literature by \citet{BMS82,BMS85,Sak87} extending a previous study on
unambiguous rational sets in totally commutative monoids by
\citet{ES69}.  In particular, it is known that $\cR_1=\cR$ iff the
independence relation $I$ is transitive \citep{BMS85,Sak87}; on the
other hand, if $I$ is not transitive, then we get the following chain
of strict inclusions:
\[
\cR_1\subsetneq\cR_2\subsetneq\cdots\subsetneq
\bigcup_{k=1}^\infty\cR_k \subsetneq \cR.
\]

We say that a rational trace language $T$ is \esidx{polynomially
  ambiguous}{\irTl} \mdidx{polynomially ambiguous}{\irTl} if it is
represented by a regular language $L$ such that, for some polynomial
$p(n)$, we have $d_L(n)\leq p(n)$ for every $n>0$.  Observe that there
exist examples of polynomially ambiguous rational trace languages that
are not finitely ambiguous.  For instance, fixing $I=\{(a,b),(b,c)\}$,
if $L=(a^*c)^*(ab)^*c(a^*c)^*$, then it turns out that $[L]$ does not
belong to $\bigcup_{k=1}^\infty\cR_k$ \citep{BMS82}: however, $[L]$ is
polynomially ambiguous with $d_L(n)=O(n)$ since, for every $x$ of the
form $x=a^{k_1}ca^{k_2}c\ldots (ab)^{k_s}c a^{k_{s+1}}c\ldots
a^{k_t}c$ with $k_1,\ldots, k_s,\ldots k_t\in\bN$, it holds
\[
L\cap[x]=
\{ a^{k_1}c\ldots (ab)^{k_i}c \ldots a^{k_t}c : k_i=k_s, 1\leq i\leq t\}.
\]

\medskip 

\enlargethispage{3em} 

Now, let us go back to our problem: here, we want to use $L$ as an
ambiguous description of $[L]$. Also in this case, we first have to
design two routines: one for generating a word in $L$ uniformly at
random and the other for determining the number of representatives of
a trace in a given regular language $L$. The first routine is already
discussed in Section~\ref{ssec:examplereg}. The other algorithm is
obtained by adapting a procedure for solving the membership problem
for rational trace language \citep{BMS89,AG98}:
\begin{pro}
  Given an independence alphabet $(\Sigma,I)$ and a regular language
  $L\subseteq\fldom$, the problem of determining the cardinality of
  $L\cap[x]$, given $x\in L$ as input, can be solved on a \mRAM under
  unit cost criterion in $O(n^\alpha)$ time, where $\alpha$ is the
  maximum clique size in $(\Sigma,I)$.
\end{pro}

As a consequence, we can apply the general paradigm for uniform random
generation and approximate counting presented in
Chapter~\ref{cha:ambparadigm}.  This leads to the following
\begin{teo}
  \label{teo:rtl}
  Let $T\subseteq\bbM(\Sigma,I)$ be a finitely ambiguous rational
  trace language and assume $I\neq\emptyset$.  Then, $T$ admits a \urg
  working in $O(n^\alpha\log n)$ time and using $O(n^2\log n)$ random
  bits on a \mPrRAM under logarithmic cost criterion, where $\alpha$
  is the size of the maximum clique in $(\Sigma,I)$.  Moreover, there
  exists a \ras for the census function of $T$ of the same time
  complexity. On the other hand, if $T$ is polynomially ambiguous,
  then it admits a polynomial time \urg and a fully polynomial time
  \ras for its census function.
\end{teo}



\RCSfooter$Id: c5.tex,v 2.2 1999/10/29 10:24:12 santini Exp $

\chapter{Some Remarks}
\label{cha:boolean}

\mycite{italian}{%
  \begin{verse}
    Spesso c'è bonaccia sulla pagina.\\
    Inutile girarla per cercare\\
    l'angolo del vento.\\
    Si sta fermi,\\
    il pensiero oscilla,\\
    si riparano le cose\\
    che la navigazione ha guastato.
  \end{verse}
}%
{Valerio Magrelli}{Ora serrata retinae}

In this brief chapter we investigate how to extend some result of the
two previous chapters about combinatorial structures back to the more
general case of $p$-relations. In particular, we shortly investigate
some closure property of some classes of $p$-relations under union and
complement boolean operators, once suitably adapted to the case of
$p$-relations.

\bigskip
\RCSfooter$Id: c5s1.tex,v 2.2 1999/09/23 14:43:06 santini Exp $

\section{From Combinatorial Structures to $p$-Relations}
\label{sec:backtoprel}

As we have stated at the beginning of Chapter~\ref{cha:ambparadigm},
for the sake of brevity and clarity in the definitions and proofs, we
have restricted our attention to combinatorial structures instead of
$p$-relations. Here we want to briefly and informally suggest that the
results obtained so far can be generalized back to the case of
$p$-relations.

To this aim, one needs a definition of \emph{description} suitable to
the case of relation. A proposal can be the following: a $p$-relation
$S\subseteq\reldom$ is an \emph{(ambiguous)
  description} of a $p$-relation $R\subseteq\reldom$
via the function $f:S\to R$ iff, for every $\alpha,\beta\in\fldom$
such that $\alpha\rR\beta$, there exists some $\gamma\in S(\alpha)$
such that $f(\alpha,\gamma)=(\alpha,\beta)$. In this case, the
\emph{ambiguity} of the description is the function $d:R\to \bN$
defined by $d((\alpha,\beta))=\#\{\gamma\in S(\alpha):
f(\alpha,\gamma)=(\alpha,\beta) \}$ and the description is said to be
polynomial whenever $f$ and $d$ are computable in time polynomial in
$|\alpha|$ and there exists some $D\in\bN$ such that
$d((\alpha,\beta))=O(|\alpha|^D)$.

The results of Chapter~\ref{cha:ambparadigm} can be rephrased
according to the following
\begin{cor}
  \label{cor:preldes}
  If a $p$-relation $R$ has a polynomially (ambiguous) description $S$
  which admits a polynomial time \urg, then $R$ admits a polynomial
  time \urg; moreover, if, for every $\alpha\in\fldom$, the
  cardinality of $S(\alpha)$ can be computed in time polynomial in
  $|\alpha|$, then $R$ admits a fully polynomial \ras.
\end{cor}

In particular, Theorem~\ref{teo:pda} of Section~\ref{sec:npda} can be
generalized to extended one-way nondeterministic auxiliary pushdown
automata (see Section~\ref{ssec:pda}), obtaining the following
stronger version of Corollary~\ref{cor:perlupdaurg}
\begin{cor}
  \label{cor:prelpdaurg}
  If a $p$-relation $R\subseteq\reldom$ is accepted by a polynomially
  ambiguous extended one-way nondeterministic auxiliary pushdown
  automaton working in polynomial time, then $R$ admits a polynomial
  time uniform random generator.
\end{cor}
\begin{proof}
  The proof follows from an argument similar to the beginning of the
  proof of Theorem~\ref{teo:pda}: given an extended \moNAuxPDA $M$,
  with state set $Q$, and some $\alpha\in\fldom$, it is always
  possible to build (in time polynomial in $|\alpha|$) a \moNAuxPDA
  $M^{(\alpha)}$, with state set $Q\times\{1,\ldots,|\alpha|\}
  \times\Sigma$, such that the computation of $M^{(\alpha)}$ on input
  $\beta$ efficiently simulates the computation of $A$ on input
  $(\alpha,\beta)$ for every $\beta\in\fldom$ and has the same
  ambiguity as $M$.  The result then follows by applying
  Theorem~\ref{teo:npda} to $M^{(\alpha)}$.
\end{proof}


\RCSfooter$Id: c5s2.tex,v 2.4 1999/11/01 11:30:53 santini Exp $

\section{The $p$-Union}
\label{sec:union}

In this section we want to investigate the closure with respect to the
union of some of the classes of $p$-relations we have encountered in
previous chapters.

First of all, we need the following
\begin{dfn}
  Let $S\subseteq\reldom$ and
  $R\subseteq\reldom$ be two $p$-relations. Their
  $p$-union $R \cup^p S$ is defined as $\alpha \mathrel{(R\cup^p S)}
  \beta$ iff $\alpha \rR \beta$ or $\alpha \mathrel S \beta$, whenever
  the polynomials which relate the length of the elements in the
  domain to the length of the elements in the codomain are identical,
  otherwise is undefined.
\end{dfn}

\clearpage 

Then, by means of Corollary~\ref{cor:preldes} and
Theorem~\ref{teo:unionurg}, one can obtain the following
\begin{pro}
  Let $S\subseteq\reldom$ and $R\subseteq\reldom$ are two
  $p$-relations such that $R\cup^p S$ is defined. If $r_R\in\cFP$ and
  $r_S\in\cFP$, then $R\cup^p S$ admits a polynomial time \urg.
\end{pro}

On the other hand, it is known \citep[Theorem 3.1]{Huy90} that there
exist two unambiguous context free languages $\hat{L}_1, \hat{L}_2$
such that their union is inherently ambiguous and its rank is not
polynomial time computable unless $\cP=\cP^{\cnP}$.  Moreover, it is
also known \citep{GS91} that unambiguous context free languages are
rankable in polynomial time and hence the slice relations\footnote{see
  Section~\ref{sec:prel}.} of $\hat{L}_1, \hat{L}_2$ are both rankable
in polynomial time.

We can summarize this discussion in the following
\begin{pro}
  The class of $p$-relations that are rankable in polynomial time is
  not close under $p$-union, hence it is a proper subclass of the
  class of $p$-relations admitting a polynomial time \urg (unless
  $\cP=\cP^{\cnP}$).
\end{pro}


\RCSfooter$Id: c5s3.tex,v 2.3 1999/09/23 14:43:06 santini Exp $

\section{The $p$-Complement}
\label{sec:complement}

Now we consider the closure with respect to the complement of some of
the classes of $p$-relations. For this reason, we suggest the
following
\begin{dfn}
  Given a $p$-relation $R\subseteq\reldom$, its
  $p$-complement $R^{pc}$ is the $p$-relation
  $R^{pc}\subseteq\reldom$ defined as $\alpha \rR^{pc}
  \beta$ iff not $\alpha \rR \beta$ and $|\beta|=p(|\alpha|)$.
\end{dfn}

Now, we show that there exists a $p$-relation $\hat{R}$ such that it
admits a polynomial time \urg, but the same does not hold for its
$p$-complement. Let $\hat{R}=(G,U)$ be the $p$-relation such that
$U\subseteq V$ is an \emph{independent set} of the graph $G=\langle V,
E\rangle$, \ie for every pair of nodes $u,v\in U$ the edge $(u,v)$
does not belong to $E$; if $V=\{v_1,\ldots,v_n\}$, assume to represent
$U$ over $\{0,1\}^*$ by a string $u_1\ldots u_n$ such that $u_i$ is
$1$ iff $v_i\in U$, for $1\leq i\leq n$.

It is then clear that $\hat{R}^{pc}$ is the relation $(G,U)$ where
either $G$ is not (a correct encoding of) a graph and $U$ is any
string of length $p(|G|)$, or $G=\langle V, E\rangle$ is a graph and
$U\subseteq V$ is such that $(u,v)\in E$ for some $u,v\in U$.

As proved by \citet[Theorem~1.17]{Sin88}, if $\hat{R}$ admits a
polynomial time \urg, then $\cNP=\cRP$.  However, using the algorithm
for generating uniformly at random the satisfying assignment to a
boolean formula in disjunctive normal form (DNF) given by
\citet{JVV86}, we can prove the following
\begin{lem}
  There exists a polynomial time \urg for $\hat{R}^{pc}$.
\end{lem}
\begin{proof}
  Let $G$ be an element of the domain of $R$. If $G$ is not a graph,
  it is easy to generate uniformly at random an element of
  $\{0,1\}^{p(|G|)}$. On the other hand, let $G=\langle V, E\rangle$
  be a graph and consider the DNF
  \[
  \bigvee_{(v,v')\in E} x_v \wedge x_{v'}
  \]
  in the variables $X_G=\{x_v : v\in V\}$. It is easy to verify that
  every truth assignment $\tau:X_G\to \text{true},\text{false}\}$
  satisfies the DNF iff $U=\{ v : \tau(x_v)=\text{true}\}$ is such
  that $(G,U)\in \hat{R}^{pc}$. Hence, the result follows by applying
  the \urg for the DNF given by \citet{JVV86}.
\end{proof}

\medskip

On the other hand, by observing that
$r_{R^{pc}}(\alpha,\beta)=\#\{\gamma : \gamma\lex
\beta\}-r_R(\alpha,\beta)$, one can immediately conclude the following
\begin{pro}
  The class of $p$-relation that are rankable in polynomial time is
  closed under $p$-complement, hence it is a proper subclass of the
  class of $p$-relations admitting a polynomial time \urg (unless
  $\cNP=\cRP$).
\end{pro}

The situation discussed in these last two section it then illustrated
by Figure~\ref{fig:class}, where the endpoints of the dotted arrows are
as shown (from top to bottom) if $\cNP\not=\cRP$ and
$\cP\not=\cP^{\cnP}$.

\bigskip 

\begin{figure}[ht]
  \begin{center}
    \input{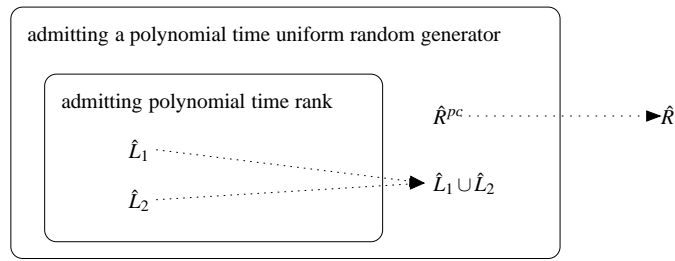}
    \caption{Some classes of $p$-relations.}
    \label{fig:class}
  \end{center}
\end{figure}



\RCSfooter$Id: c6.tex,v 2.5 1999/11/01 11:30:53 santini Exp $

\chapter{Applications to Combinatorial Optimization}
\label{cha:combopt}

\mycite{spanish}{%
  \begin{verse}
    Hay que cansar los números.\\
    Que cuenten sin parar,\\
    que se embriaguen contando,\\
    y que no sepan ya\\
    cuál de ellos será el último:\\
    !`qué vivir sin final!\\
    Que un gran tropel de ceros\\
    asalte nuestras dichas\\
    esbeltas, al pasar,\\
    y las lleve a su cima.\\
    Que se rompan las cifras,\\
    sin poder calcular\\
    ni el tiempo ni los besos.
  \end{verse}
}%
{Pedro Salinas}{La voz a ti debida}

In this chapter we discuss an application of uniform random generation
to \emph{combinatorial optimization}; notice that the research field
of combinatorial optimization is so relevant to the theory and
practice of computer science that part of the contents of this chapter
has been the very inspiration originally motivating this entire work.

After a short general introduction, we focus on the \emph{local
  search} paradigm which is probably the most used technique to solve
combinatorial optimization problems since it combines simplicity and
empirical success in many applications. As studied by \citet{Gro99},
one way to improve and fully exploit the power of local search consist
in the use of uniform random generators for producing an initial
solution; in connection to this fact, it has been possible to define
the class of \emph{expectation-guaranteed} problems: a wide and
natural class whose member admit efficient deterministic approximation
algorithms obtained by \emph{derandomizing} such a process of random
generation.  In the last section of this chapter we show a negative
result concerning this approach by proving that not every problem
admits an efficient derandomization, unless $\cP=\cnP$.

\bigskip
\RCSfooter$Id: c6s1.tex,v 2.4 1999/11/01 11:30:53 santini Exp $

\section{Some Basic Definitions and Properties}
\label{sec:codefs}

This section is intended to be a very short introduction to the basic
definitions and known facts about \idx{combinatorial optimization}; a
complete and exhaustive treatment of this subject if far behind the
scope of this work; for a more comprehensive treatment of this topic,
see for instance \citep{GJ79,BC93,Pap94,Hoc97}.

\subsection{The class \cNPO}
\label{ssec:npo}

We begin by recalling the definition of a particular class of
combinatorial optimization problems which captures the most part of
problems of practical interest; we observe that our definition is
slightly different from the standard one because of our use of
$p$-relations
\begin{dfn}
  A \emph{\cNP optimization} (\cNPO) problem \midx{\irNPO} is a
  $3$-tuple $(R,g,\goal)$ where $R\subseteq\reldom$ is
  a $p$-relation, the \esidx{objective function}{\irNPO}
  $g:\reldom\to \bN$ is a polynomial time
  computable function and $\goal\in\{\min,\max\}$.  Moreover, the
  problem is said to be \esidx{polynomially bounded}{\irNPO} whenever
  the value of $g(\alpha,\beta)$ is bounded by a polynomial in
  $|\alpha|$, for every $\alpha,\beta$.
\end{dfn}
As we have anticipated in Section~\ref{sec:prel}, the $p$-relation $R$
serves both to express, via its domain, the set of
\esidx{instances}{\irNPO} of the problem $\dom(R)=\{\alpha\in\fldom:
\alpha\rR\beta \;\text{for some}\; \beta\in\fldom \}$ and, for each
instance $\alpha\in\dom(R)$, the set of its \esidx{feasible
  solutions}{\irNPO} $\sol(\alpha)= \{\beta\in\fldom :
\alpha\rR\beta\}$. The objective function, also called
\esidx{value}{\irNPO}, or \esidx{measure}{\irNPO} of the solution,
together with the goal of the problem, defines its
\esidx{optimum}{\irNPO} as
\[
\opt(\alpha)=\goal_{\beta\in\sol(\alpha)} g(\alpha,\beta).
\]
Hence, the \emph{solution} of a \cNPO problem consists in finding, for
every instance $\alpha\in\dom(R)$, an \esidx{optimum solution}{\irNPO}
$\tilde{\beta}\in\sol(\alpha)$ such that $g(\alpha,\tilde{\beta})=
\opt(\alpha)$. Finally, recall that with every \cNPO problem it can be
associated its \esidx{decision version}{\irNPO} whose solution is to
decide, for every pair $(\alpha,t)$ with $\alpha\in\dom(R)$ and
$t\in\bQ$, whether $\opt(\alpha)\leq t$ (if $\goal=\min$), or
$\opt(\alpha)\geq t$ (if $\goal=\max$).

\bigskip

Some example of problems in this class are the ``Max Cut'' and ``Max
Clique'' problems\footnote{for a thorough compendium of \cNPO
  problems, see the survey of \citet{CK98}.}. In both cases, the
instances are graphs $G=(V,E)$.  For ``Max Cut'' case, for each graph
$G$, the feasible solutions are partitions of its vertex set $V$ into
disjoint sets $\langle V_1,V_2 \rangle$ and the goal is to maximize
the measure of a partition, defined as the cardinality of the induced
cut, \ie the number of edges with one endpoint in $V_1$ and the other
in $V_2$.  On the other hand, in ``Max Clique'' problem, for each
graph $G$, the feasible solutions are subset of vertices $V'\subseteq
V$ which are all pairwise adjacent, \ie every pair of vertices in $V'$
are joined by an edge in $E$; in this case the goal is to maximize the
measure of such subsets, defined as their cardinality.

\subsection{Approximation algorithms and approximation classes}
\label{ssec:approxalgclass}

It is well known \citep{Kar72,GJ79} that, under the conjecture $\cP
\not=\cNP$, every \cNPO problem whose corresponding decision version
is \cNP-complete is not solvable in polynomial time. As a consequence,
if we are interested in ``efficient'' algorithms, we have to restrict
our attention only to polynomial time computable ``approximate
solutions'' of the problem.  More formally
\begin{dfn}
  Let $(R,g,\goal)$ be a \cNPO problem. Given an instance
  $\alpha\in\dom(R)$ and a feasible solution $\beta\in\sol(\alpha)$,
  the \esidx{performance ratio}{\irNPO} $\cR(\alpha,\beta)$ of $\beta$
  with respect to $\alpha$ is
  \[
  \cR(\alpha,\beta)=
  \max\left\{
    \frac{\opt(\alpha)}{g(\alpha,\beta)},
    \frac{g(\alpha,\beta)}{\opt(\alpha)}\right\}
  \]
  and the \esidx{relative error}{\irNPO} $\cE(\alpha,\beta)$ of $\beta$ with
  respect to $\alpha$ is
  \[
  \cE(\alpha,\beta)=
  \frac{|\opt(\alpha)-g(\alpha,\beta)|}{\max\{\opt(\alpha),g(\alpha,\beta)\}}.
  \]
\end{dfn}
Obviously, the closer to $1$ (respectively $0$) is the performance
ratio (respectively relative error), the better are the solutions.
There is a strict relationship between the performance ratio and the
relative error:
\[
\cR(\alpha,\beta)=\frac 1{1-\cE(\alpha,\beta)}.
\]
The concepts expressed so far allow us to define what an
\emph{approximation algorithm} is
\begin{dfn}
  Given a function $r:\bN\to [1,\infty)$, an algorithm $A$ is
  a $r(n)$-\esidx{approximation algorithm}{\irNPO} for a \cNPO problem
  $(R,g,\goal)$ iff, for every instance $\alpha\in\dom(R)$, it returns
  a feasible solution $A(\alpha)\in\sol(\alpha)$ such that
  \[
  \cR(\alpha,A(\alpha))\leq r(|\alpha|).
  \]
\end{dfn}
Similarly, one can define an \emph{approximation scheme} as follows
\begin{dfn}
  An algorithm $A$ is an \esidx{approximation scheme}{\irNPO} for a
  \cNPO problem $(R,g,\goal)$ iff, for every instance
  $\alpha\in\dom(R)$ and $r\in(1,\infty)$, it returns a feasible
  solution $A(\alpha,r)\in\sol(\alpha)$ such that
  \[
  \cR(\alpha,A(\alpha,r))\leq r.
  \]
  An approximation scheme is said to be \emph{polynomial}
  \msidx{approximation scheme!polynomial}{\irNPO} if it works in time
  polynomial in $|\alpha|$ and \emph{fully polynomial}
  \msidx{approximation scheme!fully polynomial}{\irNPO} if it works in
  time polynomial in $|\alpha|$ and $r/(r-1)$.
\end{dfn}
This notion gives rise in a natural way to various classes of \cNPO
problems according to the existence of approximation algorithms
satisfying specific performance bound. As an example, we just recall
some of such classes we will refer to later on in this chapter.  The
class of \cNPO problems approximable within a constant factor is
defined as follows
\begin{dfn}
  A \cNPO problem belongs to the class \cAPX if, for some $r>1$, it
  exists a polynomial time $r$-approximation algorithm for it.
\end{dfn}
Then we can look for problems admitting efficient approximation schemes
\begin{dfn}
  A \cNPO problem belongs to the class \cPTAS if it admits a
  polynomial time approximation scheme; similarly, it belongs to the
  class \cFPTAS if it admits a fully polynomial time approximation
  scheme.
\end{dfn}
Observe that these classes do not trivially coincide since, for
instance, it is known that every problems in \cNPO whose decision
version is strongly \cNP-complete cannot have a fully polynomial time
approximation scheme \citep{GJ79}. As it is obvious to conclude from
the previous definitions
\begin{pro}
  $\cFPTAS \subseteq \cPTAS \subseteq \cAPX$.  
\end{pro}

\subsection{Logically definable approximation classes}
\label{ssec:maxsnp}

Other classes of \cNPO problems can be described by means of the so
called \esidx{logical definability}{\irNPO} (for a survey, see the
work of \citet{KT94}), a concept inspired by Fagin's characterization
of \cNP in terms of second-order logic on finite structures
\citep{Fag74}.  More precisely, according to such approach, a class of
finite structures is \cNP-computable iff it is definable by an
\emph{existential second-order formula}, \ie an expression of the form
$\exists\bS\phi(\bG,\bS)$, where $\bG$ is a \emph{finite structure},
$\bS$ is a finite sequence of predicate symbols of bounded arity, and
$\phi(\bG,\bS)$ is a first-order formula. It is well known that every
such formula is equivalent to a formula in \emph{prenex normal form},
\ie one of the form $\exists\bS\forall
\bx\exists\by\varphi(\bx,\by,\bG,\bS)$, where $\varphi$ is a
quantifier-free formula and $\bx=(x_1,\dots,x_j)$,
$\by=(y_1,\dots,y_k)$ are finite sequences of first-order variables
(for an introduction to the logic details, see, for instance,
\citep{Bar77}).

According to this framework, \citet{PY91} introduced the class \cMaxNP
of maximization problems whose optimum can be defined as
\[
\max_{\bS}|\{\bx:\exists\by\varphi(\bx,\by,\bG,\bS)\}|,
\]
where $\varphi$ is a quantifier-free first order formula.
Intuitively, in a \cNP decision problem one seeks the predicates $\bS$
that satisft some existential second-order sentence
$\forall\bx\exists\by\varphi(\bx,\by,\bG,\bS)$, while in the
corresponding maximization version in \cMaxNP one seeks those predicates
$\bS$ that maximize the number of tuples $\bx$ satisfying the
existential first-order sentence $\exists\by\varphi(\bx,\by,\bG,\bS)$.

\bigskip

A canonical example in such class is the ``Max Sat''
problem\footnote{For a more formal definition, see \citep{CK98}.}, the
instances of which are collections $C$ of clauses, \ie a disjunction
of literals, where a literal is a variable or a negated variable;
feasible solutions are truth assignments to the variables and the goal
is to maximize the number of clauses satisfied by the truth
assignment. The optimum of such problem can be written as
\[
\max_T |
\{\omega\in C:\exists x 
[(T(x)\wedge P(\omega,x))\vee (\neg T(x)\wedge N(\omega,x))]
\}|
\]
where $T=\{x\in X: \tau(x)=1\}$ is the set of variables which are true
under the truth assignment $\tau:X\to \{0,1\}$, and the binary
predicates $P(\omega,x)$ and $N(\omega,x)$ express the fact that the
literal $x$ appears as positive, respectively negative, in the clause
$\omega$.

\bigskip

The syntactic view has proven useful not only in obtaining structural
complexity results, but also in developing paradigms for designing
efficient approximation algorithms as, for instance, stated by
\citet{PY91}.
\begin{pro}
  For every problem in \cMaxNP there exists a polynomial time
  approximation algorithm whose performance ratio is bounded by a
  constant,therefore
  \[
  \cMaxNP\subseteq\cAPX.
  \]
\end{pro}
The same clearly holds for a subclass of \cMaxNP, called \cMaxSNP
(``S'' stands for \emph{strict}), consisting of those maximization
problems that are defined by quantifier-free formulae, such as
\[
\max_{\bS}|\{\bx:\varphi(\bx,\bG,\bS)\}|,
\]
where $\varphi$ is quantifier-free. The class \cMaxSNP contains several
natural maximization problems such as ``Max $3$-Sat'' (the version of
``Max Sat'' with exactly three literals per clause).

\subsection{Approximation preserving reductions}
\label{ssec:appred}

A basic notion in complexity theory is that of \emph{reducibility}.
Roughly speaking, a reduction between two decision problems allows one
to transforms every algorithm for solving a problem in an algorithm to
solve the other. More formally, if the languages $L,L'\subseteq\fldom$
encode two decision problems, we say that there is a \emph{many-one}
reduction from $L$ to $L'$ if there exists a polynomial time
computable function $f:\fldom\to\fldom$ such that $\omega\in L$ iff
$f(\omega)\in L'$.

Unfortunately, in the context of approximation, these reductions do
not guarantee to preserve the objective function, and in general, the
quality of the solution.  Thus, new \esidx{approximation-preserving
  reductions}{\irNPO} have been introduced, \ie reductions that not
only map instances of one problem into instances of the other, but
that are also able to map back ``good'' solutions into ``paragonably
good'' solutions. As an example of such idea, we only recall the
definition of AP-reduction \citep{CKPT95}, that will be used later in
this chapter.

\clearpage 

\begin{dfn}
  A \cNPO problem $(R,g,\goal)$ is said to be AP-\emph{reducible}
  \msidx{AP-reduction}{\irNPO} to a \cNPO problem $(R',g',\goal')$ iff
  there exist two polynomial time computable functions $f$ and $b$,
  and a constant $\epsilon$ such that, for every $\alpha\in\dom(R)$,
  $r>1$ and $\beta'\in \sol'(f(\alpha,r))$,
  \begin{enumerate}
  \item $f(\alpha,r)\in\dom(R')$;
  \item $b(\alpha,\beta',r)\in\sol(\alpha)$;
  \item $\cR(f(\alpha,r),\beta')\leq r$ implies
    $\cR(\alpha,b(\alpha,\beta',r))\leq 1+\epsilon(r-1)$.
  \end{enumerate}
\end{dfn}


Given a concept of reduction, it is possible to define the concepts of
\emph{hardness}, \emph{completeness} and \emph{closure} for a given
approximation class \cC of \cNPO problems as in the case of decision
problems
\begin{dfn}
  A \cNPO problem is said to be \cC-\esidx{hard}{\irNPO} (under a
  given approximation preserving reduction) if all \cNPO problems in
  \cC can be reduced to it; in addition, the problem is said to be
  \cC-\esidx{complete}{\irNPO} if it also belongs to \cC.  The
  \esidx{closure}{\irNPO} of \cC under such reduction is the class
  $\bar{\cC}$ of all \cNPO problems reducible to some problem in \cC.
\end{dfn}

These ideas are widely studied, for instance by \citet{PY91,KMSV94}.
We observe that one more reason of interest for the study of the class
\cMaxNP is that the first example of a complete problem has been given
for it \citep{KMSV94}.
\begin{pro}
  The \cNPO problem ``Max Sat'' is complete for \cMaxNP.
\end{pro}


\RCSfooter$Id: c6s2.tex,v 2.3 1999/09/23 14:43:06 santini Exp $

\section{Local Search}
\label{sec:localsearch}

In this section we introduce the basic idea of the \emph{local search
  paradigm} with a discussion of some of its strengths and limits.
Then we discuss in which sense randomization can help to improve such
a paradigm. Here, for the sake of brevity, we focus only on
maximization problems; a similar discussion can be easily carried out
for the minimization case.

\subsection{The basic idea}
\label{ssec:lsbasic}

In this section, to simplify the notation, we restrict our attention
to a very ``simple version'' of a combinatorial optimization problem.
Let $X$ be a discrete search space and let $f:X \to \bN$ be the goal
function that must be maximized on a set of feasible solutions
$S\subseteq X$. The \eidx{local search paradigm} is a general scheme
to design approximation algorithms based on stepwise improvement on
the value of the function $f$ by exploring a ``neighborhood'' of the
current solution.

A \eidx{neighborhood structure} is a map $S\to 2^S$ that defines, for
each $x\in S$, a set $S_x\subseteq S$ of solutions that are, in some
sense, ``close'' to $x$ (we assume the map is such that $y\in S_x$ iff
$x\in S_y$). The set $S_x$ is then called the
\esidx{neighborhood}{\irNS} of the solution $x$, and each $y\in S_x$
is called \esidx{neighbor}{\irNS} of $x$.

As an example, if $X=\{0,1\}^n$, one can define, for every $h>0$, the
so called \esidx{$h$-bounded}{\irNS} neighborhood structure as the map
$x\mapsto\{y\in X:d_H(x,y)\leq h\}$, where $d_H:\bX\times\bX
\to \bN$ is the \emph{Hamming distance}\footnote{We recall
  that $d_H(x,y)$ is defined as the number of positions in which $x$
  differs from $y$.}.

Once a neighborhood structure is fixed, an example of the local search
paradigm is given by the schema of Algorithm~\ref{alg:lsex}.

\begin{alg}[ht]
  \caption{Local Search: schema of algorithm.}
  \label{alg:lsex}
  \begin{myprg*}
    \sinput $x_0\in S$ \nl
    $x\stv x_0$ \nl
    \srepeat \tnl
    \emph{pick} $y\in S_x$ \nl
    \sif $f(y)>f(x)$ \sthen $x\stv y$ \unl
    \suntil $f(y)\leq f(x)$ for all $y\in S_x$ \nl
    \soutput $x$.
  \end{myprg*}
\end{alg}

With the term ``schema'' we refer to the fact that various methods can
be used to implement the \emph{pick} statement, selecting one neighbor
of the current solution, such as choosing at random, following some
predefined order, or taking the neighbor of maximum value.

\bigskip

The fundamental concept in the analysis of a local search algorithm is
that of local, as opposed to global, optimality: $x\in S$ is a
\esidx{local maximum}{\irLS} (with respect to a given neighborhood
structure) iff $f(x) \geq f(y)$ for all $y\in S_{x}$ while $x\in S$ is
a \esidx{global maximum}{\irLS} iff $f(x) \geq f(y)$ for all $y\in S$.
It is then clear that one of the main drawbacks of the local search
paradigm is that in general it offers algorithms that terminate in a
local optimum and, unfortunately, in general it is not easy to
estimate how far it is from a global one.

There is a clear trade-off between the quality of the local optima and
the neighborhood structure: the larger are the neighborhoods the
better are the local optima, but it may be harder (slower) to compute
them.  Moreover, the quality of the local optima is also strictly
related to the choice of the \esidx{initial solution}{\irLS} $x_0$;
usually, to reduce the sensibility to it, local search algorithms are
repeated on many different initial solutions (and then the best output
obtained is adopted). Whenever it is hard to come out with a feasible
initial solution, clearly another trade-off happens between the number
of initial solution considered and the quality of the final local
optimum obtained.

Therefore, designing a good local search algorithm has remained up to
now an experimental art, even if some guiding principles and
techniques have been identified. Nonetheless, as experimentally shown,
the local search paradigm usually leads to fast algorithms that find
reasonably good solutions in a reasonable amount of time.  For this
reason, they cover a wide application area ranging from mathematical
problems, to graph problems, to logical problems and so on.

\subsection{Local search and \cNPO problems}
\label{ssec:lsnpo}

Thanks to its generality and its empirical success, the local search
paradigm represents certainly one of the most used approaches in
approximating difficult optimization problems. The complexity of
finding locally optimal solutions for those problems has been mainly
investigated by \citet{JPY88,AP95}, showing the strengths and the
limits of such a paradigm.

In particular, \citet{AP95} introduce the class \cGLO of \cNPO
problems that have \esidx{guaranteed local optima}{\irNPO}: a
maximization problem $(R,g,\max)$ has guaranteed local optima iff
there exists an integer constant $h$ such that the value of all local
optima is at least $1/h$ times the value of a global optimum.  If a
problem has guaranteed local optima, then there exists an $h$-bounded
neighborhood structure such that, for every instance
$\alpha\in\dom(R)$, every local optimum $\beta$ of $\alpha$ has the
property that $\opt(\alpha)\leq h\cdot g(\alpha,\beta)$ (a similar
definition can be given when $\goal=\min$). The fundamental assumption
here is that, given an instance $\alpha$ of a (polynomially bounded)
\cNPO problem, it is possible to determine whether a given solution
$\beta$ of $\alpha$ is a local optimum, and in the negative case, to
find a better solution in its neighborhood in polynomial time.
Various problems\footnote{For a definition of such problems, see
  \citep{CK98}.}  can be shown to belong to \cGLO, such as ``Max
Cut'', ``Max Sat'', ``Max $3$-Sat''.  Moreover, the class \cGLO is a
subclass of \cAPX, but the inclusion is proper as shown for instance
by the problem ``Max $k$-GSat'' which is well known to be approximable
within a constant factor, but that does not have guaranteed local
optima \citep{Ali96}.  A further result is that
$\overline{\cGLO}=\cAPX$, where the closure is taken with respect to
the AP-reduction.


\RCSfooter$Id: c6s3.tex,v 2.4 1999/11/01 11:30:53 santini Exp $

\section{The ``Expectation-Guaranteed'' Class}
\label{sec:ge}

Different techniques have been suggested to overcame some of the
drawbacks of the local search paradigm.  Some of these are, for
instance, based on the modification of the neighborhood structure, or
of the goal function, such as in the \emph{non-oblivious local search}
\citep{KMSV94,Ali96} where a ``non-oblivious'' version of the goal
function is suitably constructed allowing to escape from local optima.

A completely different approach is taken by \citet{Gro99}, which focus
the attention on the choice of a ``good'' initial solution as a
starting point for applying the standard local search. A first and
simple idea is to generate uniformly at random some different feasible
solutions then taking as initial solution that one with the best
value.  Hence, a second step toward a deterministic approximation
algorithm consist in investigating when it is possible to derandomize
such initialization phase.

\subsection{Randomized choice of initial solutions}
\label{ssec:lsrnd}

We start by considering a very simple randomized algorithm. Let
$(R,g,\max)$ be a \cNPO problem, and let $N\geq 1$ be a parameter
whose value will be determined in the following. Then, RadomSearch is
Algorithm~\ref{alg:rsex}.

\begin{alg}[ht]
  \caption{RandomSearch.}
  \label{alg:rsex}
  \begin{myprg*}
    \sinput $\alpha\in\dom(R)$ \nl
    \sfor $i=1,\ldots, N$ \sdo \tnl
    \emph{generate} $\beta\in\sol(\alpha)$ \emph{uniformly at random} \nl
    $b_i\stv \beta$ \unl
    $\tilde{\beta}\stv \argmax_{i\in\{1,\ldots, N\}} g(\alpha,b_i)$ \nl
    \soutput $\tilde{\beta}$.
  \end{myprg*}
\end{alg}

Since the focus of this section is on the approximation performance
aspects of combinatorial optimization, the statement ``\emph{generate}
$\beta\in\sol(\alpha)$ \emph{uniformly at random}'' will be simply
understood as a call to a subroutine which returns, on every call, an
(independent and uniformly distributed) element $\beta$ of
$\sol(\alpha)$.  A discussion of the details concerning such
subroutine, in the spirit of what is studied in the other chapters of
this work, is deferred to the end of this section.

\medskip

A very simple analysis shows that this algorithm finds a solution
whose value is ``close'' to the expected value of $g(\alpha,\beta)$
(when $\beta$ is uniformly distributed on $\sol(\alpha)$) with high
confidence. More formally, define
\[
\Ex[g(\alpha,\cdot)] = 
\frac 1{\#\{ \beta: \beta\in\sol(\alpha)\}} 
\sum_{\beta\in\sol(\alpha)} g(\alpha,\beta);
\]
then, by an application of Hoeffding's inequality (see
Appendix~\ref{app:tech}), one can easily obtain the following
\begin{pro}
  For every $0<\delta< 1$ and $\epsilon>0$, if $N\geq 1/2\epsilon^2
  \cdot \log (1/\delta)$, the RandomSearch algorithm, on input
  $\alpha\in\dom(R)$, gives in output $\beta$ such that
  $g(\alpha,\beta) \geq \Ex[g(\alpha,\cdot)]-
  \epsilon\cdot\opt(\alpha)$, with probability at least $1-\delta$.
\end{pro}

Hence, if we restrict our attention to the class of problems for which
the expectation of the goal function is a reasonable approximation of
the optimum, the RandomSearch algorithm could be used as a ``good''
randomized approximation algorithm. We restrict again our attention to
the case $\goal=\max$; \citet{Gro99} introduces the following class
\begin{dfn} 
  A \cNPO problem $(R,g,\max)$ is \emph{RS-good} iff there exists a
  real constant $\rho>0$ such that, for every instance $\alpha\in
  \dom(R)$, it holds $\Ex[g(\alpha,\cdot)]\geq\rho\cdot\opt(\alpha)$.
\end{dfn}
In view of the analysis of the RandomSearch algorithm, we can conclude
the following
\begin{pro}
  If a given \cNPO problem $(R,g,\max)$ is RS-good, then, for every
  $0<\delta< 1$ and $\epsilon>0$, there exists a randomized
  approximation algorithm such that, on input $\alpha\in\dom(R)$,
  returns a solution $\beta\in\sol(\alpha)$ such that, with
  probability at least $1-\delta$,
  \[
  g(\alpha,\beta)\geq(\rho-\epsilon)\opt(\alpha),
  \]
  where $\rho>0$ is some fixed constant independent of $\alpha$. 
\end{pro}
\enlargethispage{3em} 
Observe that similar definitions and results can be given for the case
$\goal=\min$.

\clearpage 

Two solutions arise naturally. On one side, we can actually implement
a uniform random generator for the $p$-relation $R$ characterizing
the \cNPO problem which is RS-good, hence obtaining a randomized
approximation algorithm with a constant performance ratio. One of the
motivation of this thesis has been to investigate which classes of
relations (or combinatorial structures) admit efficient \urg in order
to define classes of \cNPO problem for which the approach taken in
this section can lead to efficient randomized approximation algorithm
with a constant performance ratio. Thanks to the result of
Chapter~\ref{cha:ranking} and \ref{cha:ambparadigm} (in particular,
from Corollary~\ref{cor:prelpdaurg} and~\ref{cor:prelsbtmurg}), one
can hence conclude,
\begin{cor}
  \label{cor:combopt}
  Let $(R,g,\goal)$ be a \cNPO problem that is RS-good and defined in
  terms of a $p$-relation that either
  \begin{enumerate}[(1)]
  \item belongs to the class $\cB(s(n),i(n),d(n))$ of $p$-relations
    recognized by extended Turing machines with simultaneous
    complexity bound $s(n)\cdot i(n)\cdot d(n)=O(\log n)$, or
  \item is recognized by a polynomial time extended one-way
    nondeterministic auxiliary pushdown automaton with polynomial
    ambiguity.
  \end{enumerate}
  Then, $(R,g,\goal)$ admits a polynomial time randomized
  approximation algorithm with a constant performance ratio.
\end{cor}

\medskip

On the other hand, in the next section, we see that for some \cNPO
problems, under suitable hypotheses, it is possible to
deterministically obtain a solution whose measure is as good as the
expected value for it. For those problems, as discussed in the
following, it will be possible to have (deterministic) approximation
algorithm with a constant performance ratio.

\subsection{Derandomization: the \cEG class}
\label{ssec:derand}

It is well known that, for every random variable $X:\Omega\mapsto
\bR$, it always holds $\Ex[X]\leq \sup_{\omega\in\Omega} X(\omega)$:
this is the so called \emph{internality} property of the expectation.
We want to exploit this fact in relation to the analysis of the
RandomSearch algorithm, considering the case in which it is possible
to efficiently (and deterministically) compute a $\beta$ for which
$g(\alpha,\beta)$ is greater than or equal to the expectation. More
formally and restricting again our attention to maximization problems,
\citet{Gro99}, introduces the following class
\begin{dfn}
  A \cNPO problem $(R,g,\max)$ is \emph{RS-derandomizable} iff there
  exists a polynomial time algorithm that, having in input an instance
  $\alpha\in\dom(R)$, outputs a solution $\beta\in\sol(\alpha)$ such
  that $g(\alpha,\beta)\geq\Ex[g(\alpha,\cdot)]$.
\end{dfn}

Then, by joining this property with the RS-good one, \citet{Gro99}
defines a class of problem for which there exists a good deterministic
approximation algorithm
\begin{dfn}
  The \emph{Expectation-Guaranteed}
  \msidx{expectation-guaranteed}{\irNPO}
  \mridx{expectation-guaranteed}{\irNPO} class (\cEG) is the class of
  \cNPO problems that are both RS-good and RS-derandomizable.
\end{dfn}

It is straightforward to see that, if $(R,g,\max)$ is in \cEG, there
exist a constant $\rho>0$ and a polynomial time algorithm \sA such
that, for every $\alpha\in\dom(R)$,
\[
g(\alpha,\sA(\alpha)) \geq\Ex[g(\alpha,\cdot)] \geq \rho\cdot\opt(\alpha)
\]
where the first inequality follows since $(R,g,\max)$ is
RS-derandomizable and the second since it is RS-good. Hence, since
similar definitions and results can be given for the case
$\goal=\max$, we have the following
\begin{pro}
  $\cEG\subseteq\cAPX$.
\end{pro}

As proved by \citet{Gro99}, several natural \cNPO
problems\footnote{For a definition of such problems, see
  \citep{CK98}.}  belong to this class,
\begin{pro}
  The \cNPO problems: ``Max Sat'', ``Max $k$-CSP'', ``Max $k$-Cut'',
  ``Max Bisection'', ``Min TSP-$(1,2)$'' and ``Max TSP-$(1,2)$'' all
  belong to \cEG.
\end{pro}

In particular, as a consequence of the fact that every problem in
\cMaxSNP can be viewed (for a suitable $k$) as a ``Max $k$-CSP''
problem \citep{KMSV94}, and by the previously recalled proposition,
one can conclude that
\begin{pro}
  $\cMaxSNP\subseteq\cEG$.
\end{pro}

To conclude this subsection, we recall that \citet{Gro99} has also
proven the following
\begin{pro}
  The inclusions $\cMaxSNP\subsetneq\cEG\subsetneq\cAPX$ are strict.
  Moreover, $\overline{\cEG}=APX$, where the closure is taken with
  respect to the AP-reduction.
\end{pro}

\subsection{The method of conditional expectation}
\label{ssec:condexp}

In this subsection we discuss the \eidx{method of conditional
  expectation} \citep{ES74}: a particular technique which allows to
show that some \cNPO problem, under suitable hypotheses, is
RS-derandomizable.

Let $(R,g,\max)$ be a \cNPO problem over some alphabet $\Sigma$, where
$R$ is a $p$-relation.  For every instance $\alpha\in\dom(R)$ and
$1\leq k\leq p(|\alpha|)$, define the subset of feasible solutions
$\sol(\alpha;b_1,\ldots,b_k)=\{ \beta_1\ldots\beta_{p(|\alpha|)}\in
\sol(\alpha) : \beta_j=b_j \;\text{for}\; 1\leq j\leq k\}$ having
prefix $b_1,\ldots,b_k\in\Sigma$. Then, the expectation of
$g(\alpha,\beta)$ when $\beta$ is uniformly distributed on
$\sol(\alpha)$, conditionally to $\beta$ having prefix
$b_1,\ldots,b_k\in\Sigma$, can be defined as
\[
\Ex[g(\alpha,\cdot) \mid b_1,\ldots, b_k ] =
\frac 1{\#\{ \beta: \beta\in\sol(\alpha; b_1,\ldots,b_k)\}} 
\sum_{\beta\in\sol(\alpha;b_1,\ldots,b_k)} g(\alpha,\beta).
\]

By computing such conditional expectation on prefixes of increasing
length, one can incrementally compute the $\beta$ for which
$g(\alpha,\beta)$ is greater than or equal to $\Ex[g(\alpha,\cdot)]$,
selecting for each ``position'' of the prefix the best possible value.
More formally, consider Algorithm~\ref{alg:ceex}.

\begin{alg}[ht]
  \caption{Conditional expectation.}
  \label{alg:ceex}
  \begin{myprg*}
    \sinput $\alpha\in\dom(R)$ \nl
    \sfor $k=1,\ldots, p(|\alpha|)$ \sdo \tnl
    $a_k\stv \argmax_{b\in\Sigma} \Ex[g(\alpha,\cdot) \mid a_1,\ldots, a_{k-1}, b]$ \unl
    $\beta\stv a_1\ldots a_{p(|\alpha|)}$ \nl
    \soutput $\beta$.
  \end{myprg*}
\end{alg}

By the properties of conditional expectation\footnote{for a detailed
  discussion of the probabilistic aspects of such method, see
  Appendix~\ref{assec:condexp}.}, one can verify that, for every
$b_1,\ldots,b_k\in\Sigma$,
\[
\Ex[g(\alpha,\cdot) \mid b_1,\ldots, b_k ]
\leq \max_{b\in\Sigma}
\Ex[g(\alpha,\cdot) \mid  b_1,\ldots, b_k, b];
\]
hence, a simple analysis shows that, if $\beta$ is the output of the
previous algorithm on input $\alpha$, then
$g(\alpha,\beta)\geq\Ex[g(\alpha,\cdot)]$.

Let now $T(n,k)$ be the maximum time required, by a suitable
algorithm, to compute $\Ex[g(\alpha,\cdot) \mid b_1,\ldots, b_k]$ for
all $\alpha\in\dom(R)$ such that $|\alpha|=n$. Then, since the
$\argmax$ is computed over the alphabet $\Sigma$ that is finite by
definition, the computation time of the above algorithm is
$O(\sum_{k=1}^{p(n)} T(n,k))$. Hence, we can conclude with the
following sufficient condition for establishing whether a \cNPO
problem is RS-derandomizable \citep{Gro99}
\begin{pro}
  If a \cNPO problem $(R,g,\goal)$ is such that $\Ex[g(\alpha,\cdot)
  \mid b_1,\ldots, b_k]$ is computable in time polynomial in
  $|\alpha|$ and $k$, then $(R,g,\goal)$ is RS-derandomizable.
\end{pro}

\subsubsection{When all solutions are feasible}

Now we investigate, in the particular case when all solutions are
feasible, a sufficient condition for establishing that
$\Ex[g(\alpha,\cdot) \mid b_1,\ldots, b_k ]$ is computable in
polynomial time. Assume for simplicity\footnote{Observe that this is
  not a limitation, since one can always encode over $\{0,1\}$ every
  finite alphabet $\Sigma$.} $\Sigma=\{0,1\}$ and, as already stated,
consider a \cNPO problem $(R,g,\goal)$ such that $\sol(\alpha)=
\Sigma^{p(|\alpha|)}$, for all $\alpha\in\dom(R)$ (where $p$ is the
polynomial for which $R$ is a $p$-relation).  Fix now some
$\alpha\in\dom(R)$ and let $n=p(|\alpha|)$; the goal function $g$ can
be written as
\begin{equation}
  \label{eq:Galpha}
  g(\alpha, \beta_1\ldots\beta_n) =
  \sum_{(\lambda_1,\ldots,\lambda_n)\in\{0,1\}^n}
  \hat{g}(\alpha, \lambda_1, \ldots, \lambda_n ) 
  \beta_1^{\lambda_1}\cdots\beta_n^{\lambda_n} =
  G_\alpha( \beta_1,\ldots,\beta_n)
\end{equation}
where $G_\alpha( \beta_1,\ldots,\beta_n)$ is a $n$-th variate
polynomial over $\{0,1\}^n$ whose variables have degree at most one
(and $\hat{g}(\alpha, \lambda_1, \ldots, \lambda_n )$ are considered
as coefficient in $\bN$).

In the next section a more detailed discussion on polynomials and on
their relation with pseudo-boolean functions will be carried on. Here,
instead, we only want to sketch how this simple idea can lead to
compute $\Ex[g(\alpha,\cdot) \mid b_1,\ldots, b_k ]$. First of all,
recall that the expectation is a linear functional, such a relevant
property that Erd\"os ironically used to call it ``Adam's theorem''!
Moreover, since the expectation is taken with respect to the uniform
distribution on $\beta\in\Sigma^n$, we can actually consider the
$\beta_i$ as ``independent''. From this very simple observations, it
follows that
\begin{equation*}
  \begin{split}
    \Ex[g(\alpha,\cdot) \mid b_1,\ldots, b_k ] 
    &= \Ex[ G_\alpha( \beta_1,\ldots,\beta_n) \mid b_1,\ldots, b_k ] \\
    &= \sum_{(\lambda_1,\ldots,\lambda_n)\in\{0,1\}^n} 
       g(\alpha, \lambda_1\ldots\lambda_n ) 
       \Ex[ \beta_1^{\lambda_1}\cdots\beta_n^{\lambda_n} \mid b_1,\ldots, b_k ] \\
    &= \sum_{(\lambda_1,\ldots,\lambda_n)\in\{0,1\}^n}
       g(\alpha, \lambda_1\ldots\lambda_n ) 
       b_1^{\lambda_1}\cdots b_k^{\lambda_k}
       \prod_{j=k+1}^n \Ex[ \beta_j^{\lambda_j} ] \\
    &= \sum_{(\lambda_1,\ldots,\lambda_n)\in\{0,1\}^n}
       g(\alpha, \lambda_1\ldots\lambda_n ) 
       b_1^{\lambda_1}\cdots b_k^{\lambda_k} \left(\frac 12\right)^{n-k} \\
    &= G_\alpha( b_1,\ldots, b_k, 1/2, \ldots, 1/2 ).
  \end{split}
\end{equation*}

\bigskip

We can summarize this discussion by the following \citep{Gro99}
\begin{pro}
  \label{pro:galpha}
  Let $(R,g,\goal)$ be a \cNPO problem on $\Sigma=\{0,1\}$ such that
  $\sol(\alpha)=\Sigma^{p(|\alpha|)}$ for every $\alpha\in\dom(R)$ and
  let $G_\alpha$ be defined as in equation~(\ref{eq:Galpha}). If there
  exists an algorithm that, for every $\alpha\in\dom(R)$, evaluates
  $G_\alpha$ on $\{0,1,1/2\}^{p(|\alpha|)}$ in time polynomial in
  $|\alpha|$, then $(R,g,\goal)$ is RS-derandomizable.
\end{pro}

\subsection{Derandomization and the \cEG class}
\label{ssec:drndeg}

As suggested by \citet{Gro99}, if a \cNPO problem is in \cEG, then by
combining derandomization and local search, one can obtain an
efficient approximation algorithm. As an example, assuming the
condition of Proposition~\ref{pro:galpha}, for a given neighborhood
structure $\beta\mapsto S_\beta$ (see Section~\ref{ssec:lsbasic}), we
can have Algorithm~\ref{alg:egex}.

\begin{alg}[ht]
  \caption{An approximation algorithm for \cEG problems.}
  \label{alg:egex}
  \begin{myprg*}
    \sinput $\alpha\in\dom(R)$ \\ \\
    \emph{Step $1$}\nl
    \sfor $k=1,\ldots, p(|\alpha|)$ \sdo \tnl
    $a_k\stv \argmax_{b\in\{0,1\}} G_\alpha( a_1, \ldots, a_{k-1}, b, 1/2, \ldots, 1/2)$ \unl
    $\beta\stv a_1\ldots a_{p(|\alpha|)}$ \\ \\
    \emph{Step $2$}\nl
    \srepeat \tnl
    \emph{pick} $\beta'\in S_\beta$ \nl
    \sif $g(\alpha,\beta')>g(\alpha,\beta)$ \sthen $\beta\stv \beta'$ \unl 
    \suntil $f(\beta')\leq f(\beta)$ for all $\beta'\in S_\beta$\\ \nl
    \soutput $\beta$.
  \end{myprg*}
\end{alg}

Following this idea, but in a more general fashion, \citet{Gro99}
proved the existence of a ``universal'' schema of algorithm for all
the problems in \cEG; more formally
\begin{pro}
  For every \cNPO problem $(R,g,\goal)$ in \cEG, (with respect to a
  suitable $h$-bounded neighborhood) there exists a
  $\rho$-approximation algorithm for $(R,g,\goal)$ where
  \[
  \rho=\sup_{\alpha\in\dom(R)} \frac{\opt(\alpha)}{\Ex[g(\alpha,\cdot)]}.
  \]
\end{pro}
This schema of algorithm is based on the computation of
$\Ex[g(\alpha,\cdot) \mid b_1,\ldots, b_k ]$, hence, in order to
obtain polynomial time algorithms following such a schema,
$\Ex[g(\alpha,\cdot) \mid b_1,\ldots, b_k ]$ should be computable in
polynomial time.

\bigskip

It is then clear that, in order to apply the derandomized
initialization phase to local search algorithm as discussed in this
section, it is of fundamental importance to determine whether
$\Ex[g(\alpha,\cdot) \mid b_1,\ldots, b_k ]$, or $G_\alpha( a_1,
\ldots, a_{k-1}, b, 1/2, \ldots, 1/2)$ (as defined by
equation~(\ref{eq:Galpha})) are computable in polynomial time. This is
exactly the aim of the next section.


\RCSfooter$Id: c6s4.tex,v 2.5 1999/11/01 11:30:53 santini Exp $

\section{Derandomization and Arithmetic Circuits}
\label{sec:negres}

With respect to the results about derandomization discussed in
Subsection~\ref{ssec:condexp}, here we investigate, for a given \cNPO
problem $(R,g,\goal)$ over $\Sigma=\{0,1\}$, whenever it is possible
to design algorithms to efficiently evaluate $G_\alpha$ having as
input some specification of $g$ (and $\alpha$). To this end, we use
the notion of \emph{arithmetic circuit} as a way to specify
pseudo-boolean functions (such as $G_\alpha$ and $g$). The semantic of
arithmetic circuits, and some relevant properties about them, are
expressed using polynomials, which are here formally defined.

\subsection{Polynomials and arithmetic circuits}
\label{ssec:pbdef}

\subsubsection{Polynomials}

We begin by recalling some elementary facts about polynomials; for a
thorough survey see, for example, \citep{Lan94}. Given a ring $R$ we
denote with $\pRxn$ the (commutative) $R$-algebra free over $\{\lin
xn\}$ of \emph{polynomials}\midx{\irP} in the \emph{indeterminates}
\msidx{indeterminate}{\irP} $\lin xn$.  The \esidx{degree}{\irP}
$\dop(p)$ of a polynomial $p\in\pRxn$ is the maximum degree of its
monomials which, in turn, is simply the sum of the exponents of the
variables appearing in the monomial.  Given a $R$-algebra $\cA$ and
$\ba\in\cA^n$ we denote with $p(\ba)\in\cA$ the
\esidx{evaluation}{\irP} of $p$ in $\ba$ obtained by ``replacing''
each $x_i$ with $a_i$ in $p$; if $\cA=\pRn Rym$ and $\lin qn\in\pRn
Rym$ we denote with $p[\lin qn]\in\pRn Rym$ the
\esidx{substitution}{\irP} of $\lin xn$ with $\lin qn$.  We now give
some definitions needed in the following:
\begin{dfn}
  Given a polynomial $p\in\pRxn$ and a monomial $m$ we denote by
  $[m](p)$ the coefficient of $m$ in $p$ ($0$ if $m$ is not in $p$)
  and with $\fL(p)$ the polynomial obtained by substituting every
  $x_i^k$ in $p$ with $x_i$, for each $1\leq i\leq n$ and $k>0$.  We
  denote by $\sfRxn$ the set of polynomials in $\pRxn$ whose monomials
  are square-free.
\end{dfn}
It is well known that, in some cases, a polynomial can be uniquely
determined given a suitable number of its evaluations; in particular,
we prove the following:
\begin{lem}
  \label{lem:uniq}
  Any $p\in\sfRxn$ is uniquely determined by its evaluations over
  $\bool^n$.
\end{lem}
\begin{proof}
  Let us introduce some notation to allow a more concise proof.  For
  $1\leq i\leq n$, define $\be^i=(\lin {e^i}n)\in\bool^n$ such that
  $e^i_j=1$ iff $i=j$; for every $I\subseteq\{1,\ldots, n\}$, let
  $\be^I= \sum_{i\in I}\be^i\in \bool^n$ and define $\be^\emptyset=
  (0,\ldots,0)$. Finally, denote by $\bx^I$ the monomial $\prod_{i\in
    I}x_i $, where $\bx^\emptyset$ denotes the unit of $\pRxn$. It
  is straightforward to verify that $p(\be^I)=\sum_{J\subseteq
    I}[\bx^J]p= [\bx^I]p+\sum_{J\varsubsetneq I}[\bx^J]p$, that is
  $[\bx^I]p= p(\be^I)-\sum_{J\varsubsetneq I}[\bx^J]p$. Hence, by
  induction on the cardinality of $I$, is possible to verify that the
  coefficient of every monomial $\bx^I$ of $p$ is uniquely determined by
  evaluating $p(\be^J)$ for $J\varsubsetneq I$.
\end{proof}

\subsubsection{Arithmetic circuits}

We observe that various
definitions of arithmetic circuits are present in literature (see, for
instance, the works of \citet{vzGS91} or \citet{VSBR83}) and that the
differences between those definitions mainly concern the in-degree or
out-degree.
\begin{dfn}
  An \eidx{arithmetic circuit} over the ring $R$ with
  \esidx{inputs}{\irAC} $\lin xn$ and \esidx{constants}{\irAC} from
  $S\subseteq R$ is a node-labelled direct acyclic graph such that:
  nodes with in-degree $0$ are labeled with elements of $S\cup\{x_1,
  \ldots, x_n\}$, nodes with positive in-degree are labeled with
  elements of $\{+,\times\}$ and there is a distinguished node, the
  only one with out-degree $0$, called \esidx{output}{\irAC} node.
  The set of all arithmetic circuits over the ring $R$ with inputs
  $\lin xn$ and constants from $S$ will be denoted by $\cC^S_R[\lin
  xn]$. The \esidx{size}{\irAC} $\size(c)$ of a circuit $c$ is the
  number of nodes and its \esidx{depth}{\irAC} $\depth(c)$ is the
  length of the longest path ending in the output node of $c$.
\end{dfn}
  
With every node $v$ of an arithmetic circuit over $R$ we recursively
associate a polynomial $p\in\pRxn$ according to its label $\ell$:
\begin{itemize}
\item if $\ell\in S\cup\{x_1, \ldots, x_n\}$, then $p=\ell$;
\item if $\ell\in\{+,\times\}$, then $p= p_1 \,\ell\, p_2 \ldots
  \ell\, p_k$ where $p_1, \ldots, p_k$ are the polynomials
  associated to the predecessors of $v$.
\end{itemize}
We also denote with $c$ the polynomial associated to the output node
of a circuit $c$, so by degree (maximum degree) of a circuit $c$ we
mean the degree (respectively, maximum degree) of the polynomial $c$.

Since the main statements of this section are concerned with circuits
of size polynomial in the number of inputs, it can be easily verified
that even when restricting the in-degree and out-degree to $2$ in the
previous definition, none of the presented results would be
substantially altered.

\subsubsection{Circuits representing polynomials}

In the following we will restrict our attention to the class of
circuits $\cC^{\{-1,0,1\}}_\bZ[\lin xn]$ which we will simply denote
with $\cC_n$, using $\cC^{\fL}_n$ to denote the subset of $\cC_n$ whose
circuits have their associated polynomial in $\sfbZxn$.  These
circuits are interesting since they can be effectively encoded over a
finite alphabet (in size $\size(c)^{O(1)}$ if the alphabet is at least
binary) and then some computations over them can be performed by means
of algorithms.  We give now an example of such a computation:
\begin{lem}
  \label{lem:coeff}
  Given inputs $c\in\cC^{\fL}_n$ and a monomial $m$, $[m]c\in \bZ$ can
  be computed in $O(\size(c)^2d\log d)$ time, where $d=
  \min\{\dop(m),n\}$.
\end{lem}
\begin{proof}
  If $m$ is not square-free, then $[m]c=0$. Otherwise $d(m)\leq n$ and
  $m$ can be written as $x_1^{\lambda_1} \cdots x_n^{\lambda_n}$ with
  $\lin \lambda{n}\in\bool$: define $p = c[\lambda_1z, \ldots,
  \lambda_nz] \in\bZ[z]$. It is straightforward to observe that
  $d(p)=d$ and $[z^d]p=[m]c$. The coefficients of $p$ are in $\bZ$ and
  can be computed using the (encoding of) circuit $c$: the computation
  requires, for each of the $\size(c)$ nodes of $c$, the sum or
  product of the polynomials associated to its predecessors that are
  at most $\size(c)$. Since all the involved polynomials are in
  $\bZ[z]$, their sum or product requires at most $d\log d$ steps;
  given $p$ it is then immediate to get the coefficient of $z^d$,
  hence the coefficient of $m$ in $c$.
\end{proof}
In the following, as we have done in this lemma with a small abuse of
notation, we will identify a circuit and its encoding. We are
interested in circuits of $\cC_n$ ``representing'' polynomials; more
precisely:
\begin{dfn}
  A circuit $c\in\cC^S_R[\lin xn]$ \emph{represents}
  \msidx{representing a polynomial}{\irAC} $p\in\pRxn$ over some
  subset $A$ of a $R$-algebra whenever $c(\ba)=p(\ba)$ for all $\ba\in
  A^n$.  A circuit $c\in\cC^S_R[\lin xn]$ \emph{exactly represents}
  \msidx{representing a polynomial!exactly}{\irAC} $p\in\pRxn$
  whenever $c=p$ (as polynomials).
\end{dfn}

Some interesting facts are known about the representation of
polynomials by means of arithmetic circuits in $\cC_n$ and about the
relationships between circuits representing the same polynomial:
\begin{pro}
  For every $p\in\pbZxn$ there exists $c\in\cC_n$ exactly representing
  $p$ (and thus representing $\fL(p)$ over $\bool$) and
  $c'\in\cC^{\fL}_n$ exactly representing $\fL(p)$.
\end{pro}
It is clear that each polynomial $p$ (or $\fL(p)$) immediately
``suggests'' a circuit; moreover, noting that $x^k=x$ when $x\in\bool$
(for $k\geq 1$), it is also evident that if $c$ exactly represents
$p$, then $c$ also represents $\fL(p)$ over $\bool$.  Since the number
of monomials in $n$ indeterminates of maximum degree at most $1$ is
$2^n$, the size of the circuits na\"ively obtained from $\fL(p)$ can,
in principle, be exponential in $n$. By a simple application of
previous observations and of Lemma~\ref{lem:uniq} it is simple to
notice that:
\begin{pro}
  \label{pro:uniq}
  If $c\in\cC^{\fL}_n$ represents $p\in\pbZxn$ over $\bool$, then $c$
  exactly represents $\fL(p)$.
\end{pro}

\subsection{Back to our problem}
\label{ssec:coperm}

Recall that our aim was to investigate when it is possible to design
algorithms to efficiently evaluate $G_\alpha$ having as input some
specification of $g$ (and $\alpha$), for some \cNPO problem
$(R,g,\goal)$ where $R$ is a $p$-relation.

We have chosen arithmetic circuits as a way to specify pseudo-boolean
functions: by this we mean that $g$ is specified, for every
$\alpha\in\dom(R)$, by a family of circuits $\{c_n\}_{n>0}$ with
$c_n\in \cC_n$ such that, for every $\beta\in\Sigma^{m}$ with
$m=p(|\alpha|)$, one has that $g(\alpha,\beta)=c_m(\beta_1, \ldots,
\beta_m)$. It is also clear from the definition of $G_\alpha$ that
this is a polynomial in $\sfbZxn$, hence it should be specified by a
family of circuits $\{c'_n\}_{n>0}$ with $c'_n\in \cC^{\fL}_n$ such
that $G_\alpha(\beta_1,\ldots, \beta_m)=c'_m(\beta_1,\ldots,
\beta_m)$.

Hence, we are interested in the relationship between circuits in
$\cC_n$ and $\cC^{\fL}_n$; in particular we would like to find
efficient (\ie polynomial-time) algorithms to transform circuits from
one class to the other somehow preserving their evaluation. Some
similar results are known: for instance \citet{VSBR83} have proven
\begin{pro}
  For every $c\in\cC_n$ a circuit $c'\in\cC_n$ can be computed in time
  $\size(c)^{O(1)}$ such that $c'$ exactly represents $c$ and has size
  $O(\size(c)^3)$ and depth $O(\log\size(c)\log\dop(c))$.
\end{pro}

\bigskip

We now introduce a polynomial that is central to the rest of this
note:
\begin{dfn}
  Let $A=(a_{ij})$ be a $n\times n$ matrix with values in $\bool$ and
  consider the function $f_A(\bx)=\prod_{i=1}^n \sum_{j=1}^n
  a_{ij}x_j$; let $p_A \in\pbZxn$ be the polynomial whose evaluation
  coincides with $f_A$ over $\bZ^n$.
\end{dfn}
The polynomial $p_A$ has some interesting properties investigated in
the next: \midx{permanent}
\begin{lem}
  \label{lem:rapp}
  $\perm(A)= [x_1\cdots x_n]\fL(p_A)$ and, for every $s\in\bN$ such
  that $\perm(A)<2^s$, $\perm(A)= 2^s\{2^{s(n-1)}\fL(p_A)(2^{-s},
  \ldots ,2^{-s})\}$ (where $\{x\}$ is the fractional part of
  $x\in\bR$).
\end{lem}
\begin{proof}
  Since $p_A=\sum a_{1i_1}\cdots a_{ni_n} x_{i_1}\cdots x_{i_n}$,
  where the summation extends over all the $n$-tuples $(i_1, \ldots,
  i_n)\in\{1, \ldots, n\}^n$, it is easy to conclude that $[x_1\cdots
  x_n]p_A= \perm(A)$, given that the monomial $x_1\cdots x_n$
  ``selects'' from the summation only those terms corresponding to
  tuples that are a permutation of $\{1, \ldots, n\}$. Then, observing
  that the only monomial of degree $n$ in $\fL(p_A)$ is $x_1\cdots
  x_n$, it is clear that $[x_1\cdots x_n]p_A= [x_1\cdots
  x_n]\fL(p_A)$.  Moreover, letting $q_A= \fL(p_A)[z,\ldots,z]
  \in\bZ[z]$, it is straightforward to verify that $q_A(z)=
  z^n\perm(A)+ q'_A(z)$ with $q'_A\in \bZ[z]$ such that $d(q'_A)\leq
  n-1$.  Hence, for every $s$ such that $\perm(A)/2^s<1$, we have
  $2^{s(n-1)} q_A(2^{-s})= \perm(A)/2^s + 2^{s(n-1)}q'_A(2^{-s})$
  thus, since $2^{s(n-1)}q'_A(2^{-s})\in\bN$, $\perm(A)$ can be
  obtained from the fractional part of $2^{s(n-1)} q_A(2^{-s})$.
\end{proof}

We begin with an introductory result about algorithms transforming
circuits $c\in\cC_n$ to $c'\in\cC^{\fL}_n$ such that $c'$ represents $c$
over $\bool$. It is straightforward to observe that:
\begin{pro}
  There exists $c_A\in\cC_n$ exactly representing $p_A$ of size
  $n^{O(1)}$ and constant depth.
\end{pro}
The circuit $c_A$, which clearly represents $p_A$ over $\bool$, can be
obtained from $f_A$ in a straightforward way.  Nonetheless we will
show that:
\begin{teo}
  \label{teo:weak}
  If there exists $c\in\cC^{\fL}_n$ of size $n^{O(1)}$ representing
  $p_A$ over $\bool$, then $\perm(A)$ can be computed in time
  $n^{O(1)}$ having $c$ as input.
\end{teo}
\begin{proof}
  Since $c$ exactly represents $\fL(p_A)$ by
  Proposition~\ref{pro:uniq}, then $[m]c=[m]\fL(p_A)$ for every
  monomial $m$; hence, if $\size(c)= n^{O(1)}$, by
  lemmas~\ref{lem:coeff} and~\ref{lem:rapp}, $\perm(A)$ can be
  computed, having $c$ as input, in $O(C(c)^2n\log n)= n^{O(1)}$ time,
  where $n=\dop(x_1\cdots x_n)$.
\end{proof}

Given that the computation of the permanent is a
\cnP-complete\footnote{here, as in \citep{GJ79}, \emph{completeness}
  and \emph{hardness} for \cnP are given with respect to \emph{Turing
    reductions}.}  problem \citep{Val79}, we can conclude that:
\begin{pro}
  The problem of mapping a circuit $c\in\cC_n$ to a circuit
  $c'\in\cC^{\fL}_n$ of size $n^{O(1)}$ representing $c$ over $\bool$,
  is \cnP-hard.
\end{pro}

\bigskip

We now give a stronger result by weakening the hypotheses of
Theorem~\ref{teo:weak}; in the transformation from $c\in\cC_n$ to
$c'\in\cC^{\fL}_n$, it will now be required that $c'$ represents
$\fL(c)$ only over $\{1/2\}$.

We need introduce a new polynomial, depending on $p_A$. Fix some
$s\in\bN$ and consider the substitution replacing $x_i$ with
$q_i=x_{i1}\cdots x_{is}$ for $1\leq i\leq n$; define the function
$\tilde{f}_{A,s}(x_{11}, \ldots, x_{ns})=f_A(q_1(x_{11}\cdots x_{1s}),
\ldots q_n(x_{n1}\cdots x_{ns}))$ and let $\tp_{A,s}\in\pR
\bZ{x_{11},\ldots,x_{ns}}$ be the polynomial whose evaluation
coincides with $\tilde{f}_A$ over $\bZ^{ns}$. It is straightforward to
note that $p_A[\lin qn]= \tp_{A,s}$ and that $\fL(\tp_{A,s})=
\fL(p_A[\lin qn])$, since $q_i$ and $q_j$ have no common
indeterminates if $i\not=j$. In analogy with the case of $p_A$, an
efficient representation for the polynomial $\tp_{A,s}$ can be
obviously deduced from its definition:
\begin{pro}
  There exists $\tc_A\in\cC_{ns}$ exactly representing $\tp_{A,s}$ of
  size $(ns)^{O(1)}$ and constant depth.
\end{pro}
Nonetheless, having a small circuit representing $\fL(\tp_{A,s})$ even
on only one value can lead to the computation of $\perm(A)$:
\begin{teo}
  \label{teo:main}
  If there exists $c\in\cC^{\fL}_{n^3}$ of size $n^{O(1)}$  representing
  $\fL(\tp_{A,n^2})$ over $\{1/2\}$, then $\perm(A)$ can be computed
  in $n^{O(1)}$ time having $c$ as input.
\end{teo} 
\begin{proof} 
  By the definition of $\tp_{A,s}$ it should be clear that
  $\fL(p_A)(2^{-s}, \ldots, 2^{-s})=$ \linebreak 
  $\fL(\tp_{A,s})(1/2, \ldots, 1/2)$ and, by the hypotheses of the
  theorem, $\fL(\tp_{A,s})(1/2, \ldots, 1/2)=c(1/2, \ldots, 1/2)$ and
  it can be computed, having $c$ as input, in $n^{O(1)}$ steps. Since
  $a_{ij}\in\bool$, $\perm(A)<2^{n^2}$ so, by Lemma~\ref{lem:rapp},
  $\perm(A)$ can be computed taking the fractional part of
  $2^{n^2(n-1)}c(1/2, \ldots ,1/2)$ and then multiplying by $2^{n^2}$
  which can be done in overall time $n^{O(1)}$.
\end{proof} 

\clearpage 

We are finally able to conclude with the main statement of this
chapter:
\begin{pro}
  The problem of mapping a circuit $c\in\cC_n$, exactly representing
  \linebreak 
  $p\in\pbZxn$, to a circuit $c'\in\cC^{\fL}_n$ of size $n^{O(1)}$
  representing $\fL(p)$ over $\{1/2\}$ is \cnP-hard.
\end{pro}


\RCSfooter$Id: c6s5.tex,v 2.2 1999/09/23 14:43:06 santini Exp $

\section{Conclusions}
\label{sec:coconcl}

In this chapter we have discussed an application of uniform random
generation to combinatorial optimization. As we have seen, for \cNPO
problems which are RS-good, the existance of a polynomial time \urg
leads to a polynomial time randomized approximation algorithm with a
costant performance ratio. The results of this thesis about the
existence of polynomial time \urg for several classes of $p$-relations
and languages can be applied to obtain such approximation algorithms.

On the other hand, for \cNPO problems that are also RS-derandomizable,
\ie for \cEG problems, there also exists the possibility to obtain
polynomial time deterministic approximation algorithm with a costant
performance ratio. In particular, whenever the method of conditional
expectation can be efficently applied, a ``universal'' polynomial time
approximation algorithm can be designed for every problem in \cEG.

Nonetheless, in the last section, we have shown that given a \cNPO
problem $(R,g,\goal)$ over $\Sigma=\{0,1\}$ where $g$ is a
pseudo-boolean function specified by means of arithmetic circuit of
small size, it is in general a \cnP-hard problem to obtain a
corresponding circuit to evaluate $G_\alpha$ (as defined by
equation~(\ref{eq:Galpha}) of Section~\ref{ssec:condexp}) over
$\{0,1\}$, or even over $1/2$. This negative result shows a limit to
the derandomization approach to combinatorial optimization and gives
some evidence that the abovementioned ``universal'' approach for
solving problems in \cEG cannot efficiently solve every such problem
(unless $\cFP=\cnP$).



\RCSfooter$Id: c7.tex,v 2.5 1999/11/05 10:43:31 santini Exp $

\chapter{Conclusions and Open~Problems}
\label{cha:end}

\mycite{english}{%
  \begin{verse}
    \dots\\
    so that there is no sun and no unveiling\\
    and no host\\
    only I and then the sheet\\
    and bulk dead
  \end{verse}
}%
{Samuel Beckett}{Poems in English}

\bigskip

The aim of this thesis has been to investigate, from the structural
complexity point of view, the relationships between (exact and
approximate) \emph{counting}, \emph{ranking} and \emph{uniform random
  generation} and to determine some new classes of combinatorial
structures admitting efficient approximate counting and uniform random
generation.

We have chosen the \mPrRAM as a \emph{formal model of computation} for
two main reasons.  On one side, the \mPrRAM model is polynomially
related to the classical probabilistic Turing machine model so that it
can be considered, in the usual structural complexity framework, as a
\emph{feasible} model of randomized computation. On the other side,
the \mPrRAM showed to be of sufficient \emph{high level} to let us
design the algorithms of this thesis avoiding the boaring
specifications that usually arise when dealing directly with the
Turing machine model.

In describing formally the various problems we dealt with, we have used
the notion of \emph{$p$-relation}, \emph{combinatorial structure} and
\emph{formal language} in a somehow decreasing order of generality to
suggest the broad applicability of the presented results avoiding at
the same time unnecessary technicalities in some of the given
definitions and proofs.

\bigskip

To solve the problem of random generation, the first simple idea is
related to the ranking of formal languages: if we are able to
determine a string in a formal language given its position (\ie to
solve the unranking problem for that language), the problem of random
generation of strings in such a language can be reduced to the random
generation of (bounded) integers. We have exploited this idea in the
general setting of $p$-relations, where the unranking problem is
efficiently solvable whenever the rank can be efficiently computed. We
have then extended some results about the ranking of formal languages
based on particular acceptor devices to the case of $p$-relations thus
obtaining two new classes of $p$-relations admitting an efficient
solution to the uniform random generation problem.

With regard to this approach, it remains as an open problem to
investigate other classes of $p$-relations admitting efficient
ranking. For instance, we are now interested in solving the ranking
problem for languages accepted by automata endowed with both stack and
queues, as introduced by \citet{BCC91,BCCR96}, used to describe
formally, for instance, various operating system scheduling
strategies.

\bigskip

We have also noted that there are cases in which the problem of uniform
random generation is efficiently solvable when the ranking problem is
computationally intractable. This happens, for instance, by taking the
union of some unambiguous context-free languages.  In such cases, a
very different approach still yields to an efficient solution of the
random generation and approximate counting problems. This is for
instance the case when the combinatorial structure whose elements we
want to generate uniformly at random can be, in some precise sense,
described by means of another structure which in turn admits efficient
uniform random generation.  We have applied such general framework of
(ambiguous) descriptions to some classes of formal languages, in
particular to the case of rational trace languages, languages accepted
by polynomial time nonderministic auxiliary pushdown automata (and
hence context-free languages). In each case, we have solved in an
efficient way the problem of uniform random generation and approximate
counting whenever the ambiguity of the given language is bounded
(finite, or polynomial).

It remains as an open problem to investigate further examples of
applications of the (ambiguous) descriptions framework. We are in
particular interested to approximate solutions for difficult counting
problems obtained via the use of (ambiguous) descriptions. Some
examples of the kind of problems we are interested in can be given,
for instance, by the coefficients of algebraic power series arising in
the study of formal languages~\cite{CG95}.

\bigskip

Finally, we have applied our results about random generation to an
heuristic for improving local search algorithms used in combinatorial
optimization. This application is in particular motivated from the
fact that we have proved that the derandomization of such an heuristic
can, in some case, be a \cnP-hard problem.

With regard to the latter issue, it remains as an open problem to
relate the formal results obtained here in terms of acceptor devices
to various examples of natural combinatorial optimization problems as
presented, for instance, in~\cite{CK98}.


\nocite{RI-DSI-236-99,stacs00}



%
%
\appendix
\RCSfooter$Id: a1.tex,v 2.3 1999/10/29 10:23:50 santini Exp $

\chapter{Technicalities}
\label{app:tech}

\mycite{english}{%
  It was as if these depths constantly bridged over by a structure
  that was firm enought in spite of its lightness and of its
  occasional oscillation in the somewhat vertiginous air, invited on
  occasion, in the interest of their nerves, a dropping of the plummet
  and a measurement of the abyss.}%
{Henry James}{The Beast in the Jungle}

The aim of this appendix is to separately discuss some technical and
probabilistic aspects of this thesis in order not to crowd with
details its main part.

\RCSfooter$Id: a1s1.tex,v 2.4 1999/11/12 14:14:23 santini Exp $

\section{Some Lemmas}
\label{asec:techlem}

In this section we give the proof of some technical results that are
needed in the thesis.  We start by an inequality we have often used in
proving several upper bounds:
\begin{lem}
  \label{lem:expineq}
  For every $\alpha>0$ and $\beta>0$, there exists a constant
  $\kappa>0$ such that, for every $0<\epsilon<1/\beta$ and $\delta>0$,
  it holds that
  \[
  \alpha (1-\beta\epsilon)^{\frac\kappa\epsilon \max\{1,\log(1/\delta)\}}<\delta.
  \]
\end{lem}
\begin{proof}
  By taking the logs, the above inequality becomes
  $\kappa\max\{1,\log(1/\delta)\}/\epsilon\log(1-\beta\epsilon)<
  \log(\delta/\alpha)$. Since $\log(1-x)\leq -x$, for $x\leq 1$, the
  inequality holds whenever 
  \[
  -\kappa\beta \max\{1,\log(1/\delta)\}<
  \log(\delta/\alpha).
  \] 
  Therefore, if $\kappa>(1 + \max\{0,\log\alpha\})/\beta>0$, we have
    \begin{equation*}
      \begin{split}
        \kappa &> \frac 1\beta \left(
          \frac{\log(1/\delta)}{\max\{1,\log(1/\delta)\}}
          +\frac{\max\{0,\log\alpha\}}{\max\{1,\log(1/\delta)\}}
        \right) \geq \\
        &\geq \frac 1\beta
        \frac{\log(1/\delta)+\log\alpha}{\max\{1,\log(1/\delta)\}} =
        -\frac{\log(\delta/\alpha)}{\beta\max\{1,\log(1/\delta)\}}.
        \qed
\end{split}
\end{equation*}
\end{proof}

\medskip

Now we consider two algorithms that, respectively, compute the least
common multiple (\lcm) of the first $n$ integers and the integer part
of the logarithm of $n$:
\begin{lem}
  \label{lem:lcm}
  The least common multiple ($\lcm$) of $\{1, \ldots, n\}$ is an
  integer of $O(n)$ bits and it can be computed in $O(n^2)$ time by a
  \mRAM under logarithmic cost criterion.
\end{lem}
\begin{proof}
  First, we recall that $\lcm\{1, \ldots, n\}\leq n^{\pi(n)}$,
  where $\pi(n)$ is the $n$-th prime~\citep[Lemma~4.1.2.]{Sze94}. As a
  consequence, $\lcm\{1, \ldots, n\}$ has $O(n)$ bits since it is
  well known that $\pi(n)\sim n/\log n$.  Moreover, a na\"ive
  iterative procedure, based on Euclid's algorithm for computing the
  g.c.d.\, can be designed to compute $\lcm\{1, \ldots, n\}$ in
  $O(n^2)$ time.
\end{proof}

\begin{lem}
  \label{lem:bit}
  The number $\lceil\log N\rceil$ of bits required to represent
  $\{1,\ldots,N\}$ can be computed in $O(\log N\log\log N)$ by a
  \mRAM under logarithmic cost criterion.
\end{lem}
\begin{proof}
  We compute $\lceil\log N\rceil$ by the call Size$(N-1)$, where
  ``Size'' is the recursive procedure given by
  Algorithm~\ref{alg:bit}.

  \begin{alg}[ht]
    \caption{Computing the bit size of an integer.}
    \label{alg:bit}
    \begin{myprg*}
      \sprocedure Size$(n)$ \nl
      \sif $n\leq 1$ \sthen \sreturn $1$ \nl
      \selse \tnl
      $h\stv 1$, $h_0\stv h$ \nl
      $k\stv 2$, $k_0\stv k$ \nl
      \swhile $k\leq n$ \sdo \tnl
      $h_0\stv h$, $h\stv h+h$\nl
      $k_0\stv k$, $k\stv k\cdot k$ \unl
      \sreturn $h_0+$Size$( n/k_0 )$.
    \end{myprg*}
  \end{alg}
  
  A simple analysis shows that the time complexity of Size$(n)$ is
  given by $O(\sum_{i=1}^k i2^i)=O(k2^k)$, where $k=\log\log n$.
\end{proof}


\RCSfooter$Id: a1s2.tex,v 2.3 1999/09/23 14:43:06 santini Exp $

\section{Probabilisitc Detour}
\label{asec:prob}

This section is intended to give a separate discussion of the
probabilistic aspects involved in the thesis. In this way, we hope to
give a more precise and detailed evidence of some of the properties
used in order to prove the correctness of algorithms we have presented
in the previous chapters.

\subsection{Hoeffding's inequality}
\label{assec:hoeff}

\emph{Hoeffding's inequality}~\citep{Hof63} is the main tool used in
this work to prove concentration results, \ie to show that the sum of
certain random variables has, with high probability, a value very near
to its expectation. This inequality is based on the so called
\emph{Chernoff's bounding} technique~\citep{Che52} that uses an
exponential version of \emph{Markov's inequality}. Here, we want to
give a concise proof of such an inequality since is based on a very
simple and intuitive idea we would like to elucidate.

\bigskip

Let $\left<\Omega, \cF, P\right>$ be a probability space, and consider
a random variable $X$ on it, \ie a measurable function
$X:\Omega\to\bR$. The \emph{expectation} of $X$, denoted by $\Ex[X]$,
is defined as
\[
\Ex[X]=\int_\Omega XdP
\]
where the integral is understood in the Lebesgue sense (for a detailed
discussion of these details, see, for instance~\citep{Bil79,Fel68}).
If, for every $\omega\in\Omega$, it holds that $X(\omega)>0$, then,
for every $t>0$, 
\[
\int_\Omega XdP \geq 
\int_{X\geq t} tdP + \int_{X\leq t} XdP 
\geq t P\{X\geq t\} + 0.
\]
Hence, we have proven \emph{Markov's inequality}:
\[
P\{X\geq t\}\leq \frac{\Ex[X]}t.
\]

Now, by taking the exponential, we can apply this inequality for every
random variable $Y$, even those assuming negative values: for every
$\alpha>0$, we have
\[
P\{Y\geq t\}=
P\left\{ e^{\alpha Y}\geq e^{\alpha t} \right\}\leq
\frac{\Ex[ e^{\alpha Y} ]}{e^{\alpha t}};
\]
Chernoff's bounding idea consists in chooseing $\alpha$ such that the
right hand side of the previous inequality is minimized. We can apply
the inequality to a sum of independent random variables $X_1, \ldots,
X_n$:
\begin{equation}
  \label{eq:hoeff}
  \begin{split}
  P\Bigl\{ \bigl(\sum_{i=1}^n X_i -\Ex\bigl[\sum_{i=1}^n X_i\bigr]\bigl) \geq t \Bigr\} 
  &\leq e^{-\alpha t}\Ex\Bigl[ \exp\bigl( \alpha\sum_{i=1}^n (X_i -\Ex[X_i]) \bigr)\Bigr] \\
  &= e^{-\alpha t}\prod_{i=1}^n\Ex\Bigl[ e^{\alpha(X_i -\Ex[X_i])} \Bigr],
  \end{split}
\end{equation}
where the equality follows from the independence of the $X_i$'s.  In
order to choose the right $\alpha$, we need an upper bound for
$\Ex[e^{\alpha Y}]$ (where, in this case, $Y=X_i -\Ex[X_i]$).  Assume
that $Y$ is bounded, \ie that there exist some $a,b\in\bR$ such that
$P\{ a\leq Y\leq b \}=1$; then, by convexity of the exponential
function,
\[
\Ex\left[ e^{\alpha(Y-\Ex[Y])} \right] \leq \frac{b}{b-a}e^{\alpha
  a}-\frac{a}{b-a}e^{\alpha b}= e^{g(u)}
\]
where $u=\alpha(b-a)$ and $g(u)=-pu+\log(1-p+pe^u)$ with
$p=-x/(b-a)$. Is easy to verify that $g(0)=g'(0)=0$ and that
$g''(u)\leq 1/4$. Hence by Taylor's expansion, for a suitable
$\theta\in\bR$,
\[
g(u)=g(0)+ug'(0)+\frac{u^2}{2}g''(\theta)\leq\frac{u^2}{8}
\]
and we are able to conclude that
\[
\Ex\left[ e^{\alpha Y} \right] \leq e^{\alpha\Ex[Y]+\alpha^2(b-a)^2/8}.
\]

We are ready to give a proof of Hoeffding's inequality. Let $X_1,
\ldots, X_n$ be independent random variables such that $P\{X_i\in
[a_i,b_i]\}=1$ for $1\leq i\leq n$; then, for every $\epsilon>0$, by
applying equation~(\ref{eq:hoeff}) and taking $\alpha=
4\epsilon/\sum_{i=1}^n(b_i-a_i)^2$, we have
\[
P\Bigl\{\sum_{i=1}^n X_i -\Ex\bigl[\sum_{i=1}^n X_i\bigr] \geq \epsilon \Bigr\}\leq
\exp\Bigl({-\frac{2\epsilon^2}{\sum_{i=1}^n(b_i-a_i)^2}}\Bigr).
\]
In a similar way, we can obtain that
\[
P\Bigl\{\sum_{i=1}^n X_i  \Ex\bigl(\sum_{i=1}^n X_i\bigr) \leq -\epsilon \Bigr\}\leq 
\exp\Bigl({-\frac{2\epsilon^2}{\sum_{i=1}^n(b_i-a_i)^2}}\Bigr).
\]

\bigskip

We conclude this section by stating two lemmas that restate
Hoeffding's inequality in a form which is used in some of the proofs
of this thesis. Here, we use the notation $\Pr\{A\}$ to denote the
probability of an event $A$ whenever we do not want to formally
specify the probability space over which the event itself is defined.
\begin{lem}
  \label{lem:hoefbinom}
  Let $X,X_1, \ldots ,X_n$ be independent, identically distributed,
  binomial random variables such that $\Pr\{ X=1 \}=1/2+\delta$, for
  $1/2<\delta<1$. Then,
  \[
  \Pr\Bigl\{ \sum_{i=1}^n X_i \leq n/2 \Bigr\}<e^{-2n\delta^2}.
  \]
\end{lem}
\begin{proof}
  The statement follows by taking $b_i-a_i=1$, for $1\leq i\leq n$,
  and replacing $\epsilon$ by $n\delta$.
\end{proof}
\begin{lem}
  \label{lem:hoefrel}
  Let $X,X_1, \ldots ,X_n$ be independent and identically distributed
  random variables taking values in $[0,1]$. Then, for every
  $\epsilon>0$,
  \[
  \Pr\Bigl\{
  (1-\epsilon)E(X) \leq \sum_{i=1}^n X_i/n\leq (1+\epsilon)E(X)
  \Bigr\}>1-2e^{-2n(E(X)\epsilon)^2}
  \]
\end{lem}
\begin{proof}
  The claim follows by taking $b_i-a_i=1$, for $1\leq i\leq n$, and
  replacing $\epsilon$ by $n\epsilon E(X)$.
\end{proof}

\subsection{Improving randomized algorithms}
\label{assec:impr}

As we have seen, randomized algorithms sometimes fail to give the
correct answer and somehow signal this fact, let us say by giving an
``undefined'' output denoted by the special symbol \no. A common
strategy to decrease the probability of such an event, is to iterate
the algorithm, usually for no more than a specified number of times,
and to give as output the first non-undefined output, or some other
function of the non-undefined outputs obtained in the iterations (or
to output \no if, for all the specified number of iterations, the
algorithm always outputs \no).

Here, we try to formalize this idea only with regard to probabilisty
issues.  For this reason, all the interpretations of the probabilistic
statements of this subsection in terms of randomized algorithms is
printed in a smaller font.

\bigskip

Let $X, X_1, \ldots, X_n$ be independent and identically distributed
random variables taking values in a set $\bX\subseteq\bR$ and let
$K\subseteq\bX$ be a fixed measurable set.

\sex{Clearly, our idea is to use $X$ to represent the output of a
  randomized algorithm and $K$ to mean the set of its intended output.
  Usually, in this thesis, $K=\bN$ and \bX is as simple as
  $K\cup\{\no\}$, where we assume to represent \no in \bR by some
  number not in $K$. Finally, the $X_i$'s represent the output of $n$
  iterations of the algorithm.}

Let now $Y, Y_1, \ldots, Y_n$ be independent and identically
distributed random variables taking values in \bX such that
\begin{equation}
  \label{eq:intdist}
  \Pr\{ Y\in H\}=\Pr\{ X\in H \mid X\in K \}
\end{equation}
for every measurable set $H\subseteq K$.

\sex{Again, our idea is to use $Y$ to represent the intended output of
  the algorithm, together with its desired distribution.
  Equation~(\ref{eq:intdist}), in fact, formally states that $Y$
  should be distributed like $X$ whenever (conditionally to) $X$ is
  different from \no. The same, due to the identity of the
  distribution, is true for the $Y_i$'s.}

We now show that, for every measurable function $F:K^n\to\bR^m$
and every measurable set $M\in\bR^m$, it holds that
\begin{equation}
  \label{eq:samedist}
  \Pr\{ F(X_1, \ldots, X_n)\in M \mid X_1\in K, \ldots, X_n\in K \}
  =\Pr\{ F(Y_1, \ldots, Y_n)\in M \}.
\end{equation}
For the sake of brevity, let
\[
F^{-1}_i(M)=\{ x : F(y_1, \ldots, y_{i-1}, x, y_{i+1}, 
\ldots, y_n)\in M, y_1, \ldots, y_n\in K \} 
\] 
for $1\leq i\leq n$. Then
\begin{equation*}
  \begin{split}
    \Pr\{ F(X_1, \ldots, X_n) \in M \mid  X_1\in K, \ldots, X_n\in K  \}
    &= \frac{\Pr\{ \bigcap_{i=1}^n X_i\in F^{-1}_i(M) \cap  \bigcap_{i=1}^n X_i\in K \}}
      {\Pr\{ X\in K \}^m} \\
    &= \prod_{i=1}^n \frac{ \Pr\{ X_i\in F^{-1}_i(M) \cap X_i\in K  \}}{\Pr\{ X\in K \}} \\
    &= \prod_{i=1}^n \Pr\{ X\in F^{-1}_i(M) \mid X\in K \} \\
    &= \prod_{i=1}^n \Pr\{ Y\in F^{-1}_i(M) \} \\
    &= \Pr\{ \bigcap_{i=1}^n Y_i\in F^{-1}_i(M) \} \\
    &= \Pr\{ F(Y_1, \ldots, Y_n) \in M  \}
  \end{split}
\end{equation*}

\sex{Equation~(\ref{eq:samedist}) states that the probability of every
  event $F(X_1, \ldots, X_n)\in M$ obtained taking some function $F$
  of the outputs of $n$ iterations of a probabilistic algorithm, given
  that all the iterations are non-undefined, is equal to the
  probability of the same event $F(Y_1, \ldots, Y_n)\in M$, expressed
  in terms of the intended outputs distribution.}

\subsubsection{Elementary case}

\sex{Here we investigate the first strategy mentioned at the beginning
  of this section: to iterate the algorithm, giving in output the
  first non-undefined output so obtained.}

Under the previous hypotheses, let $\tau_K=\inf\{ t : 1\leq t\leq n,
X_t\in K\}$ (assuming, as usual, $\inf\emptyset = \infty$). Then, for
every measurable set $H\subseteq K$, it holds that
\begin{enumerate}[(i)]
\item \label{it:dist} $\Pr\{ X_{\tau_K} \in H \mid \tau_K <\infty \} = \Pr\{ X\in H
  \mid X\in K\}=\Pr\{Y\in H\}$,
\item \label{it:succ} $\Pr\{ \tau_K=\infty \}=\Pr\{ X\not\in K\}^n$.
\end{enumerate}
The second property is trivially verifiable; the first comes from
\begin{equation*}
  \begin{split}
    \Pr\{ X_{\tau_K} \in H \mid \tau_K <\infty \}
    &= \frac{\Pr\{  X_{\tau_K} \in H \cap \bigcup_{i=1}^n \tau_K=i \}}
    {\sum_{i=1}^n \Pr\{ \tau_K=i\}} \\
    &= \frac{\sum_{i=1}^n \Pr\{  X_i \in H \cap \tau_K=i \}}
    {\sum_{i=1}^n \Pr\{ \tau_K=i\}} \\
    &= \frac{\sum_{i=1}^n \Pr\{ \bigcap_{j=1}^{i-1} X_j\not\in K \cap X_i\in K \cap X_i\in H\}}
    {\sum_{i=1}^n \Pr\{ \bigcap_{j=1}^{i-1} X_j\not\in K \cap X_i\in K \}} \\
    &= \frac{ \Pr\{ X\in H\}\sum_{i=1}^n \Pr\{ \bigcap_{j=1}^{i-1} X_j\not\in K \}}
    {\Pr\{X\in K\}\sum_{i=1}^n \Pr\{ \bigcap_{j=1}^{i-1} X_j\not\in K \}} \\
    &= \Pr\{ X\in H \mid X\in K\}.
  \end{split}
\end{equation*}

\sex{Property~(\ref{it:dist}) tells that if one of the iterations of
  the algorithm is ever successful, then that output has the same
  distribution as the intended one. Moreover, Property~(\ref{it:succ})
  says that the probability that such ``iteration strategy'' has an
  undefined output decreases exponentially with the number of
  iterations.}

\subsubsection{A more complex situation}

\sex{Now we assume that, after a certain number of iterations of a
  randomized algorithm, we want to return some function of the outputs
  of all the non-undefined iterations, such as, for instance, their
  median, or mean.}

Under the previous hypotheses, given a family of \bR valued measurable
functions $\{f_n\}_{n>0}$, let $\{f^K_n\}_{n>0}$ be the corresponding
family of functions defined as
\[
f^K_n(x_1, \ldots, x_n)=f_m(x_{i_1}, \ldots, x_{i_k})
\]
where $i_1< \cdots< i_k$ are exactly all the indexes for which
$x_{i_m}\in K$ with $1\leq k\leq m\leq n$, for every $n>0$. Observe
that the functions $f^K_n$ are obviously all \bR valued measurable
functions, for $n>0$.

\sex{Clearly, $f^K_n$ is the function computed at the end of the $n$
  iterations on all the outputs, that, by definition, computes the
  function $f_k$ (for instance, the mean, or the median) only on the
  $k$ non-undefined outputs of all the iterations.}

Let $\eta_K=|\{ t : 1\leq t\leq n, X_t\in K\}|$. Then for every
measurable set $H\subseteq K$ and every $k>0$, it holds that
\begin{equation*}
  \begin{split}
    \Pr\{ f^K_n(X_1, \ldots, X_n)\in H \mid \eta_K=k \} 
    &=\Pr\{ f_k(X_1, \ldots, X_k)\in H \mid X_1\in K, \ldots, X_k\in K \} \\
    &=\Pr\{ f_k(Y_1, \ldots, Y_k)\in H \}.
\end{split}
\end{equation*}

\sex{This property formally states that the above mentioned strategy
  is ``correct'', in the sense that the value of $f^K_n$ on all
  outputs, given there were $k$ non-undefined outputs, is distributed
  as $f_k$ computed over $k$ intended outputs of the algorithm.}

The second equality comes from equation~(\ref{eq:samedist}). To prove
the first inequality, define, for the sake of brevity, the events
\[
A(\Gamma)=\bigcap_{i\in \Gamma} X_i\in K \qquad\text{e}\qquad 
B(\Gamma)=\bigcap_{1\leq i\leq n, i\not\in\Gamma} X_i\not\in K;
\]
where $\Gamma=(i_1, \ldots, i_k)$ is any ordered $k$-tuple of indexes
without repetitions, \ie $1\leq i_1<\cdots<i_k\leq n$ and $1\leq
k\leq n$. Then,
\begin{equation*}
  \begin{split}
    \Pr\{ f^K_n(X_1, \ldots, X_n) \in H \mid \eta_K = k\}
    &= \frac{\Pr\{   f^K_n(X_1, \ldots, X_n) \in H \cap 
      \bigcup_{\Gamma:|\Gamma|=k}A(\Gamma)\cap B(\Gamma)  \}}
    {{n\choose k} \Pr\{ X\in K \}^k  \Pr\{ X\not\in K \}^{n-k}}    \\
    &= \frac{\sum_{\Gamma=(i_1,\ldots, i_k)}  
      \Pr\{ f_k(X_{i_1}, \ldots, X_{i_k}) \in H \cap A(\Gamma)\}\Pr\{ B(\Gamma) \}}
    {{n\choose k} \Pr\{ X\in K \}^k  \Pr\{ X\not\in K \}^{n-k}}    \\
    &= \frac{{n\choose k} \Pr\{ f_k(X_1, \ldots, X_k) \in H \cap A(\{1, \ldots, k\})\}
      \Pr\{ X\not\in K \}^{n-k} }
    {{n\choose k} \Pr\{ X\in K \}^k  \Pr\{ X\not\in K \}^{n-k}}    \\
    &= \frac{\Pr\{ f_k(X_1, \ldots, X_k) \in H \cap A(\{1, \ldots, k\})\}}{\Pr\{ X\in K \}^k} \\
    &= \Pr\{ f_k(X_1, \ldots, X_k)\in H \mid X_1\in K, \ldots, X_k\in K \}.
  \end{split}
\end{equation*}

\subsubsection{An upper bound}

\sex{Usually, we would like to prove some concentration result about
  the output obtained applying $f^K_n$. To this aim, we assume to have
  an (exponential) upper bound on the probability of the concentration
  of $f_k$ applied to $k$ intended outputs of the algorithm.  Hence,
  we show a corresponding (exponential) bound on the concentration of
  $f^K_n$.}

Under the previous hypotheses, define $s_K=\Pr\{ X\in K\}$ and, for a
given measurable set $H\subseteq K$, assume that there exists a
$0<c_H<1$ such that, for every $k>0$,
\[
\Pr\{ f_k(Y_1, \ldots, Y_k)\in H\}\leq e^{-kc_H}.
\]
Then, it holds that
\begin{equation}
  \label{eq:conc}
  \Pr\{ f^K_n(X_1, \ldots, X_n) \in H \mid \eta_K > 0\}
  < \frac 1{s_K} \left( 1- \frac {s_kc_H}2 \right)^n.
\end{equation}
Observe, in fact, that
\begin{equation*}
  \begin{split}
    \Pr\{ f^K_n(X_1, \ldots, X_n) \in H \mid \eta_K > 0\} 
    &= \frac{\Pr\{ f^K_n(X_1, \ldots, X_n) \in H \cap \eta_K > 0\}}
    { \Pr\{ \eta_K > 0 \} } \\
    &= \frac{\sum_{k=1}^n \Pr\{ f^K_n(X_1, \ldots, X_n) \in H \mid \eta_K = k\}\Pr\{ \eta_K = k\}}
    { 1-(1-\Pr\{ X\in K\})^n } \\
    &\leq \frac{\sum_{k=1}^n e^{-kc_H}{n \choose k}\Pr\{ X\in K\}^k\Pr\{X\not\in K\}^{n-k}}
    { 1-(1-\Pr\{ X\in K\})^n } \\
    &< \frac{\sum_{k=0}^n {n \choose k}(e^{-c_H}Pr\{ X\in K\})^k\Pr\{X\not\in K\}^{n-k}}
    { 1-(1-\Pr\{ X\in K\})^n } \\
    &= \frac{\left( 1-\Pr\{X\in K\}\left( 1 - e^{-c_H}\right)\right)^n}
    { 1-(1-\Pr\{ X\in K\})^n } \\
    &\leq \frac{\left( 1-\Pr\{X\in K\}c_H/2\right)^n}
    { 1-(1-\Pr\{ X\in K\}) } \\
  \end{split}
\end{equation*}
where the last inequality follows from the facts that $1-e^{-x}\geq
x/2$ and $x^n\leq x$ whenever $x\in[0,1]$ and $n>0$.

\subsection{The method of conditional expectation}
\label{assec:condexp}

We conclude this appendix with a detailed discussion, from the
probabilistic viewpoint, of the so called \emph{method of conditional
  expectation}.  Let $X_1, \ldots, X_n$ be independent random
variables (not necessarily identically distributed); moreover, to
avoid boring technical details concerning the existence of the
conditional expectation used in the following, assume that each $X_i$
takes values in some discrete set\footnote{this is usually stated by
  sayng that the $X_i$'s are \emph{simple} random variables.}
$\bX\subset\bR$, for $1\leq i\leq n$. Finally, let $f:\bR^n\to\bR$ be
some fixed function.

If we define $Z$ as the random variable $Z(\omega)=f(X_1(\omega),
\ldots, X_n(\omega))$, then, by the \emph{internality} of the
expectation,
\[
\inf_{\omega\in\Omega} Z(\omega) \leq \Ex[Z] \leq \sup_{\omega\in\Omega} Z(\omega).
\]
In our hypotheses, this means in particular that there exists some
$\tilde{\omega}\in\Omega$ such that $\Ex[Z]\leq Z(\tilde{\omega})$,
or, that there exist $\tilde{x}_i=X_i(\tilde{\omega})\in\bR$, for
$1\leq i\leq n$, such that
\[
\Ex[f(X_1, \ldots, X_n)]\leq f(\tilde{x}_1, \ldots, \tilde{x}_n ).
\]
The aim of the method of conditional expectation is to obtain such
$\tilde{x}_i$'s by computing some conditional expectations.

\medskip

For the sake of brevity define, for $1\leq k\leq n$, the functions
$g_k:\bR^k\to\bR$ such that
\[
g_k( x_1, \ldots, x_k ) = \Ex[ Z \mid X_1=x_1,\ldots, X_k=x_k]. 
\]
By a straightforward application of the definition of conditional
expectation and the elementary property of iterated expectations, one
can verify that the following statements are true, for every $x_1,
\ldots, x_n\in\bR$ and $1\leq k<n$:
\begin{enumerate}[(a)]
\item $\Ex[Z] = \Ex[g_1(X_1)]$, 
\item $g_k( x_1, \ldots, x_k ) = \Ex[ g_{k+1}( x_1, \ldots, x_k, X_{k+1} ) ]$ and 
\item \label{it:first} $g_n( x_1, \ldots, x_n ) = f( x_1, \ldots, x_n )$, 
\item $\Ex[Z]\leq\max_{x\in\bX} g_1(x)$ and 
\item \label{it:last} $g_k( x_1, \ldots, x_k )\leq \max_{x\in\bX} g_{k+1}( x_1, \ldots, x_k, x )$.
\end{enumerate}

Hence, if the $\hat{x}_i$'s are recursively defined as
\[
\hat{x}_{i}=
\begin{cases}
  \argmax_{x\in\bX} g_1(x) & \text{if $i=1$,} \\
  \argmax_{x\in\bX} g_i( \hat{x}_1, \ldots, \hat{x}_{i-1}, x) & \text{if $i>1$,}
\end{cases}
\]
from~(\ref{it:first})--(\ref{it:last}), one can conclude that
$\Ex[Z]\leq f( \hat{x}_1, \ldots, \hat{x}_n )$.



\RCSfooter$Id: a2.tex,v 2.2 1999/09/23 14:43:06 santini Exp $

\chapter*{Symbols and Abbreviations}
\label{app:symabbr}

In this appendix we summarize some notaion, together with some pointer
to relevant literature for the notions used in this thesis that are
not explicitly defined in other chapters.

\RCSfooter$Id: a2s1.tex,v 2.2 1999/09/22 17:34:19 santini Exp $

\section*{Some Complexity and Approximation Classes}
\label{asec:complclass}

\begin{list}{}{\settowidth{\labelwidth}{\cFPTAS\quad}}
  
\item[\cP] languages recognizable in polynomial time by a
  (deterministic) Turing machine~\citep{GJ79}.

\item[\cFP] functions computable in polynomial time by a
  (deterministic) Turing machine~\citep{GJ79}.

\item[\cNP] languages recognizable in polynomial time by a
  nondeterministic Turing machine~\citep{GJ79}.

\item[\cnP] functions yelding the number of accepting computations in
  nondeterministic polynomial time Turing machines~\citep{Val79}.
  
\item[\cRP] languages recongizable in probabilistic polynomial time
  with one-sided error~\citep{BDG95}.
  
\item[$\cNC^K$] languages recognized (functions computed by) log-space
  uniform boolean circuits of bounded fan-in, polynomial size and
  $O(\log^k n)$ depth~\citep{KR90,BDG90}.

\item [$\cB$] ($\cB(s(n),i(n),d(n))$) class of languages recognized by
  Turing machines with simultaneous complexity bounds
  (Section~\ref{sec:sbtm}).

\item [\cNPO] \cNP optimization problems (Section~\ref{sec:codefs}).

\item [\cGLO] \cNPO problems with guaranteed local optima
  (Section~\ref{ssec:lsnpo}).
  
\item [\cEG] (and RS-good, RS-derandomizable) expectation-guaranteed
  \cNPO problems and related subclasses (Section~\ref{sec:ge}).

\item [\cAPX] \cNPO problems approximable in polynomial time within a
  costant factor (Section~\ref{ssec:approxalgclass}).

\item [\cPTAS] \cNPO problems admitting a polynomial time
  approximation scheme (Section~\ref{ssec:approxalgclass}).

\item [\cFPTAS] \cNPO problems admitting a fully polynomial time
  approximation scheme (Section~\ref{ssec:approxalgclass}).

\item [\cMaxNP] (and \cMaxSNP) logically defined \cNPO problems
  (Section~\ref{ssec:lsnpo}).

\end{list}


\RCSfooter$Id: a2s2.tex,v 2.1 1999/09/22 17:34:19 santini Exp $

\section*{Models of Computation}
\label{asec:compmod}

\begin{list}{}{\settowidth{\labelwidth}{\moAuxPDA\quad}}

\item [\mRAM] random access machine (Section~\ref{ssec:rammodel}).
\item [\mPrRAM] probabilistic \mRAM (Section~\ref{ssec:prrammodel})
  
\item [\mTM] (\mNTM) (nondeterministic) Turing machine
  (Section~\ref{sec:sbtm}).
  
\item [\moAuxPDA] (\moNAuxPDA, \moUAuxPDA) (nondeterministic,
  unambibuous) one-way auxiliary pushdown automata (Section~\ref{ssec:pda}).

\end{list}


\RCSfooter$Id: a2s3.tex,v 2.1 1999/09/22 17:34:19 santini Exp $

\section*{Number Rings and Fields}
\label{asec:numringfld}

\bN, \bZ, \bQ and \bR denote, respectively, the ring of \emph{integer}
and \emph{natural} number and the field of \emph{rational} and
\emph{real} numbers.


\RCSfooter$Id: a2s4.tex,v 2.2 1999/09/23 14:43:06 santini Exp $

\section*{Basic notation}
\label{asec:banot}

\begin{list}{}{\settowidth{\labelwidth}{\cFPTAS\quad}}

\item [$|\omega|$] the lenght of the string $\omega$  
\item [$\# A$]  the cardinality of set $A$ 

\end{list}



%
%
\backmatter
\backmatterpagestyle

\listofalgorithms
\listoffigures
\listoftables

\bibliographystyle{abbrvnat}
\bibliography{theory,formlang,combopt,rndgen,prob,personal}

\printindex
\printindex[aut]

\iffinal{\RCSfooter$Id: b.tex,v 2.3 1999/09/23 14:43:06 santini Exp $

\thispagestyle{empty}

\mbox{}\vfill

{\small 
\noindent  This thesis was written using \emph{free} and
\emph{open source} software only.  In particular,
  \begin{enumerate}[(i)]
  \item Linux, the operating system, mainly due to \textsl{Linus
      Torvalds};
  \item \TeX, together with \LaTeX\ and \AmS-\LaTeX,  the typesetting
    software, respectively and mainly due to \textsl{Donald Knuth},
    \textsl{Leslie Lamport} and the \textsl{American Mathematical
      Society};
  \item \{X\}Emacs, together with AUCTeX, RefTeX and XFig, the
    typesetting editor and environment, respectively and mainly due to
    \textsl{Richard Stallmann}, \textsl{Kresten Krab Thorup},
    \textsl{Carsten Dominik} and \textsl{Supoj Sutanthavibul}.
  \end{enumerate}
  I would like to thank all of them and the people of the free and
  open source community for giving us such an incredible effort in
  developing extraordinary pieces of software. }


}

\end{document}
